\documentclass[aps,pra,reprint,superscriptaddress,amsmath,amssymb,
               longbibliography,floatfix]{revtex4-2}
\usepackage[T1]{fontenc}
\usepackage[utf8]{inputenc}
\usepackage{bm}
\usepackage{graphicx}
\usepackage{microtype}
\usepackage{mathtools,mathrsfs,booktabs,array}
\usepackage{xcolor}

\usepackage[
  colorlinks=true,
  linkcolor=blue,
  citecolor=blue,
  urlcolor=blue
]{hyperref}

\usepackage{amsthm}
\usepackage{placeins}
\DeclareMathOperator{\Tr}{Tr}
\DeclareMathOperator{\rank}{rank}
\DeclareMathOperator{\Span}{span}
\newcommand{\Mmap}{\mathcal M}
\newcommand{\Umap}{\mathcal U}
\newcommand{\Cmap}{\mathcal C}
\newcommand{\HS}{\mathrm{HS}}
\newcommand{\R}{\mathbb R}

\newtheorem{theorem}{Theorem}
\newtheorem{proposition}{Proposition}

\theoremstyle{plain}
\newtheorem{suppthm}{Theorem}[section]
\newtheorem{suppprop}[suppthm]{Proposition}

\theoremstyle{definition}
\newtheorem{suppassumption}[suppthm]{Assumption}

\theoremstyle{remark}

\theoremstyle{plain}

\begin{document}

\title{The power of quantum reservoir computing}

\author{Minwoo James Kim}
 \affiliation{Dept. of Computer Science and Engineering, Seoul National University, Seoul 08826, South Korea}
 \affiliation{NextQuantum Center, Seoul National University, Seoul 08826, South Korea}
\affiliation{Automation and Systems Research Institute, Seoul National University, Seoul 08826, South Korea}

\author{Jeongho Bang\textsuperscript{*}}
\affiliation{Institute for Convergence Research and Education in Advanced Technology, Yonsei University, Seoul 03722, Republic of Korea}
\affiliation{Department of Quantum Information, Yonsei University, Incheon 21983, South Korea}
\thanks{Corresponding authors: \href{jbang@yonsei.ac.kr}{jbang@yonsei.ac.kr} and \href{taehyun@snu.ac.kr}{taehyun@snu.ac.kr}}

\author{Taehyun Kim\textsuperscript{*,}}
\affiliation{Dept. of Computer Science and Engineering, Seoul National University, Seoul 08826, South Korea}
\affiliation{NextQuantum Center, Seoul National University, Seoul 08826, South Korea}
\affiliation{Automation and Systems Research Institute, Seoul National University, Seoul 08826, South Korea}
\affiliation{Institute of Computer Technology, Seoul National University, Seoul 08826, South Korea}
\affiliation{Institute of Applied Physics, Seoul National University, Seoul 08826, South Korea}
\thanks{Corresponding authors: \href{jbang@yonsei.ac.kr}{jbang@yonsei.ac.kr} and \href{taehyun@snu.ac.kr}{taehyun@snu.ac.kr}}
 
\date{November 28, 2026}
\begin{abstract}
Quantum reservoir computing is motivated by the expectation that large quantum operator spaces can provide useful temporal features for classical learning. We develop an operator-valued Volterra framework separating temporal feature generation from accessibility under finite measurement resources. For a specified temporal truncation, latent, visible and operational dimensions distinguish generated directions, algebraic transmission and finite-resolution access. Visibility geometry and an exact alignment law show how poor alignment of the latent space with the measurable operator space, together with small relative spectral spread, can concentrate visibility angles at small values and weaken measurement strengths despite preserved algebraic visibility. Bounds on linear reconstruction capacity connect this geometry to learning under specified response statistics and measurement noise, while optimal observable selection identifies the strongest geometric access through the readout register. Numerical benchmarks relate measurement access to temporal learning performance. The framework guides reservoir design toward features that remain resolvable within available resources.
\end{abstract}
\maketitle

\section{Introduction}

Characterizing the computational power of a learning model is essential for understanding which unknown functions it can approximate and what resources that approximation requires. Such a framework connects model structure to representable functions, providing a basis for comparing architectures and identifying their limitations. In classical neural networks, approximation theory has established conditions for approximating broad function classes and quantified how accuracy depends on network size and structure~\cite{Barron1993,Yarotsky2017,Gonon2023}.

This characterization is particularly important for quantum neural networks and quantum machine learning models, where a large quantum state space does not by itself establish useful predictive power. For function approximation, the relevant question is how quantum encoding, dynamics and measurement determine the input--output functions that a model can express~\cite{Schuld2021,Abbas2021}. Studies of parametrized quantum circuits have revealed constraints including exponentially vanishing gradients and concentration phenomena, which can obstruct optimization or make input-dependent information costly to resolve~\cite{McClean2018,Thanasilp2024}. For temporal quantum models, a corresponding characterization must also account for how information about past inputs is retained, transformed and exposed to measurement. Understanding these elements together remains an important challenge.

Quantum reservoir computing~(QRC) provides a framework for learning such temporal relationships. Following classical reservoir computing~\cite{Jaeger2001,Maass2002,Verstraeten2007}, an input stream drives a dynamical system whose state depends on the input history, and measured features are supplied to a trained classical output layer. In QRC, the reservoir is a quantum system, and its measured responses provide nonlinear functions of present and past inputs~\cite{Fujii2017,Nakajima2019,Ghosh2019,Chen2020}. The reservoir dynamics are held fixed during readout training, avoiding the need to optimize quantum circuit parameters through a potentially difficult quantum training landscape. This separation has motivated interest in QRC for memory and prediction tasks using quantum dynamics as a temporal feature generator.

For a linear readout, the measured features span the temporal functions available for approximating an unknown target. Universality results establish that suitable quantum reservoir families can approximate broad classes of fading-memory maps~\cite{Chen2020}, but leave open the expressive capability of a particular reservoir under fixed resources. Memory and information-processing capacities quantify the reconstruction of prescribed temporal targets~\cite{Jaeger2002,Dambre2012,MartinezPena2023b}. These capacities are evaluated under a specified input distribution, with task-resolved scores additionally depending on the chosen target functions. Resolvable expressive capacity and eigentasks further quantify the functions recoverable in the presence of sampling noise~\cite{Hu2023REC}. These results motivate a complementary structural characterization that separates feature generation from measurement access and connects this geometry to noise-limited learning.

Other approaches identify how the components of a quantum reservoir shape its computational capability. Sch{\"u}tte et al. characterize temporal expressivity through a multidimensional Fourier representation, showing how input encoding and sequential injection constrain the functions available in the readout, and connect this description to finite-shot capacity and reservoir size~\cite{Schutte2025}. Quantum Volterra theory also describes temporal responses in measurement--reset architectures~\cite{Hu2024}. Echo-state and fading-memory analyses establish conditions for stable dependence on input history~\cite{MartinezPena2023a,MartinezPena2025}, while Krylov-based methods characterize dynamical complexity and observability~\cite{Balasubramanian2022,Domingo2024,Cindrak2024,Cindrak2025a,Cindrak2025b}. These studies reveal complementary characteristics and limitations of QRC, showing how its computational capability depends on input encoding, memory and dynamical structure. They also motivate asking how these ingredients must work together for a larger reservoir to provide a richer representation to the learner.

A closely related operator-space perspective has been developed for quantum extreme learning machines (QELMs). Gross and Rieser use a Pauli transfer matrix representation to distinguish encoding-generated features from their transmission through reservoir evolution and measurement, and connect geometric and statistical decodability to nonlinear information-processing capacity~\cite{GrossRieser2026}. Their framework considers initial-state encoding followed by input-independent reservoir evolution, with possible temporal multiplexing. For recurrent reservoirs with persistent memory and repeated input injection, the temporal feature library itself is generated by the full input-dependent recurrence.

Increasing reservoir size or dynamical complexity is motivated by the large operator spaces available to quantum systems~\cite{Tran2020,Ahmed2024,Hu2024}. The computational benefit of this expansion, however, depends on whether additional temporal directions become accessible through the measured features. The learner receives a finite set of expectation values with finite precision, so directions generated within the reservoir can be erased by measurement or transmitted too weakly to be resolved. Analyses of concentration in quantum reservoirs identify related sampling limitations as measured responses become small~\cite{Sannia2025,Xiong2025}. Changing the observable family or measurement precision can therefore change the accessible representation even when the dynamics remain fixed. We seek a task-independent, measurement-resolved characterization of temporal expressivity under finite resources, connecting internal feature generation to classical access and identifying the bottlenecks that limit the use of a large operator space.

In this work, an operator-valued Volterra representation identifies the temporal directions generated before measurement, and the measurement map restricted to their span determines their classical accessibility. We define latent dimension, visible Volterra rank~(VVR) and operational Volterra dimension~(OVD) to quantify the generated, algebraically transmitted and finitely resolvable dimensions, respectively. For a specified temporal truncation, these metrics characterize the representation accessible from a fixed reservoir and observable family without requiring a task target or trained readout. We first examine how they respond to changes in reservoir dynamics, observable budget and measurement resolution, separating resources that generate temporal structure from those that make it accessible.

We then explain measurement access through the geometry of the generated temporal space relative to the readout interface. Approximate isotropic compression and an exact alignment law relate supported measurement strengths to sector coverage, alignment and spectral spread. These results explain how reservoir growth can weaken measurement access even when the algebraic rank remains unchanged. Bounds on population linear capacity connect this geometry to learning under specified response statistics and measurement noise, while optimal observable selection characterizes the strongest geometric access available through the readout register. Finally, short-term-memory, channel-equalization and Mackey--Glass benchmarks illustrate how these resource constraints are reflected in learning performance.

\section{Results}

\subsection{Operational dimension of measured temporal features in QRC}
\label{sec:operational}

\begin{figure}[t]
 \centering
 \includegraphics[width=\linewidth]{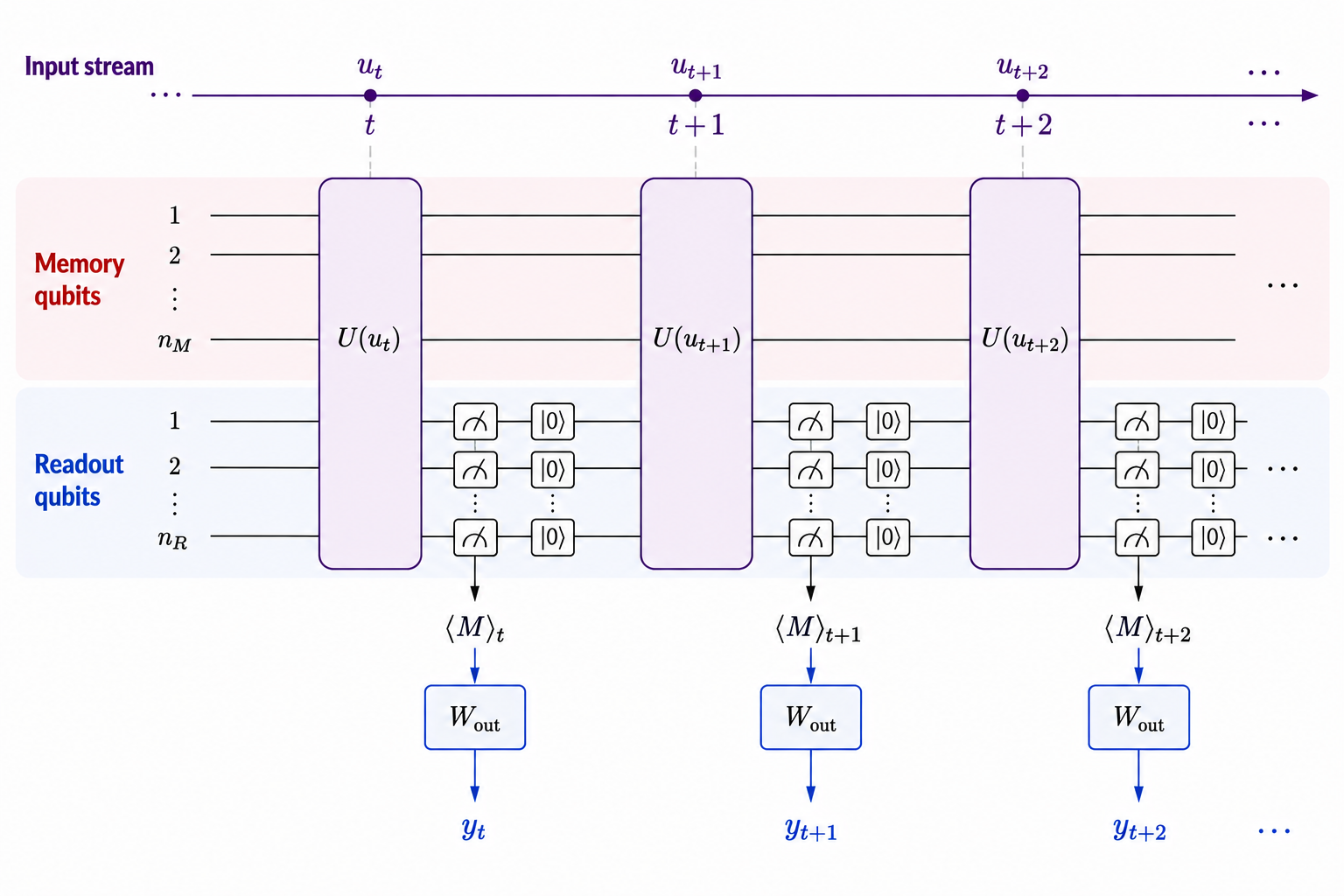}
 \caption{\textbf{Generation and measurement of temporal features.} A scalar input drives the joint evolution of persistent memory and a readout register. The readout is measured and reset at every timestep, while the memory carries the averaged recurrent state. Operator-valued Volterra kernels describe the joint response before measurement.}
 \label{fig:architecture}
\end{figure}

Quantum reservoir computing extracts classical features from a driven quantum system while retaining information about previous inputs. A subsystem description makes the relation between measurement and memory explicit. Here, a persistent memory interacts with a readout register through which information is extracted. Indirect measurements obtained from this interaction can describe both weak monitoring and projective measurement, including full projective readout with a suitable coupling and readout register~\cite{Mujal2023}. The measurement--reset architecture in Fig.~\ref{fig:architecture} provides a common formulation in which the generated temporal response and its measurement interface can be characterized separately.

Stable temporal processing requires the reservoir to forget its initialization while retaining a fading dependence on past inputs, as expressed by the echo-state and fading-memory properties~\cite{MartinezPena2023a}. Here, information is discarded by unconditionally resetting the readout register after every step, while the memory subsystem persists. Resetting makes the reduced memory evolution nonunitary and permits contraction; the stability conditions required for fading memory are stated in Supplementary Section~S1.2. Measurement outcomes are averaged without feedback, so the recurrent state follows a deterministic quantum channel.

For a readout register prepared in the fixed state $\rho_0^R$, the joint state immediately before measurement and the subsequent memory state are
\begin{equation}
 \begin{aligned}
 \sigma_t
 &=\Umap_{u_t}\!
   \left(\rho_{t-1}^M\otimes\rho_0^R\right),\\
 \rho_t^M
 &=\Tr_R\sigma_t
 =\Cmap_{u_t}(\rho_{t-1}^M),
 \end{aligned}
 \label{eq:update}
\end{equation}
where $\Cmap_u(X)=\Tr_R\Umap_u(X\otimes\rho_0^R)$. For a reference input $u_*$ and stationary memory state $\bar\rho^M=\Cmap_{u_*}(\bar\rho^M)$, define $\bar\sigma=\Umap_{u_*}(\bar\rho^M\otimes\rho_0^R)$, $Z_t=\sigma_t-\bar\sigma$ and $v_t=u_t-u_*$. Under the analytic and summability assumptions stated in Supplementary Section~S1.3, the centered response admits an operator-valued Volterra expansion. The index set
\begin{equation}
 \begin{aligned}
 \Lambda_{P,L}:=\bigl\{&(p,\tau_1,\ldots,\tau_p):1\le p\le P,\\
 &0\le\tau_p\le\cdots\le\tau_1\le L\bigr\}
 \end{aligned}
 \label{eq:volterra-index}
\end{equation}
retains nonlinear orders up to $P$ and delays up to $L$. The corresponding truncated response and latent operator space are
\begin{equation}
 \begin{aligned}
 Z_t^{(P,L)}&=\sum_{\lambda\in\Lambda_{P,L}}
 \widehat h_\lambda\,\phi_\lambda(\mathbf v_t),\\
 S&:=\Span\{\widehat h_\lambda:\lambda\in\Lambda_{P,L}\}.
 \end{aligned}
 \label{eq:volterra}
\end{equation}
For $\lambda=(p,\tau_1,\ldots,\tau_p)$, $\phi_\lambda(\mathbf v_t)=\prod_{r=1}^{p}v_{t-\tau_r}$ is the corresponding delayed-input monomial; repeated delays represent repeated powers. The latent dimension $d_{P,L}^{\mathrm{lat}}=\dim S$ counts independent operator directions generated within this truncation. It depends on encoding and dynamics, but not on the subsequently chosen observables. Operator spans are taken over $\mathbb R$, with Hilbert--Schmidt inner product $\langle X,Y\rangle_{\HS}=\Tr(XY)$.

Throughout the main text, we consider traceless Hermitian readout observables $M_1,\ldots,M_J$, normalized so that their joint extensions are Hilbert--Schmidt orthonormal $\Tr[(I_M\otimes M_j)(I_M\otimes M_k)]=\delta_{jk}$. General measurement families are treated in Supplementary Section~S2.5. The measurement map is
\begin{equation}
 \Mmap(Z)=\bigl(\Tr[(I_M\otimes M_j)Z]\bigr)_{j=1}^{J}.
 \label{eq:measurement}
\end{equation}
For a Hilbert--Schmidt orthonormal basis $\{H_\ell\}_{\ell=1}^{d_{P,L}^{\mathrm{lat}}}$ of the latent space $S$, the restricted measurement matrix is
\begin{equation}
 (O_{P,L})_{j\ell}=\Tr[(I_M\otimes M_j)H_\ell].
 \label{eq:restriction}
\end{equation}

Each column gives the measured signature of one normalized latent operator direction. The matrix therefore represents how the generated quantum features are transmitted into the measured feature space $\Mmap(S)\subseteq\mathbb R^J$. Features living in $\ker\Mmap$ are erased by the measurement interface.

The number of independent temporal feature directions that survive this restriction defines the visible Volterra rank (VVR),
\begin{equation}
 \nu_{P,L}:=\dim\Mmap(S)=\rank O_{P,L}.
 \label{eq:vvr}
\end{equation}
For linearly independent retained monomials on an input domain with nonempty interior, VVR also equals the dimension of the truncated scalar temporal function space available to a linear readout.

Algebraic survival, however, does not ensure that a direction can be resolved with finite measurement precision. A nonzero measured signature may be too weak to distinguish from measurement uncertainty. The singular values of $O_{P,L}$ quantify the measurement strengths of orthogonal, Hilbert--Schmidt-normalized latent directions. At a prescribed resolution $\epsilon\geq0$, the operational Volterra dimension (OVD) counts strengths above the separation threshold,
\begin{equation}
 \omega_{P,L}(\epsilon)
 :=\#\{j:\sigma_j(O_{P,L})>2\epsilon\}.
 \label{eq:ovd}
\end{equation}
At zero resolution, $\omega_{P,L}(0)=\nu_{P,L}$; more generally,
\begin{equation}
 \omega_{P,L}(\epsilon)\le\nu_{P,L}
 \le\min\{d_{P,L}^{\mathrm{lat}},J\}.
 \label{eq:hierarchy}
\end{equation}
The singular spectrum is independent of the chosen orthonormal latent basis. Singular strengths and resolution thresholds are expressed in the normalized measurement coordinates of Eq.~\eqref{eq:measurement}.

This finite-resolution interpretation concerns normalized latent directions. Kernel amplitudes, input statistics and target alignment remain relevant to learning and are not separately represented by OVD. The factor of two provides a worst-case separation between resolvable singular strengths and perturbations of exactly null strengths, as detailed in Section~\ref{sec:methods-resolution} and Supplementary Section~S5.1.

\subsection{Dynamical and measurement resources}
\label{sec:resources}

\begin{figure*}[t]
 \centering
 \begin{minipage}[t]{0.32\textwidth}
  \vspace{0pt}
  \noindent\textbf{a}\par\vspace{0.2ex}
  \centering
  \includegraphics[width=\linewidth]{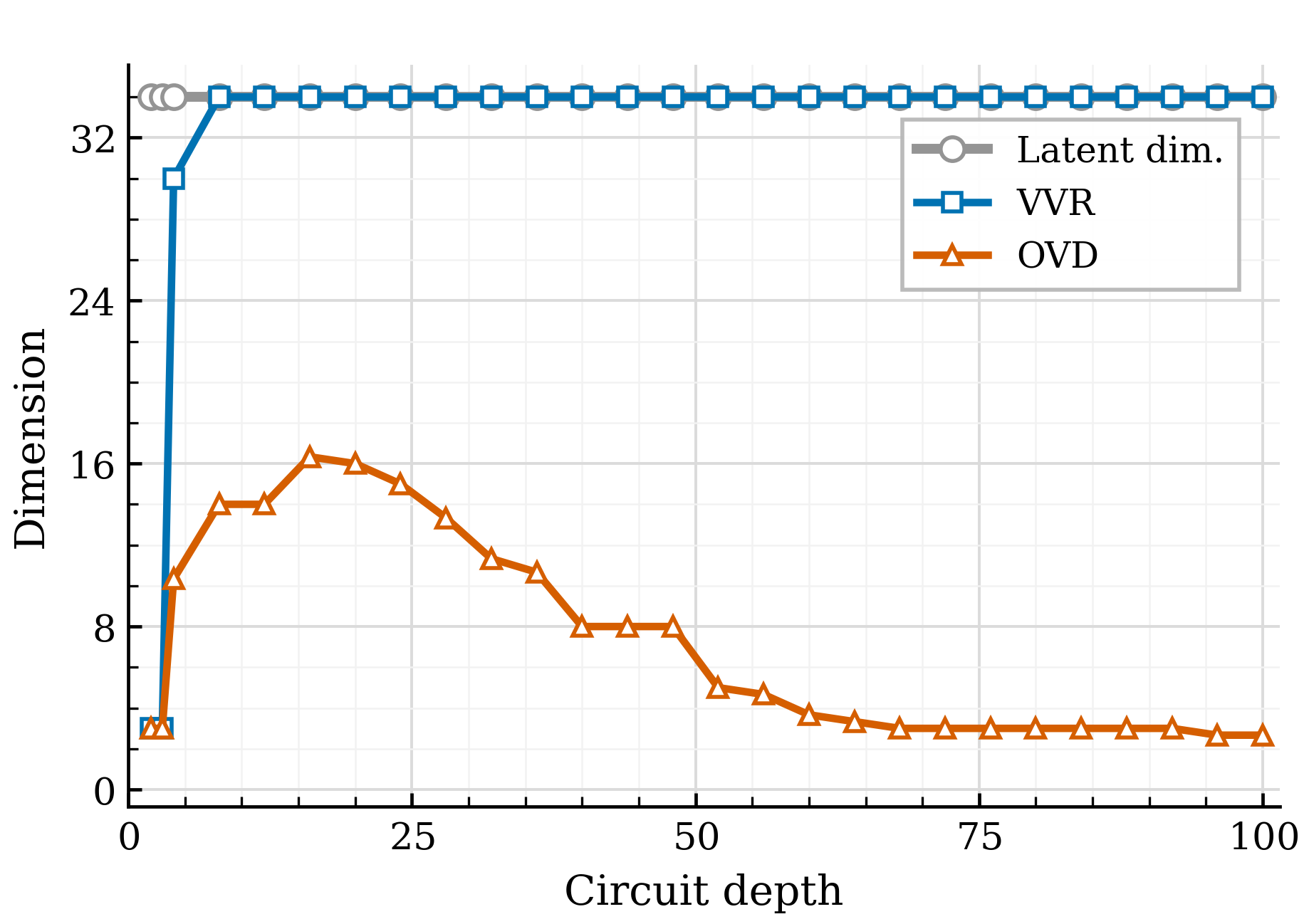}
 \end{minipage}\hfill
 \begin{minipage}[t]{0.32\textwidth}
  \vspace{0pt}
  \noindent\textbf{b}\par\vspace{0.2ex}
  \centering
  \includegraphics[width=\linewidth]{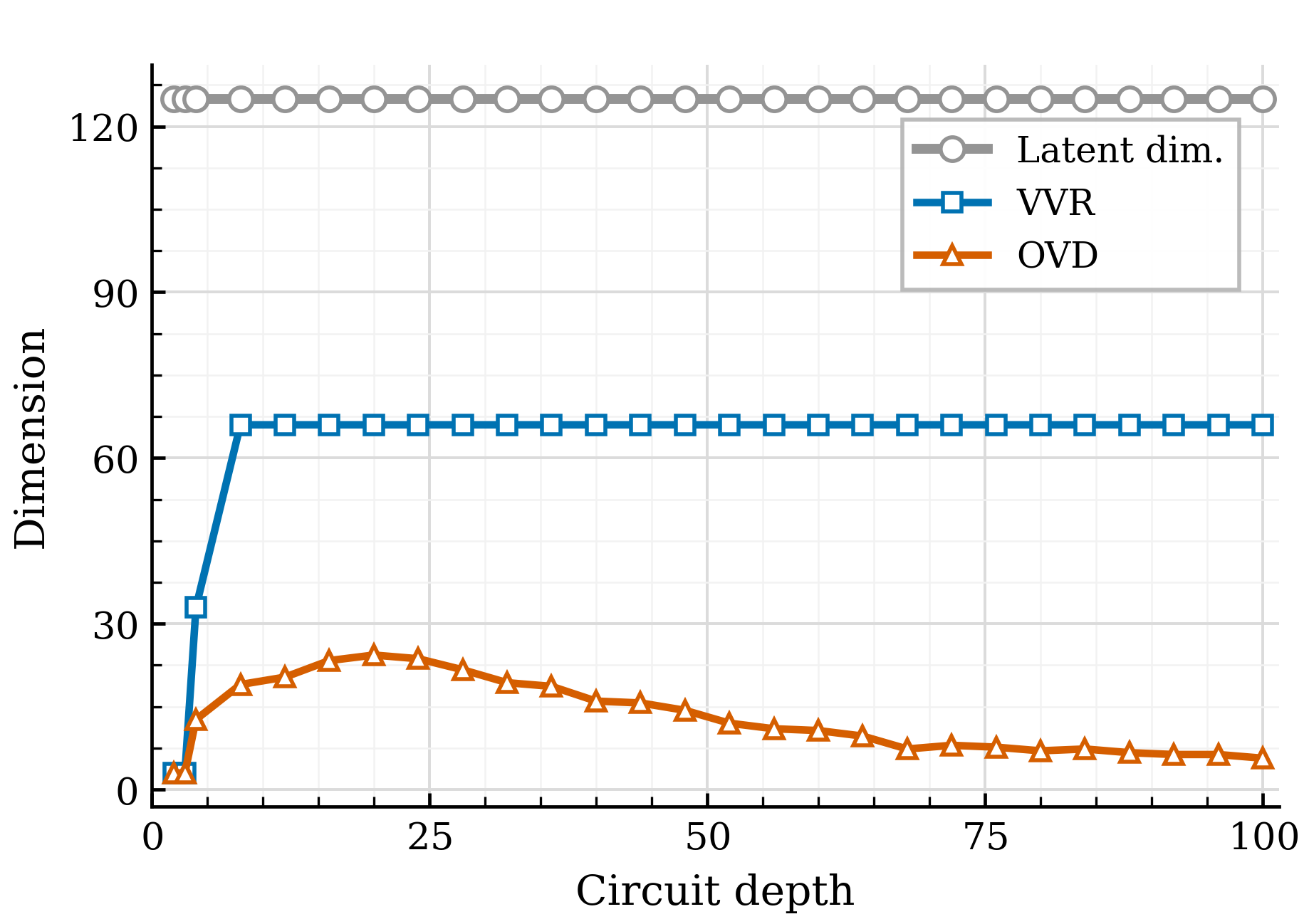}
 \end{minipage}\hfill
 \begin{minipage}[t]{0.32\textwidth}
  \vspace{0pt}
  \noindent\textbf{c}\par\vspace{0.2ex}
  \centering
  \includegraphics[width=\linewidth]{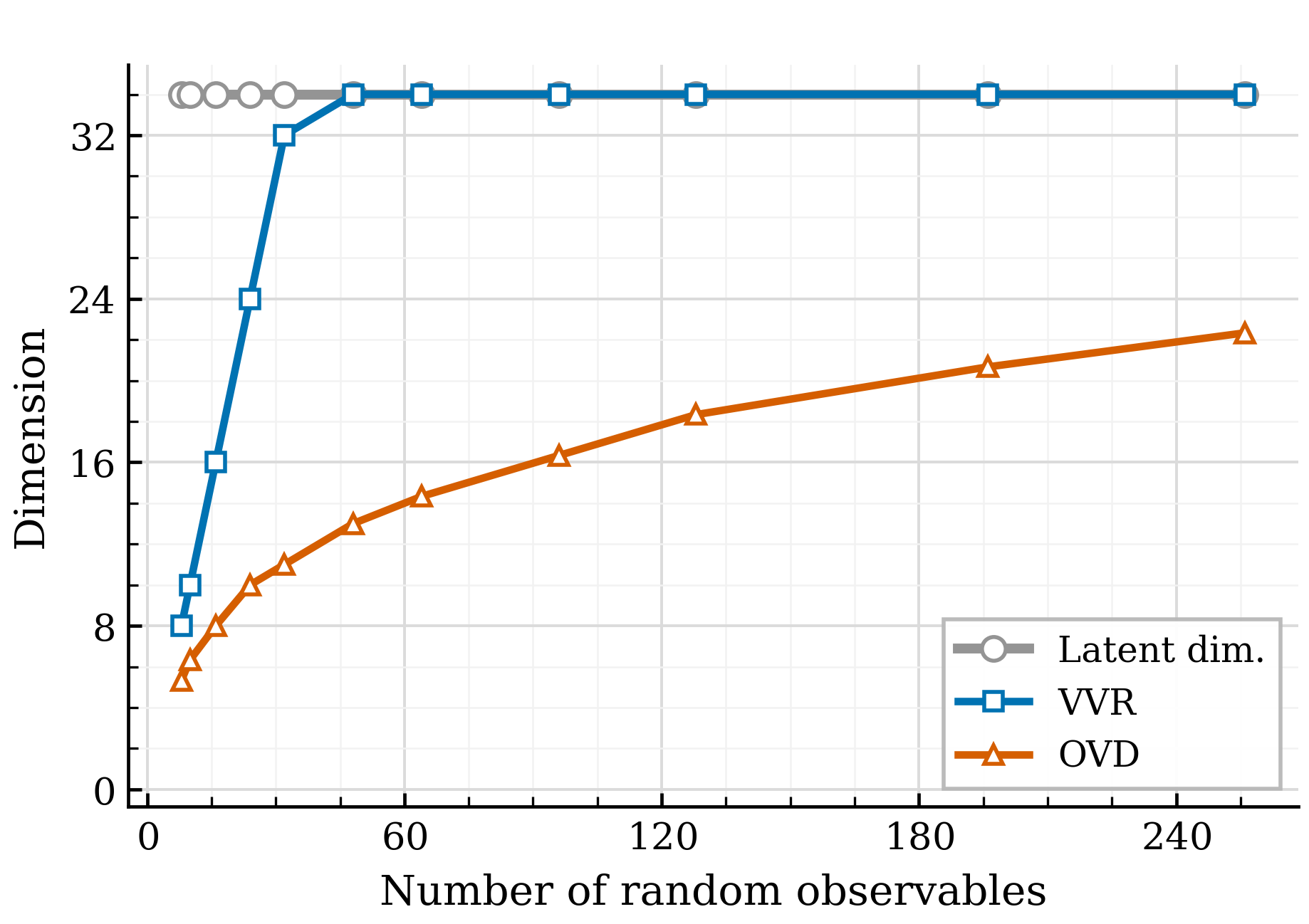}
 \end{minipage}
 \par\vspace{0.8ex}
 \begin{minipage}[t]{0.32\textwidth}
  \vspace{0pt}
  \noindent\textbf{d}\par\vspace{0.2ex}
  \centering
  \includegraphics[width=\linewidth]{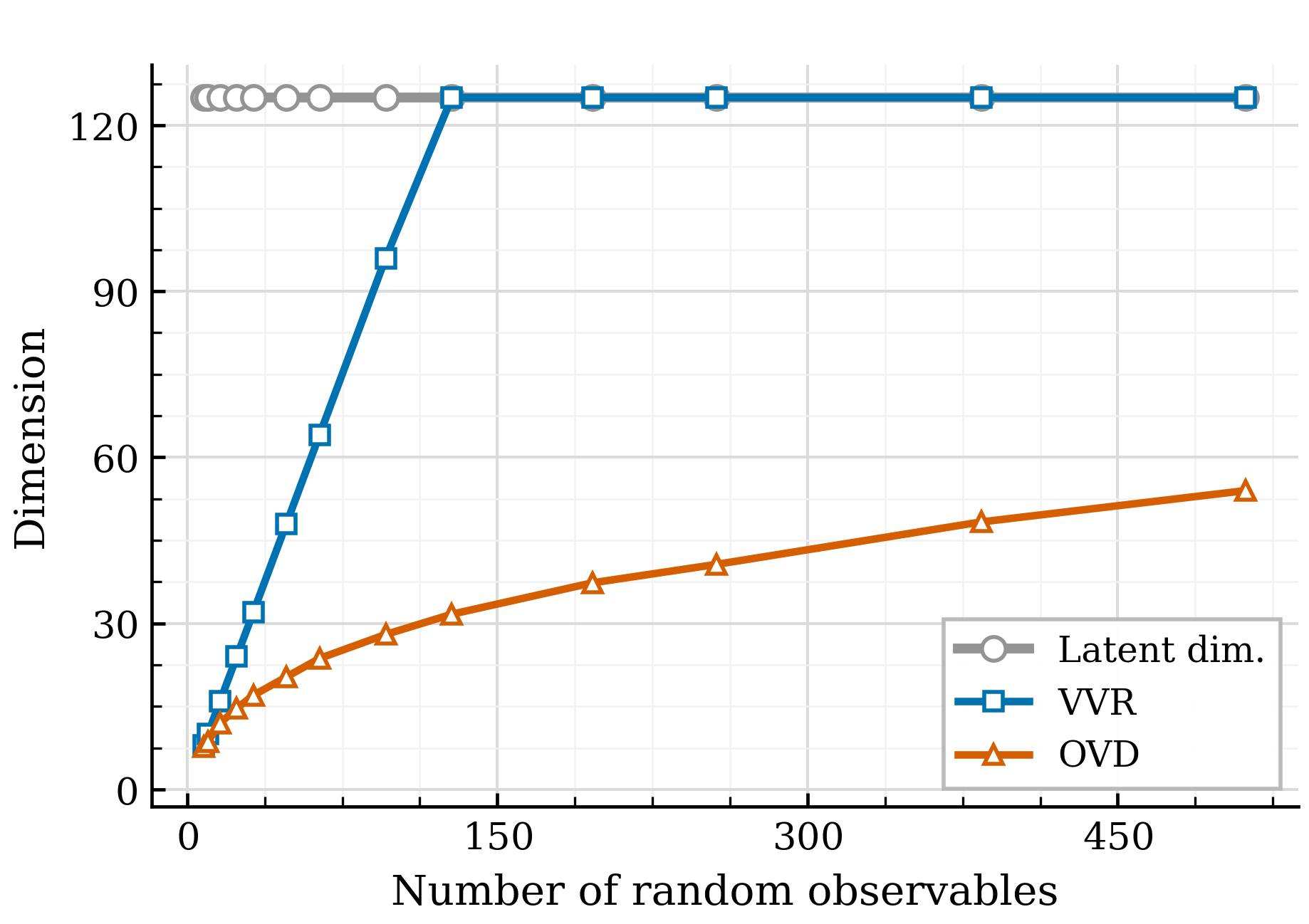}
 \end{minipage}\hspace{0.02\textwidth}
 \begin{minipage}[t]{0.32\textwidth}
  \vspace{0pt}
  \noindent\textbf{e}\par\vspace{0.2ex}
  \centering
  \includegraphics[width=\linewidth]{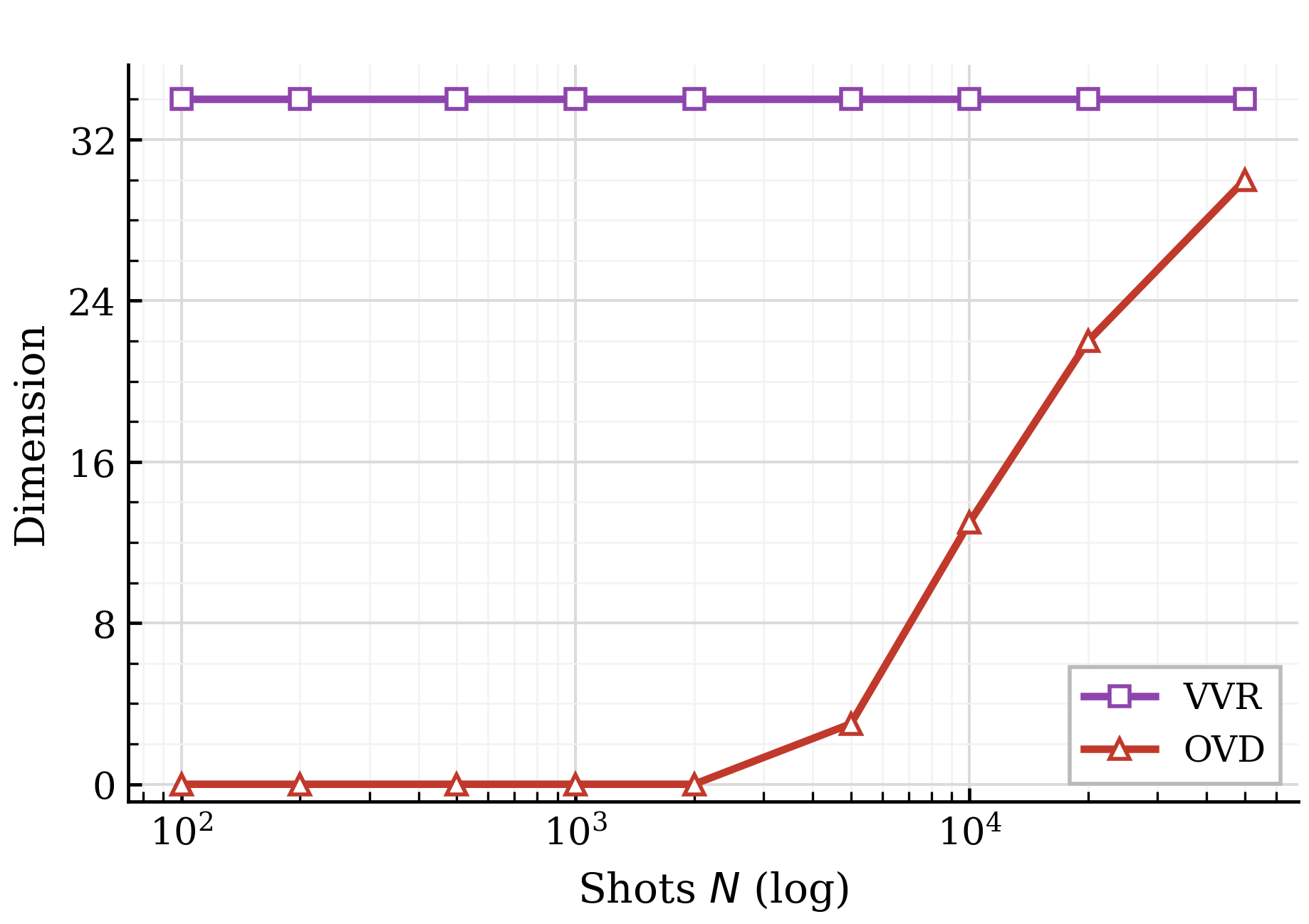}
 \end{minipage}
 \caption{\textbf{Dynamics, observable access and resolution control different aspects of the representation.} \textbf{a,b}, Latent dimension, VVR and OVD versus circuit depth for $(P,L)=(3,3)$ and $(4,4)$. \textbf{c,d}, Nested observable-budget sweeps at fixed dynamics for the corresponding truncations. \textbf{e}, Resolution sweep at fixed dynamics and observables, obtained by rethresholding the spectrum with a nominal shot-equivalent parameter. Panels a–d show averages over three circuit realizations.}
 \label{fig:resources}
\end{figure*}

The framework separates changes in feature generation from changes in measurement access and resolution. We examine these controls through circuit-depth, observable-budget and reference-resolution sweeps (Fig.~\ref{fig:resources}). Register sizes remain fixed within each sweep. The circuit-depth sweeps change the generated response, whereas the observable-budget and resolution sweeps change access to a fixed response.

Increasing the depth of a nearest-neighbor random circuit modifies the Volterra kernels and their latent span (Fig.~\ref{fig:resources}a,b). The latent dimension and VVR rapidly reach their respective ceilings, while OVD initially increases more gradually as additional measured strengths exceed the resolution threshold. At larger depths, however, OVD decreases despite the saturated latent dimension and VVR. This nonmonotonic behavior is consistent with stronger scrambling weakening the exposure of generated temporal features to the measured observables. Directions remain algebraically visible, but their measured strengths fall below the fixed operational threshold. Thus increasing depth can first enrich the resolvable representation and subsequently reduce access to it. The following sections explain this behavior through visibility geometry, separating the mean exposure of supported directions, their alignment with measurement and the spread of their visible strengths.

Enlarging a nested observable family at fixed dynamics exposes additional components of an unchanged latent space (Fig.~\ref{fig:resources}c,d). VVR rises as independent directions become visible, while OVD counts those transmitted strongly enough to exceed the prescribed threshold. Adding observables expands the measurement-visible space, or leaves it unchanged when the new observables provide no additional independent directions.

Resolution changes which already visible directions are counted without changing the dynamics or observable map. In Fig.~\ref{fig:resources}e, the threshold decreases with the inverse square root of a nominal shot-equivalent parameter, following the reference threshold stated in the Methods. Latent dimension and VVR remain fixed, while OVD increases as weaker directions cross the threshold. This sweep characterizes resolution dependence without explicitly sampling finite-shot features.

These controls show how dynamics, observable access and measurement precision shape the operational feature budget at fixed register sizes. To understand what determines the strengths of the directions that survive algebraically, we next examine the geometry of the generated temporal space relative to the measurement kernel.

\subsection{The spectrum of measurement accessibility}
\label{sec:geometry}

For generated features to be resolved through finite measurements, the corresponding singular values of the restriction matrix $O_{P,L}$ must exceed the operational threshold. Under orthonormal measurement, these strengths are determined exactly by the angles that latent operator directions make with the measurement kernel. We show how approximately isotropic compression onto the visible space confines the supported visible weights to a narrow interval around their mean, connecting this geometry directly to OVD.

Some linear combinations of the generated features lie entirely in the measurement kernel $K=\ker\Mmap$ and are therefore erased by measurement. We separate these invisible directions from their orthogonal complement within $S$ through
\begin{equation}
 \begin{gathered}
 S=N\oplus S^+,\qquad N=S\cap K,\\
 q:=\dim S^+=\nu_{P,L}.
 \end{gathered}
 \label{eq:supported}
\end{equation}
The supported space $S^+$ contains the remaining directions, each with a nonzero measurable component. If $q=0$, VVR and OVD vanish; supported angles and their summaries below are defined only for $q>0$.

For $q>0$, choose a Hilbert--Schmidt orthonormal basis $\{H_\ell^+\}_{\ell=1}^{q}$ of $S^+$. The basis map $Q^+$ takes a coordinate vector $x\in\mathbb R^q$ to the operator $Q^+x=\sum_{\ell=1}^{q}x_\ell H_\ell^+$.

Let $\Pi_{\mathrm{vis}}$ be the orthogonal projector onto $K^\perp$. Projecting the supported basis onto this visible space gives the matrix
\begin{equation}
 B^+=(Q^+)^{\mathsf T}\Pi_{\mathrm{vis}}Q^+,
 \label{eq:visibility}
\end{equation}
where $B^+_{\ell k}=\langle\Pi_{\mathrm{vis}}H_\ell^+,\Pi_{\mathrm{vis}}H_k^+\rangle_{\HS}$. This matrix describes how much of each supported direction survives the geometric projection, including overlaps between the projected basis elements.

\begin{figure}[t]
 \centering
 \includegraphics[width=\linewidth]{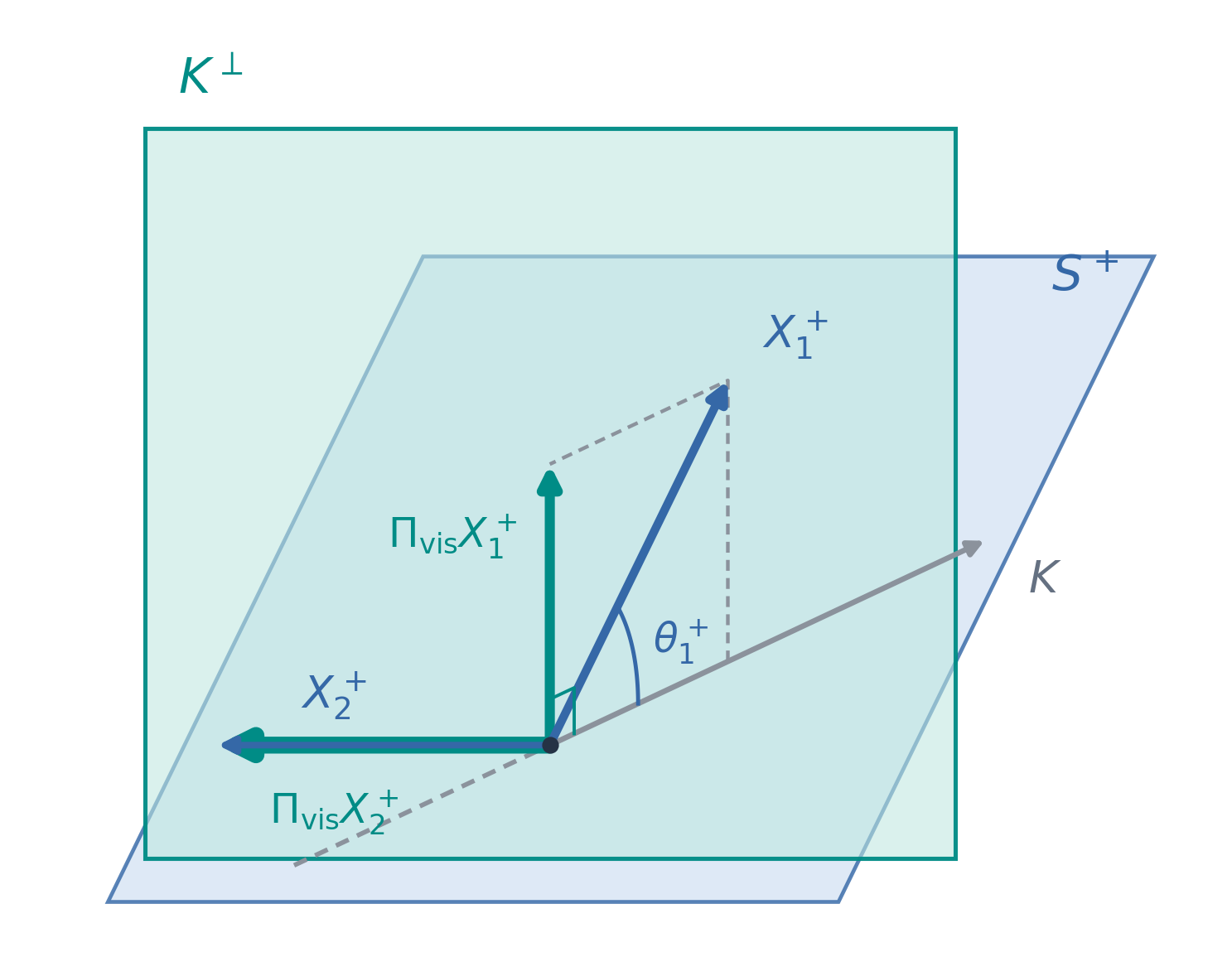}
 \caption{\textbf{Visibility geometry.} A normalized supported direction $X^+$ makes an angle $\theta^+$ with the measurement kernel $K$. Its projection onto $K^\perp$ has norm $\|\Pi_{\mathrm{vis}}X^+\|_{\mathrm{HS}}=\sin\theta^+$, which equals the measured signature norm under orthonormal measurement. Small angles therefore indicate weak measurement access. The schematic shows $K$ as a line and $S\cap K=\{0\}$; invisible intersections can be nontrivial in the full operator space.}
 \label{fig:geometry}
\end{figure}

Let $\{a_j\}_{j=1}^q$ be an orthonormal eigenbasis of $B^+$. The canonical directional operators
\begin{equation}
 X_j^+:=Q^+a_j
 =\sum_{\ell=1}^q(a_j)_\ell H_\ell^+
\end{equation}
form a Hilbert--Schmidt orthonormal basis of $S^+$. We define their visibility angles $\theta_j^+\in(0,\pi/2]$ as
\begin{equation}
 \sin^2\theta_j^+
 =\lambda_j(B^+)
 =\|\Pi_{\mathrm{vis}}X_j^+\|_{\mathrm{HS}}^2.
\end{equation}
Each angle measures the separation of a canonical direction from the measurement kernel $K$. A small angle means that most of the direction lies in $K$, leaving only a weak component in the measurement-visible space $K^\perp$ (Fig.~\ref{fig:geometry}).

Under our orthonormal measurement convention, this geometry determines measurement strength exactly. Since $\Mmap^*\Mmap=\Pi_{\mathrm{vis}}$, the restriction of $O^+$ to the supported space obeys
\begin{equation}
 \begin{aligned}
 (O^+)^{\mathsf T}O^+
 &=(Q^+)^{\mathsf T}\Pi_{\mathrm{vis}}Q^+=B^+,\\
 \sigma_j(O^+)&=\sin\theta_j^+,
 \qquad j=1,\ldots,q,
 \end{aligned}
 \label{eq:comparison}
\end{equation}
with consistently ordered spectra. The restriction to $N=S\cap K$ vanishes, so these are also the nonzero singular values of the restricted measurement matrix $O_{P,L}$. VVR counts all $q$ supported directions, whereas OVD counts those whose measured strengths satisfy $\sin\theta_j^+>2\epsilon$. A direction can therefore remain algebraically visible while becoming too weak to resolve.

To describe the distribution of these strengths, define the mean supported visible weight and its maximum spectral deviation,
\begin{equation}
 \alpha^+=\frac{\Tr B^+}{q},\qquad
 \delta_{\mathrm{iso}}^+
 =\|B^+-\alpha^+I_q\|_{\mathrm{op}}.
 \label{eq:centerwidth}
\end{equation}
The mean $\alpha^+$ sets the overall scale of measurement exposure, while $\delta_{\mathrm{iso}}^+$ bounds how much the squared measured norm of any Hilbert--Schmidt-normalized supported direction can differ from this mean. When $\delta_{\mathrm{iso}}^+\ll\alpha^+$, all supported directions have nearly the same measured strength. Measurement therefore acts approximately as an isometry scaled by $\sqrt{\alpha^+}$, preserving relative lengths and angles while uniformly reducing their magnitude. We call this approximate isotropic compression.

In this regime, the squared singular strengths cluster around $\alpha^+$, while the singular strengths cluster around $\sqrt{\alpha^+}$, with characteristic angle $\theta_+^\star=\arcsin\sqrt{\alpha^+}$. If $\alpha^+$ decreases while $\delta_{\mathrm{iso}}^+/\alpha^+$ remains small, the supported directions weaken together. The following sufficient condition identifies when this weakening eliminates finite-resolution access altogether.

\begin{proposition}[Finite-resolution collapse]
\label{prop:resolution-collapse}
For $q>0$, if
\begin{equation}
 \alpha^++\delta_{\mathrm{iso}}^+\leq4\epsilon^2,
\end{equation}
then OVD vanishes while VVR equals $q$.
\end{proposition}
\begin{proof}
Every supported measurement strength satisfies $\sigma_j(O^+)^2\leq\alpha^++\delta_{\mathrm{iso}}^+\leq4\epsilon^2$, so none exceeds the operational threshold $2\epsilon$.
\end{proof}

The geometry also specifies the best readout subspace when arbitrary Hermitian observables on the readout register are available. Let $d_M=\dim\mathcal H_M$ and $d_R=\dim\mathcal H_R$, and define the normalized partial-trace map on the latent space,
\begin{equation}
 \begin{aligned}
 T_R&:S\longrightarrow\{B=B^\dagger:\Tr B=0\},\\
 T_R(X)&=\frac{\Tr_M X}{\sqrt{d_M}}.
 \end{aligned}
 \label{eq:normalized-partial-trace}
\end{equation}
Both spaces carry their HS inner products. Write $\tau_1\geq\tau_2\geq\cdots\geq0$ for its singular values, padded with zeros when needed.

\begin{proposition}[Optimal readout subspace]
\label{prop:optimal-readout}
For $1\leq J\leq d_R^2-1$, maximize over traceless Hermitian readout observables $M_1,\ldots,M_J$ with HS-orthonormal joint extensions. At a common geometric resolution $\epsilon\geq0$,
\begin{equation}
 \begin{aligned}
 \max_{\{M_j\}}\omega_{P,L}(\epsilon)
 &=\min\{J,\#\{k:\tau_k>2\epsilon\}\}.
 \end{aligned}
 \label{eq:optimal-readout}
\end{equation}
The maximum is attained by $M_j=B_j/\sqrt{d_M}$, where $B_j$ are the first $J$ HS-orthonormal left singular operators of $T_R$, completed arbitrarily in the readout space if necessary.
\end{proposition}

This result separates information absent from the readout register from information lost by the choice of measured observables. It is an optimum at fixed geometric normalization and threshold. The corresponding physical sampling costs can differ between observables and must be included when imposing a total shot budget. The proof is given in Supplementary Section~S4.2.

We next examine how sector coverage and supported-space alignment determine the mean visibility entering the collapse criterion.

\subsection{Reservoir growth and visibility dilution}
\label{sec:scaling}

Adding qubits enlarges the operator space available to the dynamics, but a fixed measurement interface may expose a decreasing fraction of the generated temporal structure. To characterize this possibility, we relate the mean supported visibility to the accessible operator sector and the alignment of the supported temporal space with measurement.

The mean supported visibility has an exact decomposition in the dynamically accessible response sector. Let $W_{\mathrm{dyn}}$ denote the operator sector spanned by centered joint responses over admissible input histories at fixed model parameters and reservoir dynamics, and let $P_{\mathrm{dyn}}$ be its orthogonal projector. Precise definitions are given in Supplementary Section~S2.2. Every retained Volterra kernel lies in this sector. Its dimension and total visible weight are
\begin{equation}
 D_{\mathrm{dyn}}=\Tr P_{\mathrm{dyn}},\qquad
 r_{\mathrm{dyn}}=\Tr(\Pi_{\mathrm{vis}}P_{\mathrm{dyn}}).
 \label{eq:sector}
\end{equation}
The ratio $r_{\mathrm{dyn}}/D_{\mathrm{dyn}}$ measures how much of the dynamically accessible operator space is visible to measurement on average. It is obtained by averaging the squared visible norms over any HS-orthonormal basis of $W_{\mathrm{dyn}}$. The total visible weight satisfies $r_{\mathrm{dyn}}\le\rank\Mmap\le J$ and need not be an integer. Thus, if the dynamically accessible operator space grows rapidly with system size while the number of measured observables remains fixed, the average visible weight is bounded by the rapidly decreasing ratio $J/D_{\mathrm{dyn}}$.

Measurement need not expose all directions within $W_{\mathrm{dyn}}$ equally. Some directions have a large component in the measurement-visible space, while others remain weakly exposed. This variation is described by the compressed operator $C_{\mathrm{dyn}}=P_{\mathrm{dyn}}\Pi_{\mathrm{vis}}P_{\mathrm{dyn}}$. Meanwhile, the supported temporal space $S^+$ specifies the directions generated by the reservoir that survive measurement algebraically. Its orthogonal projector $P^+=Q^+(Q^+)^{\mathsf T}$ records where this space lies within $W_{\mathrm{dyn}}$. Comparing $P^+$ with $C_{\mathrm{dyn}}$ therefore reveals whether the supported temporal directions preferentially occupy parts of the dynamically accessible space that are more or less visible than average.

\begin{theorem}[Alignment law and supported visibility bound]
\label{thm:visibility-scaling}
Let $S^+\subseteq W_{\mathrm{dyn}}$ have dimension $q>0$. The mean supported visibility satisfies
\begin{equation}
 \begin{aligned}
 \alpha^+&=\frac{r_{\mathrm{dyn}}}{D_{\mathrm{dyn}}}+\chi_{\mathrm{om}},\\
 \chi_{\mathrm{om}}&=\frac1q\Tr\!\left[
 \left(C_{\mathrm{dyn}}-
 \frac{r_{\mathrm{dyn}}}{D_{\mathrm{dyn}}}P_{\mathrm{dyn}}\right)P^+\right].
 \end{aligned}
 \label{eq:alignment}
\end{equation}
Every supported visibility angle consequently obeys
\begin{equation}
 \begin{gathered}
 \left|\sin^2\theta_j^+-\frac{r_{\mathrm{dyn}}}{D_{\mathrm{dyn}}}\right|
 \le\delta_{\mathrm{iso}}^++|\chi_{\mathrm{om}}|,\\
 j=1,\ldots,q.
 \end{gathered}
 \label{eq:geometricbound}
\end{equation}
\end{theorem}

The alignment correction determines the displacement of the mean from the sector-average visibility, whereas the isotropy defect controls the width of the supported spectrum. Positive $\chi_{\mathrm{om}}$ indicates preferential occupation of more visible directions; negative $\chi_{\mathrm{om}}$ indicates preferential occupation of less visible directions. When $|\chi_{\mathrm{om}}|+\delta_{\mathrm{iso}}^+$ is small, the supported visible weights $\sin^2\theta_j^+$ lie close to the sector-average visibility $r_{\mathrm{dyn}}/D_{\mathrm{dyn}}$. Neither the dynamics nor the measurement interface must be individually isotropic for small $\chi_{\mathrm{om}}$. The relevant conditions concern their alignment and the compression onto the supported temporal space.

Under orthonormal measurement, this geometric scale directly bounds the singular strengths resolved by measurement.

\begin{theorem}[Sector bound on measurement strength]
\label{thm:operational-visibility-bound}
Under orthonormal measurement and the conditions of Theorem~\ref{thm:visibility-scaling}, the restricted measurement matrix satisfies
\begin{equation}
 \begin{aligned}
 \sigma_{\max}(O_{P,L})
 &\leq\sqrt{\alpha^++\delta_{\mathrm{iso}}^+}\\
 &\leq\sqrt{
 \frac{r_{\mathrm{dyn}}}{D_{\mathrm{dyn}}}
 +|\chi_{\mathrm{om}}|
 +\delta_{\mathrm{iso}}^+}.
 \end{aligned}
 \label{eq:finitebound}
\end{equation}
\end{theorem}

The proofs of Theorems~\ref{thm:visibility-scaling} and~\ref{thm:operational-visibility-bound} are given in Supplementary Sections~S2.3 and~S2.4, respectively. Together with Proposition~\ref{prop:resolution-collapse}, Eq.~\eqref{eq:finitebound} gives a collapse criterion in terms of sector coverage, alignment and spectral spread. Supplementary Section~S3 samples the latent space before separating visible and invisible directions. The following reference model keeps this measurement-dependent support selection explicit.

\begin{proposition}[Latent orientation]
\label{prop:latent-orientation}
Write $D=D_{\mathrm{dyn}}\geq2$, $C=C_{\mathrm{dyn}}$, $r=\Tr C$, $m=\rank C>0$ and $C^\circ=C-(r/D)P_{\mathrm{dyn}}$. Let $S_U$ be a uniformly random $s$-dimensional subspace of $W_{\mathrm{dyn}}$ under the real orthogonal Haar ensemble, with $1\leq s\leq D$. After removing its intersection with $\ker\Mmap$, the supported dimension and mean visibility obey $q=\min\{s,m\}$ almost surely, and
\begin{equation}
 \mathbb E[\alpha_U^+]=\frac{sr}{Dq}.
 \label{eq:latent-orientation-mean}
\end{equation}
The variance of the supported mean across random spaces is
\begin{equation}
 \operatorname{Var}(\alpha_U^+)
 =\frac{2s(D-s)\|C^\circ\|_{\HS}^2}
 {q^2D(D-1)(D+2)}.
 \label{eq:latent-orientation-variance}
\end{equation}
For every $\epsilon>0$, its OVD under orthonormal measurement
satisfies
\begin{equation}
 \mathbb E[\omega_U(\epsilon)]
 \leq\min\left\{q,\frac{sr}{4D\epsilon^2}\right\}.
 \label{eq:latent-orientation-count}
\end{equation}
\end{proposition}

The supported dimension and visibility moments are derived in Supplementary Sections~S3.2 and~S3.3; the expected spread within the supported spectrum is given in Supplementary Section~S3.4. For a full-sector rank-$J$ measurement projector, $r=m=J$ and Eq.~\eqref{eq:latent-orientation-mean} becomes $\mathbb E[\alpha_U^+]=\max\{s,J\}/D$. Thus $J/D$ is the supported mean in this ensemble when $s\leq J$; removing an invisible intersection when $s>J$ changes it to $s/D$. Equation~\eqref{eq:latent-orientation-variance} describes fluctuations of this mean between random spaces, rather than the spectral width within one space. The count bound requires no narrow-spectrum assumption. This is a reference ensemble of operator subspaces, not an automatic consequence of Haar-random quantum gates. Visibility dilution depends on how the latent dimension, measurement coverage and resolution scale together.

\begin{figure*}[t]
 \centering
 \begin{minipage}[t]{0.43\linewidth}
  \vspace{0pt}
  \noindent\textbf{a}\par\vspace{0.35ex}
  \centering
  \includegraphics[width=\linewidth]{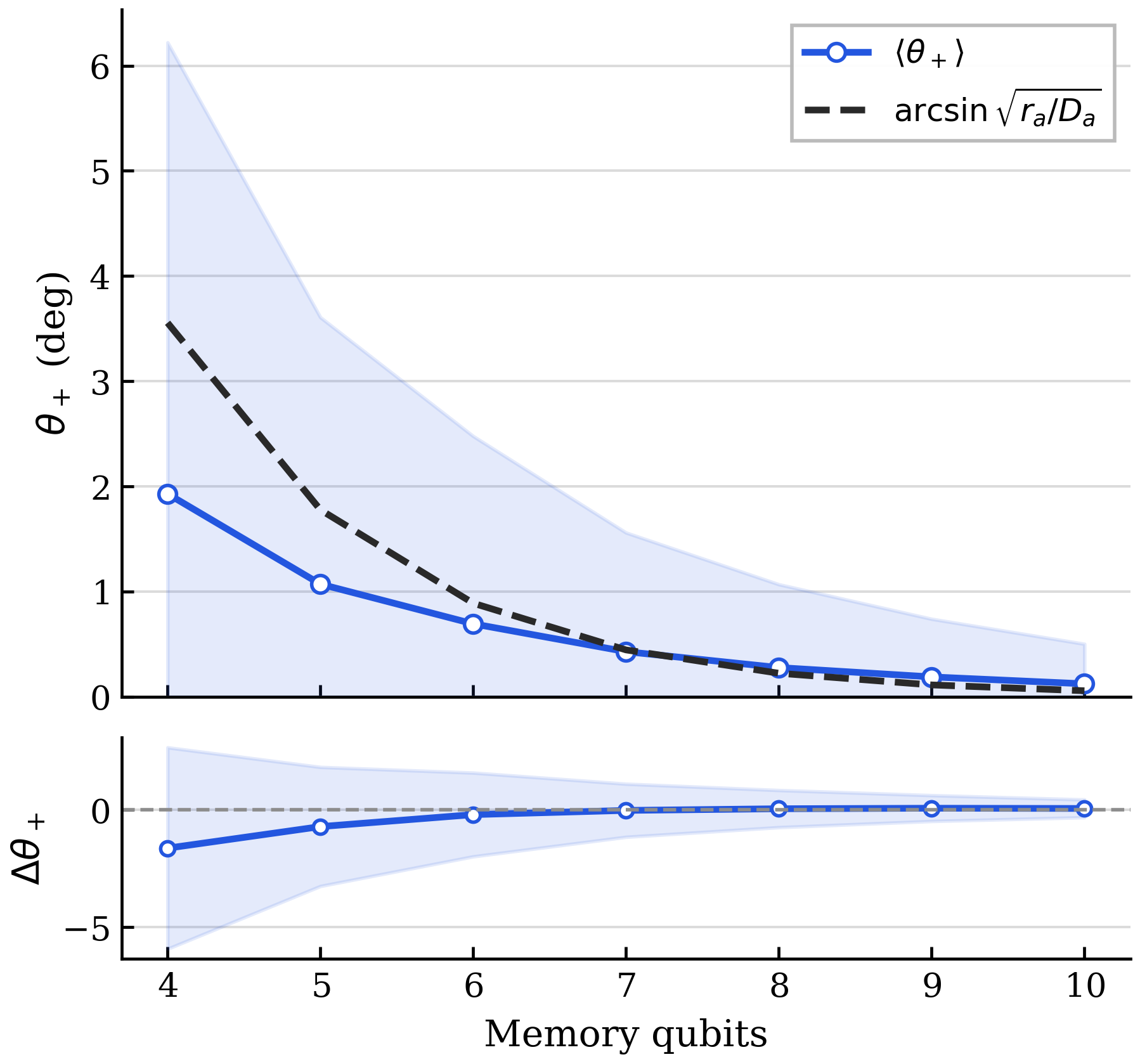}
 \end{minipage}\hfill
 \begin{minipage}[t]{0.43\linewidth}
  \vspace{0pt}
  \noindent\textbf{b}\par\vspace{0.35ex}
  \centering
  \includegraphics[width=\linewidth]{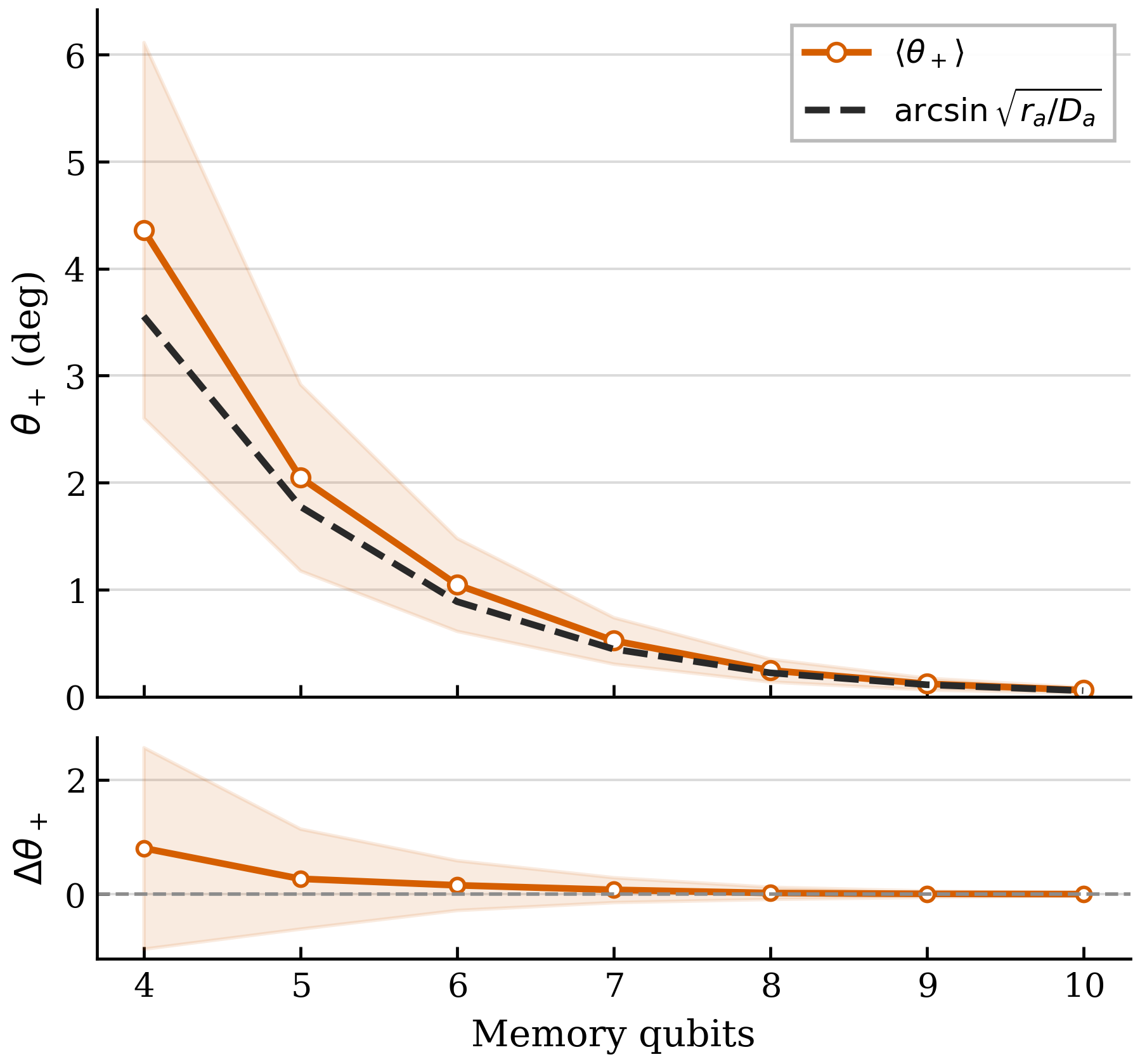}
 \end{minipage}
 \par\vspace{0.8ex}
 \begin{minipage}[t]{0.43\linewidth}
  \vspace{0pt}
  \noindent\textbf{c}\par\vspace{0.35ex}
  \centering
  \includegraphics[width=\linewidth]{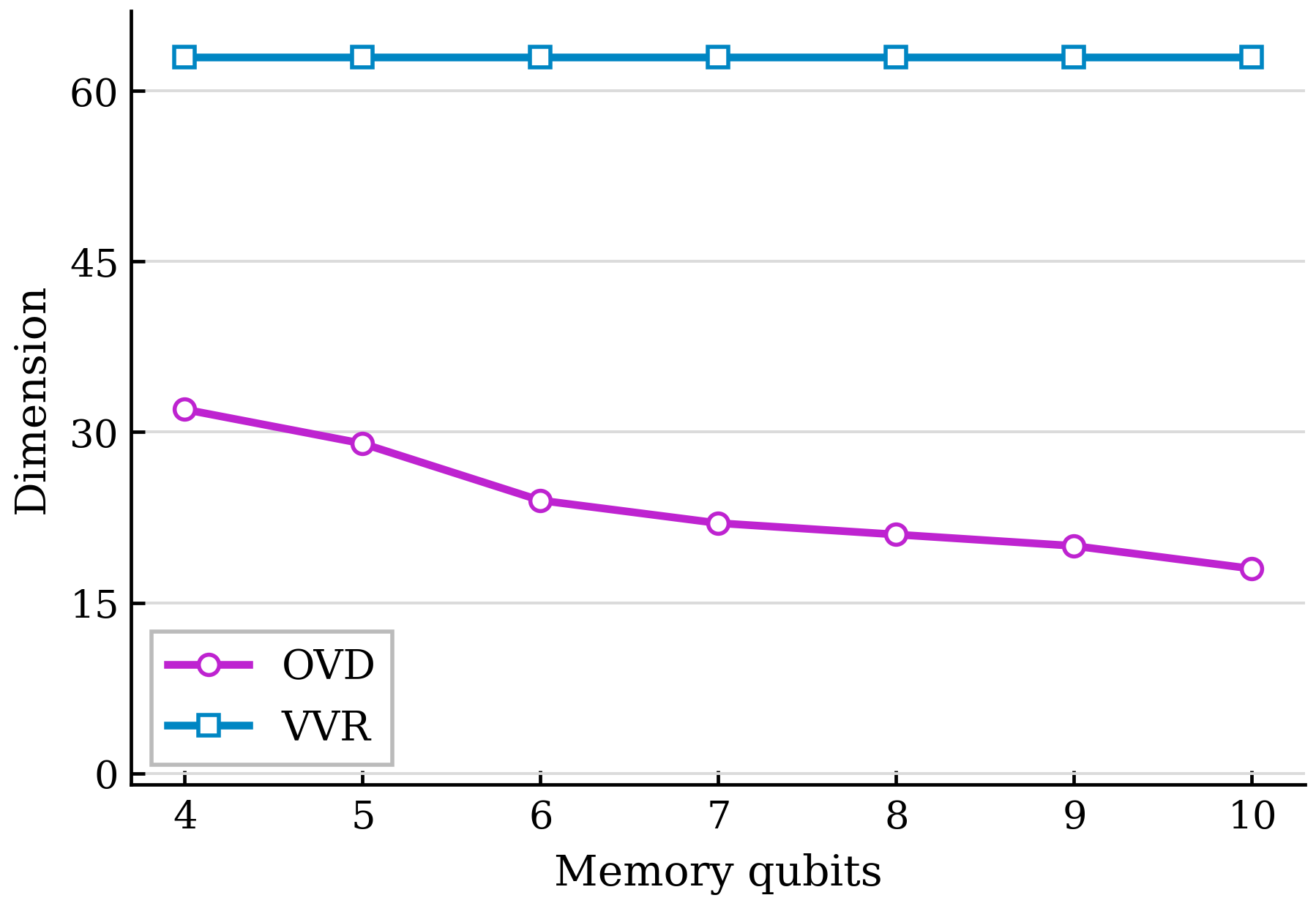}
 \end{minipage}\hfill
 \begin{minipage}[t]{0.43\linewidth}
  \vspace{0pt}
  \noindent\textbf{d}\par\vspace{0.35ex}
  \centering
  \includegraphics[width=\linewidth]{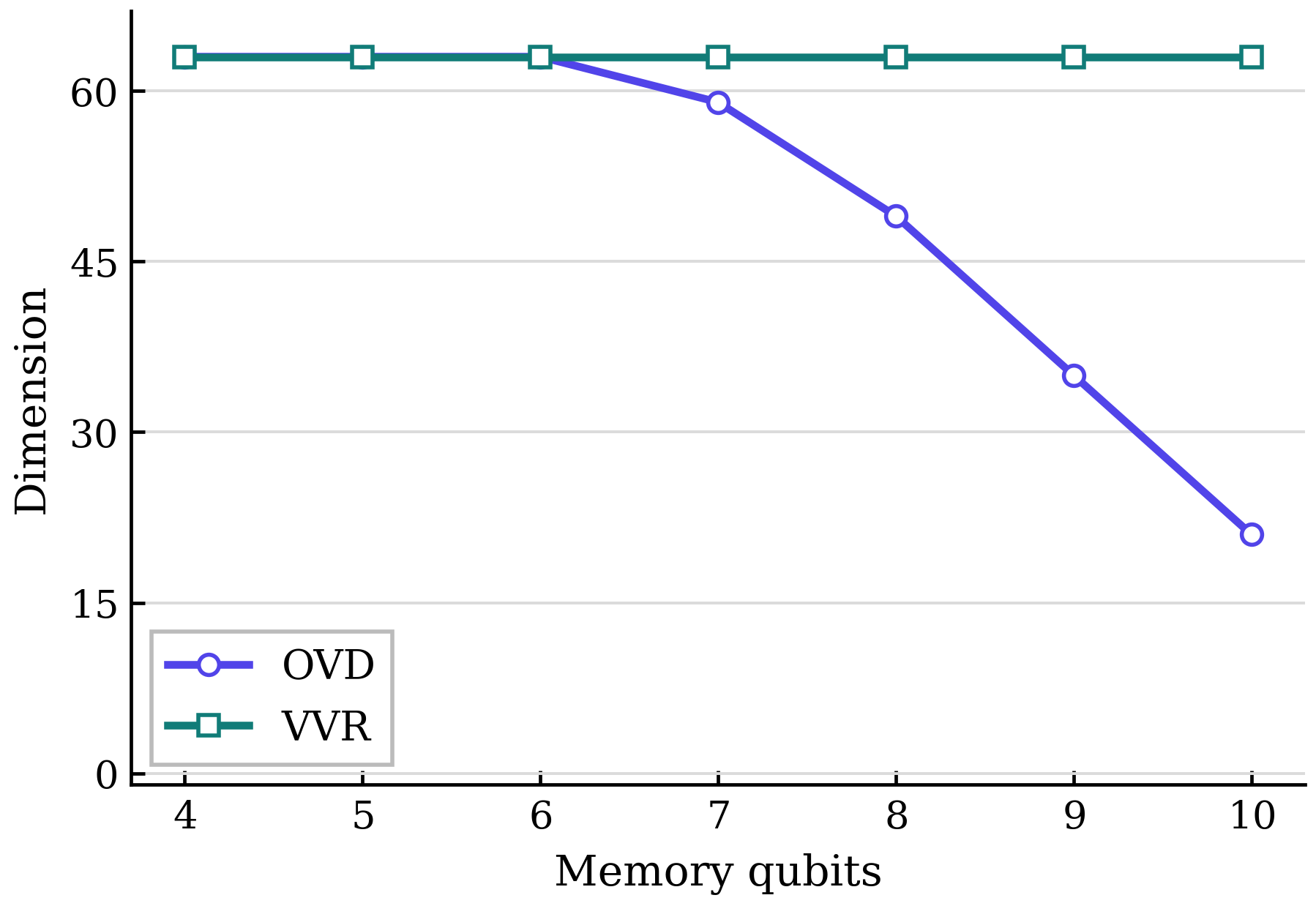}
 \end{minipage}
 \caption{\textbf{Increasing reservoir size can reduce the operational feature budget.} \textbf{a,b}, Mean supported visibility angles for Ising and random-circuit reservoirs as memory size increases. Shaded bands indicate one standard deviation across supported angles within each reservoir instance. Dashed curves show the full-space reference $\theta_a=\arcsin\sqrt{r_a/D_a}$, with $r_a=J$ and $D_a=4^{n_M+n_R}-1$. This reference derives from the sector-average visible weight; the supported angle distribution additionally depends on latent-space selection and alignment with measurement. The lower axes show the deviation $\Delta\theta_+=\langle\theta_+\rangle-\theta_a$. \textbf{c,d}, VVR and OVD for the corresponding size sweeps at the prescribed resolution. VVR remains saturated, while OVD decreases as fewer supported directions exceed the operational threshold.
 }
 \label{fig:scaling}
\end{figure*}

Figure~\ref{fig:scaling} examines this mechanism in Ising and random-circuit reservoirs as the memory size increases at fixed observable budget and reference sampling parameters. The Ising evolution duration and random-circuit depth are chosen to represent a scrambling regime, in which we approximate $W_{\mathrm{dyn}}$ by the full traceless joint operator space. Accordingly, we use $D_a=4^{n_M+n_R}-1$ and $r_a=J$ as references for the dynamically accessible dimension and its total visible weight. The corresponding dimension-ratio reference for the characteristic angle is $\arcsin\sqrt{r_a/D_a}$, which decreases exponentially with memory size at fixed observable budget and readout size. In both models, the supported angle distributions shift toward smaller values (Fig.~\ref{fig:scaling}a,b). The decrease is consistent with visibility dilution. The dashed curve is an ambient reference rather than a universal prediction for the supported mean, which also depends on latent-space selection as in Proposition~\ref{prop:latent-orientation}.

The dimension counts show the corresponding operational loss (Fig.~\ref{fig:scaling}c,d). VVR remains saturated while OVD decreases, indicating that algebraically surviving directions become unresolved at the prescribed resolution. These observations support the visibility-dilution mechanism without independently establishing dimension-ratio scaling for the complete dynamically accessible sector. Reservoir growth can therefore coexist with a smaller finite-resolution temporal feature budget. The next section examines how changes in operational access are reflected in learning performance.

\subsection{Accessible temporal features and learning performance}
\label{sec:performance}

The geometric spectrum can be connected to the total linear approximation capacity of the generated temporal functions once the input-induced response amplitudes and measurement noise are specified. Fix the truncation, write $s=d_{P,L}^{\mathrm{lat}}$ and $O=O_{P,L}$, and let $\mu$ denote the chosen distribution of input histories. Let $z\in\R^s$ be the coordinates of $Z_t^{(P,L)}-\mathbb E[Z_t^{(P,L)}]$ in its HS-orthonormal latent basis. Although the response is operator-valued, each coordinate $z_\ell(\mathbf v_t)$ is a scalar function of input history.

Consider centered measured features $y=Oz+\eta$, where $\mathbb E[\eta\mid z]=0$. Expectations include the input-history distribution and, where applicable, measurement noise. This model isolates the truncated signal and sampling noise; Section~\ref{sec:methods-truncation} controls the additional truncation error.

\begin{theorem}[Population linear capacity and geometric access]
\label{thm:population-capacity}
Let $\Sigma=\mathbb E[zz^{\mathsf T}]\succ0$ and $V=\mathbb E[\eta\eta^{\mathsf T}]\succ0$. For $x=\Sigma^{-1/2}z$, define the total optimal linear reconstruction capacity
\begin{equation}
 \begin{aligned}
 C_{\mathrm{lin}}&:=\sum_{\ell=1}^{s}
 \left[1-\min_{w\in\R^J}\mathbb E\bigl[(x_\ell-w^{\mathsf T}y)^2\bigr]\right],\\
 A&:=V^{-1/2}O\Sigma^{1/2}.
 \end{aligned}
 \label{eq:population-capacity-definition}
\end{equation}
If $\lambda_\ell$ are the eigenvalues of $A^{\mathsf T}A$, including zeros, then
\begin{equation}
 C_{\mathrm{lin}}=\sum_{\ell=1}^{s}\frac{\lambda_\ell}{1+\lambda_\ell}.
 \label{eq:population-capacity}
\end{equation}
If $aI_s\preceq\Sigma\preceq bI_s$ and $v_-I_J\preceq V\preceq v_+I_J$, where $a,v_->0$, their decreasingly ordered spectra satisfy
\begin{equation}
 \begin{gathered}
 \kappa_-\sigma_\ell(O)^2\leq\lambda_\ell
 \leq\kappa_+\sigma_\ell(O)^2,\\
 \kappa_-=\frac{a}{v_+},\qquad\kappa_+=\frac{b}{v_-}.
 \end{gathered}
 \label{eq:capacity-spectral-sandwich}
\end{equation}
Singular values are padded with zeros to length $s$.
\end{theorem}

The transformation $x=\Sigma^{-1/2}z$ combines and rescales the scalar temporal functions so that each has unit variance and distinct functions are uncorrelated under the chosen input distribution. This normalization concerns the temporal functions; the HS normalization concerns the underlying operator basis. Each summand in $C_{\mathrm{lin}}$ is the fraction of one basis function's unit variance recovered by its best linear prediction from $y$. Summing over a complete orthonormal basis makes $C_{\mathrm{lin}}$ independent of which such basis is used. The functions $x_\ell$ therefore provide a basis for evaluating the aggregate approximation capacity of the space of scalar temporal functions present in the truncated response, without selecting a particular benchmark target.

The spectral formula identifies the temporal modes contributing to this capacity. If $v_\ell$ are orthonormal eigenvectors of $A^{\mathsf T}A$, then $\psi_\ell(\mathbf v_t)=v_\ell^{\mathsf T}x(\mathbf v_t)$ are orthonormal temporal functions with individual reconstruction capacities $\lambda_\ell/(1+\lambda_\ell)$. Modes with large signal-to-noise strength $\lambda_\ell$ contribute nearly one unit, weak modes contribute a fraction, and null modes contribute zero. Thus $0\leq C_{\mathrm{lin}}\leq q$, where $q=\rank O$, is an effective number of recoverable temporal functions. The strengths $\lambda_\ell$ depend jointly on the response covariance, measurement map and noise through $A$.

Writing $G=O\Sigma O^{\mathsf T}$ for the covariance of the noiseless measured signal gives the equivalent expression
\begin{equation}
 C_{\mathrm{lin}}=\Tr\!\left[(G+V)^{-1}G\right].
 \label{eq:population-capacity-trace}
\end{equation}
This is the centered capacity trace underlying information-processing and resolvable expressive capacity, with the temporal modes above giving its eigentask interpretation~\cite{Dambre2012,Hu2023REC}. For physical sampling noise satisfying $\mathbb E[\eta\mid\mathbf v_t]=0$, functions orthogonal to $\Span\{z_1(\cdot),\ldots,z_s(\cdot)\}$ under the input-history distribution have zero linear capacity in this truncated model, so extending the orthonormal target basis beyond this space adds no contribution. The geometric comparison in Eq.~\eqref{eq:capacity-spectral-sandwich} connects this statistical capacity to the measurement strengths of HS-normalized latent directions.

For an explicit connection to OVD, let $q=\rank O$, $\omega=\omega_{P,L}(\epsilon)$, $t_\epsilon=\min\{4\epsilon^2,1\}$ and $F_\kappa(t)=\kappa t/(1+\kappa t)$. Under the covariance bounds of Theorem~\ref{thm:population-capacity}, each nonzero squared singular strength $u_\ell=\sigma_\ell(O)^2$ contributes between $F_{\kappa_-}(u_\ell)$ and $F_{\kappa_+}(u_\ell)$ to the capacity. Splitting these strengths into the $\omega$ values above $4\epsilon^2$ and the remaining $q-\omega$ values, and using the monotonicity of $F_\kappa$, gives
\begin{equation}
 \omega F_{\kappa_-}(t_\epsilon)
 \leq C_{\mathrm{lin}}\leq\omega F_{\kappa_+}(1)
 +(q-\omega)F_{\kappa_+}(t_\epsilon).
 \label{eq:ovd-capacity-bounds}
\end{equation}
These bounds require control of both response covariance and noise; OVD alone does not supply either. Unlike the threshold count in OVD, $C_{\mathrm{lin}}$ weights each mode continuously by its reconstruction quality, so OVD can vanish while a nonzero statistical capacity remains. If $V=V_1/N$ at fixed dynamics and input distribution, Eq.~\eqref{eq:population-capacity} becomes $C_{\mathrm{lin}}(N)=\sum_\ell N\lambda_\ell^{(1)}/(1+N\lambda_\ell^{(1)})$, where $\lambda_\ell^{(1)}$ are the eigenvalues of $\Sigma^{1/2}O^{\mathsf T}V_1^{-1}O\Sigma^{1/2}$. Increasing $N$ improves the recovery of existing modes, and the capacity approaches $q$ as $N\to\infty$. This population quantity evaluates aggregate approximation capacity; performance on a prescribed target additionally depends on its overlap with the generated temporal functions and on finite-training effects. The proof of Theorem~\ref{thm:population-capacity} is given in Supplementary Section~S4.1.

We now examine the relevance of operational access to short-term memory, nonlinear channel equalization and Mackey--Glass prediction. The regression tasks use affine predictions, whereas channel equalization uses affine logits followed by softmax. Both are trained on measured reservoir features. Neither the targets nor the fitted readouts enter the construction of the Volterra metrics.

\begin{figure*}[t]
 \centering

 \begin{minipage}[t]{0.32\textwidth}
  \vspace{0pt}
  \noindent\textbf{a}\par\vspace{0.2ex}
  \centering
  \includegraphics[width=\linewidth]{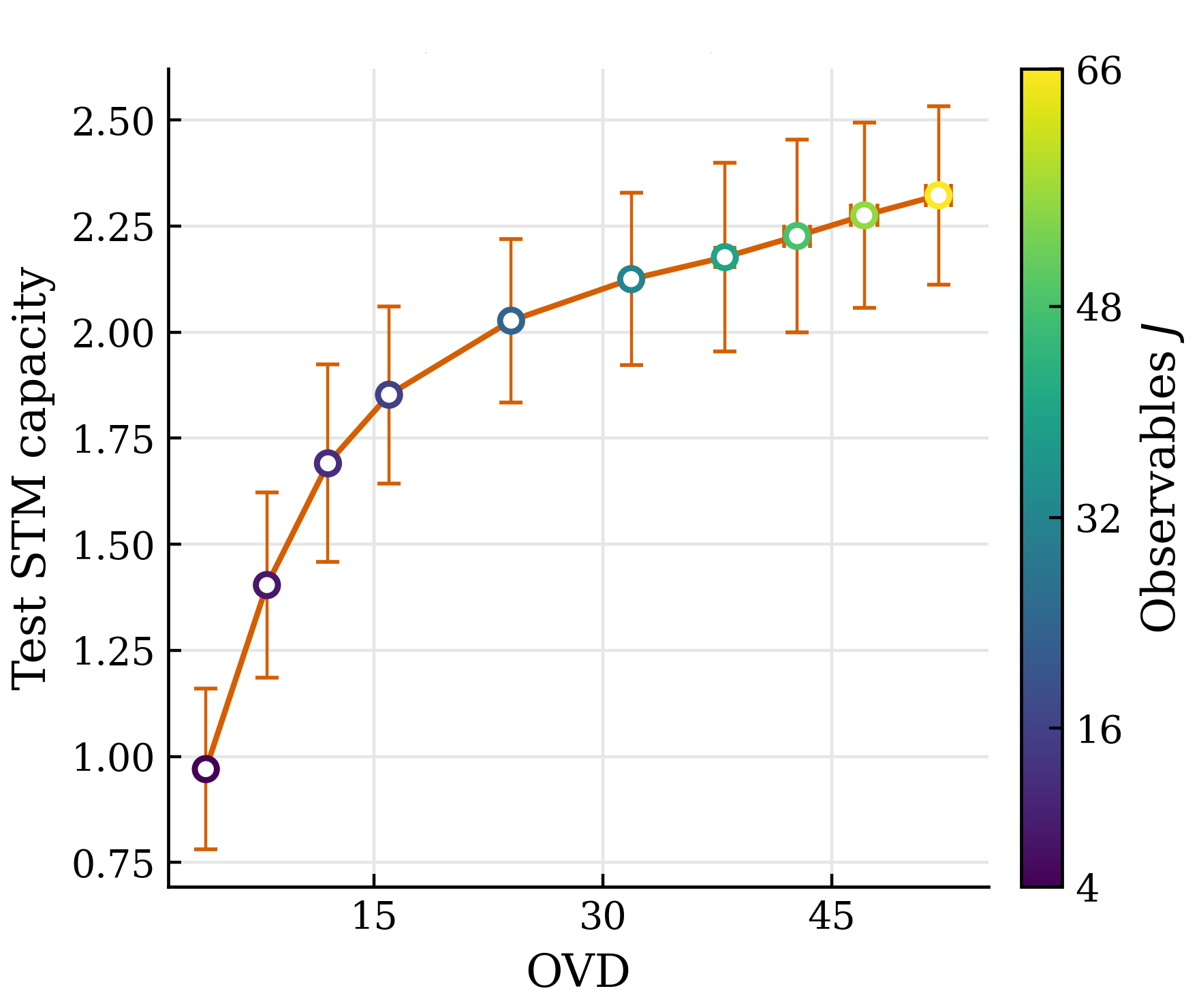}
 \end{minipage}\hfill
 \begin{minipage}[t]{0.32\textwidth}
  \vspace{0pt}
  \noindent\textbf{b}\par\vspace{0.2ex}
  \centering
  \includegraphics[width=\linewidth]{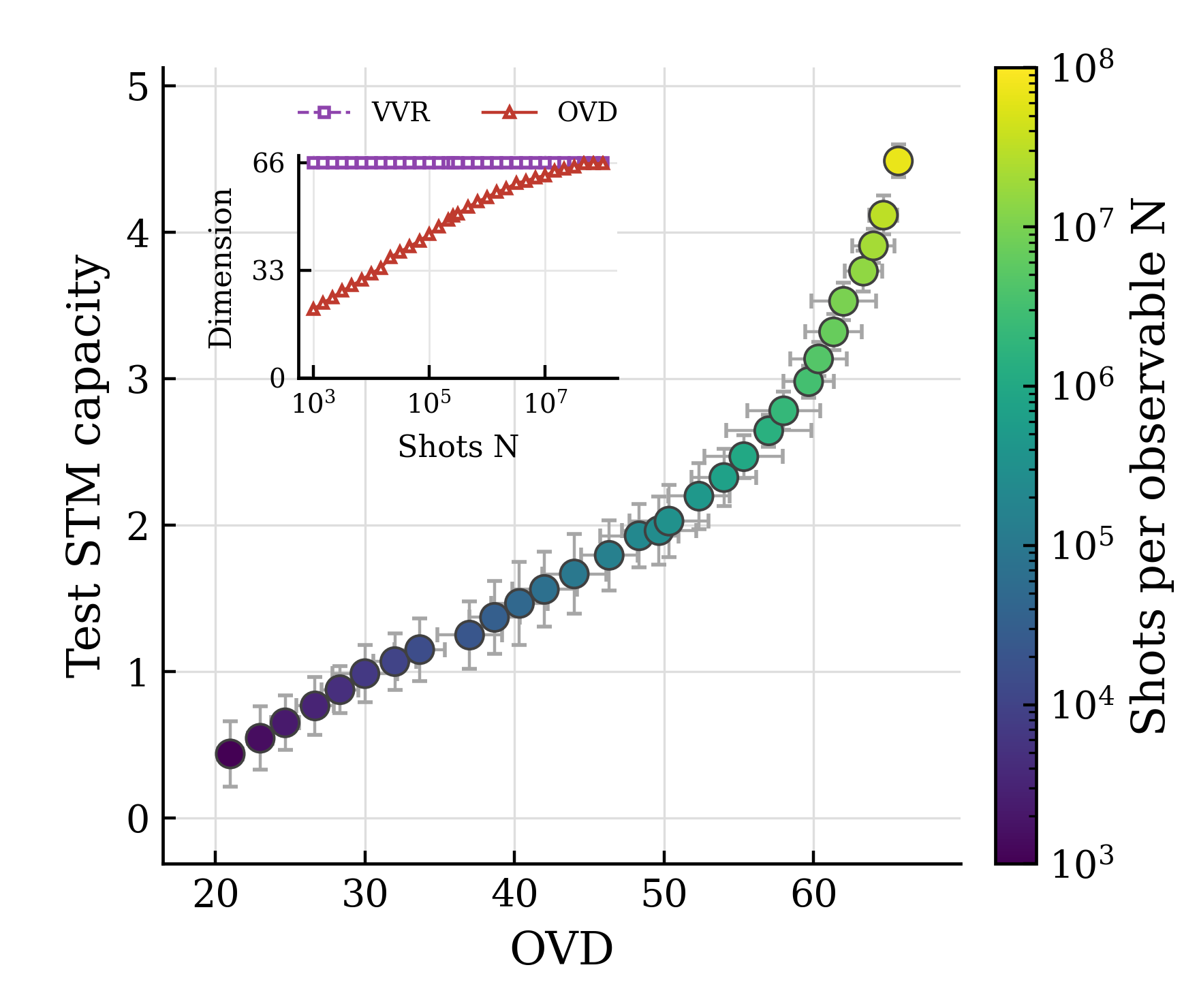}
 \end{minipage}\hfill
 \begin{minipage}[t]{0.32\textwidth}
  \vspace{0pt}
  \noindent\textbf{c}\par\vspace{0.2ex}
  \centering
  \includegraphics[width=\linewidth]{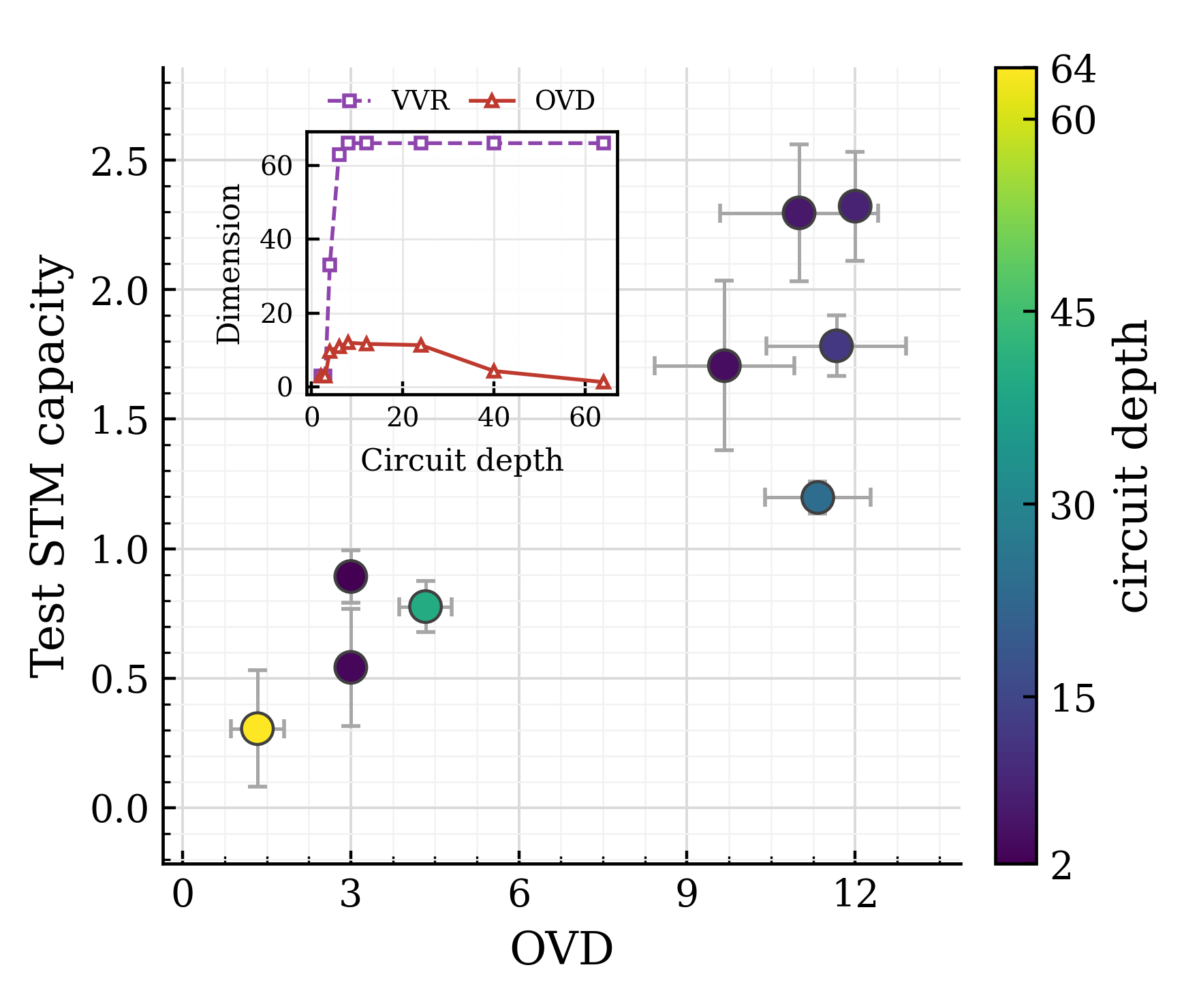}
 \end{minipage}

 \par\vspace{0.8ex}

 \begin{minipage}[t]{0.32\textwidth}
  \vspace{0pt}
  \noindent\textbf{d}\par\vspace{0.2ex}
  \centering
  \includegraphics[width=\linewidth]{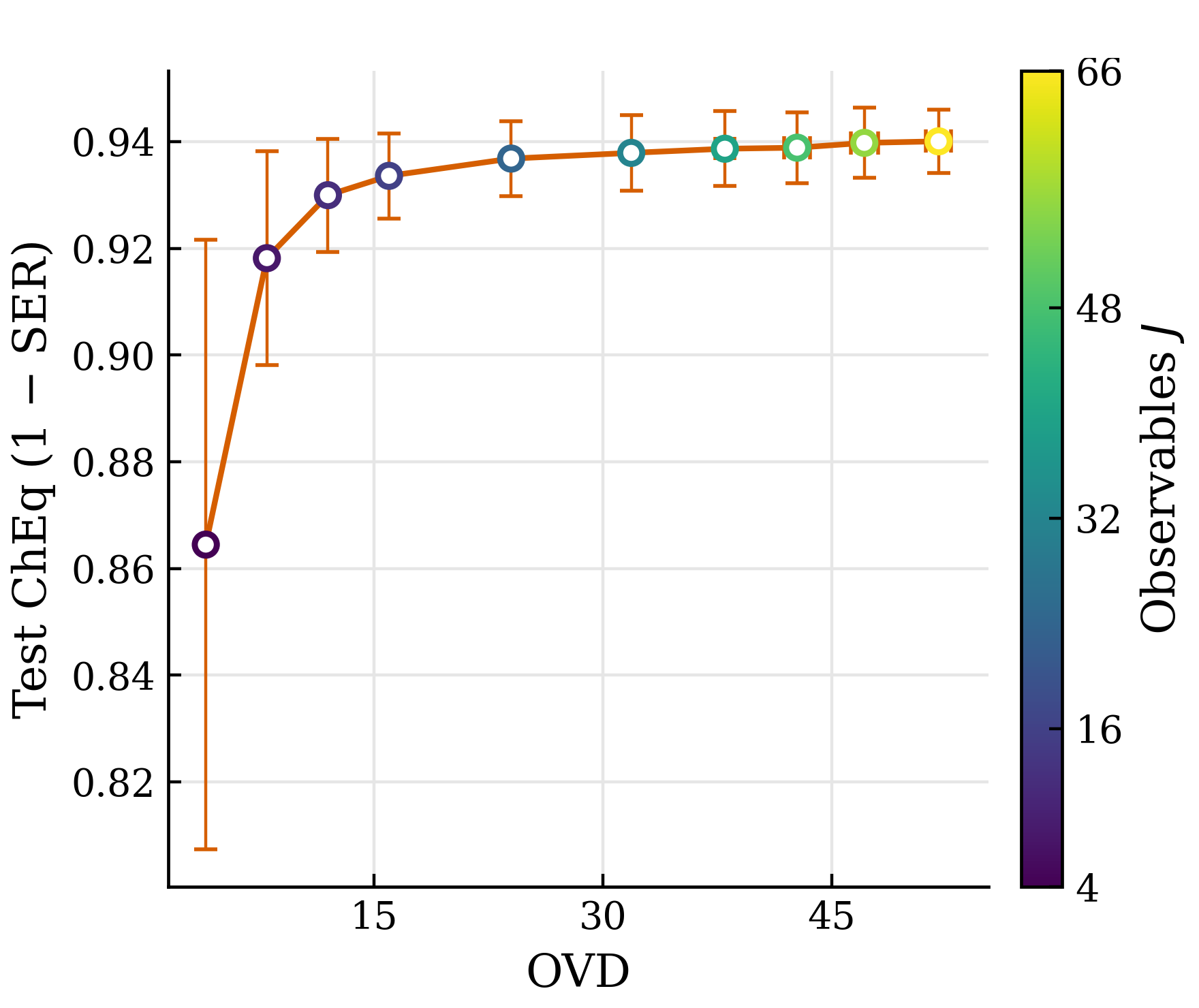}
 \end{minipage}\hspace{0.02\textwidth}%
 \begin{minipage}[t]{0.32\textwidth}
  \vspace{0pt}
  \noindent\textbf{e}\par\vspace{0.2ex}
  \centering
  \includegraphics[width=\linewidth]{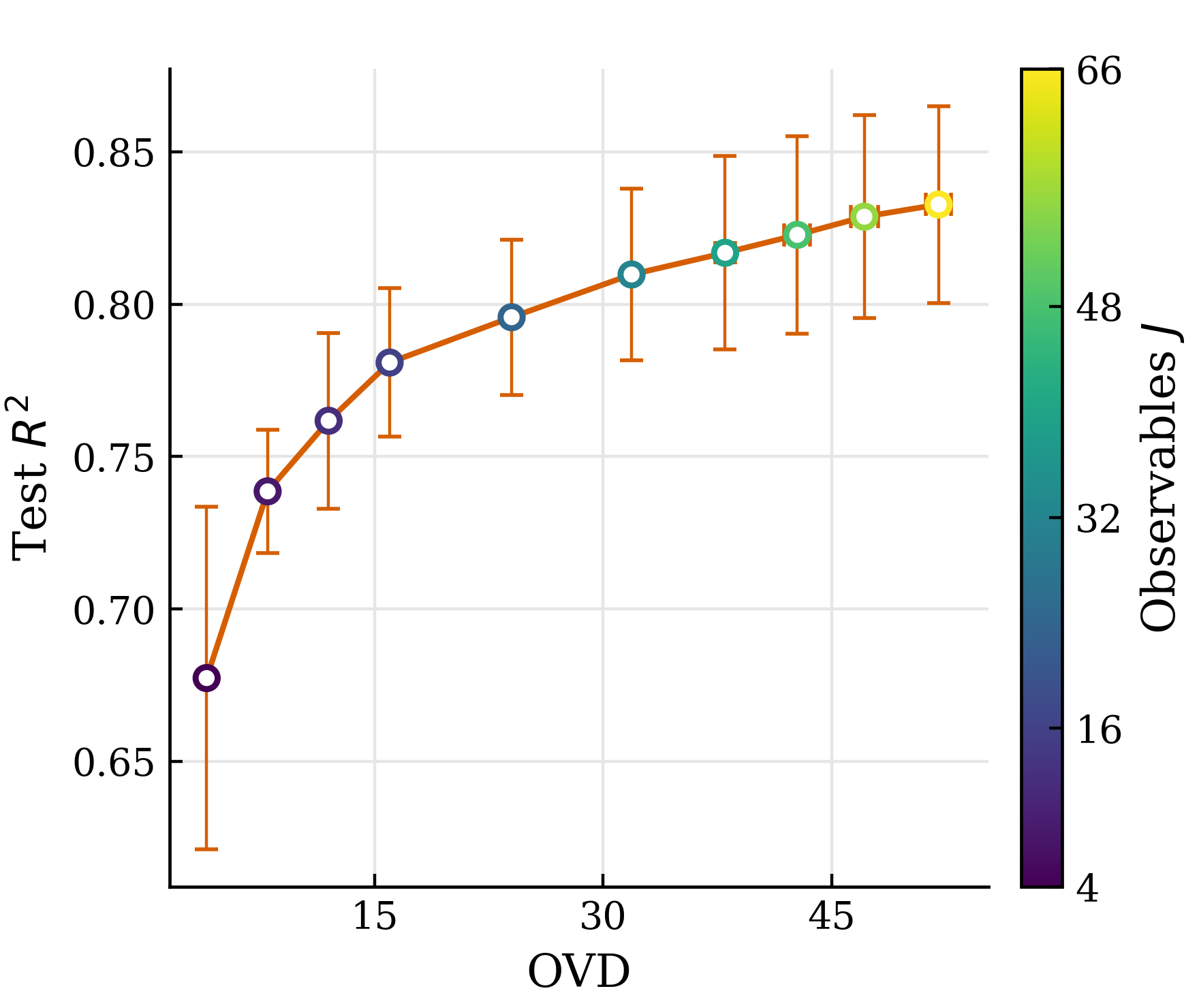}
 \end{minipage}

 \caption{\textbf{Operational access and temporal learning.} \textbf{a}, Short-term-memory score versus OVD under a nested observable-budget sweep at fixed dynamics. \textbf{b}, Short-term memory under a shot-budget sweep with fixed dynamics and observables. \textbf{c}, Short-term memory under a circuit-depth sweep. \textbf{d,e}, Channel-equalization accuracy and Mackey--Glass prediction score under observable-budget sweeps. The insets in \textbf{b,c} show VVR and OVD. In \textbf{a,d,e}, points show means over five observable families with ten independent shot-sampling repeats per family, giving 50 score evaluations per setting. In \textbf{b,c}, points show means over ten shot-sampling repeats for one fixed observable family. Vertical error bars indicate one standard deviation across the corresponding score evaluations. Horizontal error bars in \textbf{a,d,e} indicate the population standard deviation of OVD across the five observable families.}

 \label{fig:performance}
\end{figure*}

Short-term-memory performance increases with OVD as a nested observable family expands at fixed dynamics (Fig.~\ref{fig:performance}a). Channel equalization and Mackey--Glass prediction show the same positive association, with task-dependent saturation (Fig.~\ref{fig:performance}d,e). Within each sweep, the task sequence and dynamics remain fixed, so the intervention changes access to an existing temporal response. These results connect the observable-resource trends in Section~\ref{sec:resources} to useful information available to a trained readout.

The shot-budget sweep tests the role of measurement precision directly (Fig.~\ref{fig:performance}b). With dynamics and observables fixed, Pauli outcomes are sampled at increasing repetition budgets. Improved precision accompanies larger OVD and better delayed-input reconstruction. All learning experiments use finite-shot feature estimates; the observable-budget benchmarks vary the measured observable family while holding the repetition count per observable fixed, so the total number of separately sampled outcomes increases with the observable budget. These physical sampling budgets are distinct from the nominal resolution parameter in Fig.~\ref{fig:resources}e, which changes the threshold applied to a fixed response spectrum.

Changing circuit depth changes both accessibility and the temporal functions carried by the generated response (Fig.~\ref{fig:performance}c). Reservoirs with similar OVD can therefore have different scores. Kernel amplitudes, input statistics, target alignment and readout regularization also influence learning. OVD provides a task-independent budget of normalized directions accessible through measurement, whose association with performance is clearest under controlled changes in observable access or precision.

\section{Discussion}

The temporal representation generated by a quantum reservoir reaches a learner through a measurement interface. The latent dimension, VVR and OVD distinguish generation, algebraic transmission and finite-resolution access within a specified Volterra truncation. This hierarchy gives operational content to reservoir expressivity while preserving the distinction between a representation budget and its usefulness for a particular target. The capacity bounds make the roles of input-induced response amplitudes and measurement noise explicit, while optimal observable selection identifies the strongest geometric access allowed by the readout subsystem. The benchmark results associate improved access with better temporal learning under the controlled measurement interventions examined here.

The geometric analysis supports two complementary design strategies: improve the alignment of generated temporal features with the measurement interface, or preserve a subset of strongly exposed directions even when the mean visibility is low. Controlling locality and operator spreading may help, although the resulting measurement access must be considered alongside the richness and classical simulability of the dynamics.

The current numerical construction generally requires resources that grow exponentially with system size, as accurately determining the dimensions involves constructing an HS-orthonormal representation of the generated temporal operator space. Approximate surrogates or more efficient sample-based methods for estimating the relevant spectra would therefore be valuable for larger reservoirs. Extensions to trajectory-based approaches, conditional measurements, adaptive readout and feedback could further reveal how measurement backaction modifies temporal feature generation and accessibility. Non-reset architectures would provide another setting in which to examine the interplay between persistent quantum memory and classical access.

\section{Methods}

\subsection{Reservoir models and measurement protocol}
\label{sec:methods-models}

The scrambling reservoir is a local random circuit on an open chain, with memory followed by readout. Successive layers act on alternating sets of disjoint nearest-neighbor bonds, starting with $(0,1),(2,3),\ldots$ and then $(1,2),(3,4),\ldots$. One depth unit is one disjoint-bond layer. The two-qubit gates are sampled once for each circuit realization and reused across inputs and timesteps. At fixed size and realization, longer circuits share a common gate prefix. Alongside ordinary Haar-random $U(4)$ gates, we use scaled variants obtained from the complex Schur decomposition: principal eigenphases are centered by subtracting their mean, multiplied by a bond-strength factor and reconstructed in the same basis. Separate factors within memory, across the memory--readout boundary and within readout tune the dynamics at fixed circuit layout and depth.

The random-circuit reservoirs use repeated input encoding: the same input is applied through a local $Z$ rotation before every gate layer,
\begin{equation}
 \begin{aligned}
 R_D(u_t)&=\exp[-i(g_{\mathrm{in}}/D)u_tZ_{M_j}],\\
 U_{\mathrm{rep}}(u_t)&=C_{D-1}R_D(u_t)\cdots C_0R_D(u_t).
 \end{aligned}
 \label{eq:repeated-input-encoding}
\end{equation}
Here $C_\ell$ is a fixed gate layer, $D$ is the circuit depth, and the encoding duration is absorbed into $g_{\mathrm{in}}$. Dividing the rotation strength by $D$ keeps the total nominal input strength fixed as depth changes. The intervening gates mix the encoded information between injections. For the scaled six-memory, five-readout-qubit reservoir, the gate-strength factors are $0.25$ within memory, $0.60$ across the memory--readout boundary and $0.35$ within readout. Input is applied at $M_5$, with total encoding gain $g_{\mathrm{in}}=1.5$. The unscaled scrambling references use input at $M_0$ and total encoding gain $g_{\mathrm{in}}=3$ in the same convention.

The Ising reservoir used in Fig.~\ref{fig:scaling}a,c evolves according to $U(u)=\exp[-i\tau(H_0+u s_{\mathrm{in}}Z_{M_0})]$, with $\tau=3$ and $s_{\mathrm{in}}=1$. Memory and readout form an open chain. The transverse and longitudinal fields are $(0.9,0.2)$ on memory and $(0.8,0.1)$ on readout. Nearest-neighbor interactions within each register have $ZZ$ and $XX$ strengths $1$ and $0.15$, respectively; the single memory--readout interface has only a $ZZ$ coupling of strength $0.6$.

The readout is initialized and reset to its computational-basis zero state. After each update, readout Pauli expectations are evaluated from the joint state, the readout is traced out and reset, and only the unconditional reduced memory density matrix is propagated. The zero-input stationary memory reference is obtained starting from $I_M/d_M$ and accepted when $\|\Cmap_0(\bar\rho^M)-\bar\rho^M\|_{\HS}\leq10^{-10}$. Learning trajectories start from computational-basis zero memory. For finite-shot experiments, physical outcomes $\pm1$ are sampled from the Pauli observables and averaged to obtain the classical features, without feedback into the memory dynamics. Physical Pauli measurement coordinates entering the dimension metrics are divided by $\sqrt d$, together with their uncertainties, where $d=d_Md_R$ is the joint Hilbert-space dimension.

The depth sweeps in Fig.~\ref{fig:resources}a,b use the scaled six-memory, five-readout-qubit reservoir with $D\in\{2,3,4,8,12,\ldots,100\}$. The observable family contains all 66 weight-one and weight-two Paulis on $R_0,\ldots,R_3$, with identity on $R_4$. The truncations are $(P,L)=(3,3)$ and $(4,4)$, respectively. The observable-budget sweeps in Fig.~\ref{fig:resources}c,d fix $D=24$ and use nested unrestricted Pauli subsets on all five readout qubits, with observable counts ranging from 8 to 256 and from 8 to 512 for the corresponding truncations. All four panels average three circuit realizations and use nominal $N=4096$ and $c=7$.

The resolution sweep in Fig.~\ref{fig:resources}e rethresholds a fixed saved spectrum from an unscaled six-memory, four-readout-qubit circuit at depth $D=24$, using all 66 weight-one and weight-two readout Paulis and $(P,L)=(3,3)$. Nominal $N$ ranges from 100 to 50000, with $c=7$.

The memory-size sweeps in Fig.~\ref{fig:scaling} use $n_M=4,\ldots,10$, $n_R=3$, all 63 nonidentity readout Paulis and $(P,L)=(4,4)$. The Ising sweep uses nominal $N=10000$ and $c=1$. The scrambling sweep uses one unscaled depth-24 realization per size, nominal $N=512$ and $c=0.75$. The dashed reference curves use the ambient traceless joint operator-space dimension $4^{n_M+n_R}-1$ and visible rank 63.

In these structural sweeps, $N$ is a nominal shot-equivalent resolution parameter. At fixed register size and reference parameter $\delta$, the threshold defined in Section~\ref{sec:methods-resolution} scales as $c/\sqrt N$. Changing $c$ changes the criterion for counting operational directions. In finite-shot learning experiments, $N$ is instead the physical repetition count per distinct observable per timestep, so increasing it also reduces sampling noise in the task features. Circuit depth, gate strengths, register sizes, input location and encoding scheme change the reservoir dynamics.

The observable-composition diagnostic in Supplementary Section~S2.1 uses a six-memory, three-readout-qubit dephased $R_Y$ delay line. Each update shifts the six-bit memory, discards its oldest bit and inserts a bit prepared by $R_Y(0.7+u)|0\rangle$. Readout qubit $j$, initialized in $|0\rangle$, receives a memory-controlled rotation with angle $b_j+\sum_i w_{ji}z_i$, where $z_i\in\{-1,1\}$ are memory-$Z$ values. The seeded parameters satisfy $b_j\in[0.3,1.3]$ and $w_{ji}\in[-0.9,0.9]$. Complete computational-basis dephasing follows insertion and joint output.

For this diagnostic, we use $(P,L)=(4,4)$ and $J=20$ distinct readout Pauli observables with the same Hilbert--Schmidt normalization. Each family contains $20-k$ designated $Z$-only strings and $k$ draws from the unrestricted nonidentity Pauli pool, for $k=13,\ldots,20$, without duplicates. Unrestricted draws can also be $Z$-only, so the designated fraction need not equal the actual fraction. Five families are sampled at each setting. The geometry is computed from the model without finite-shot sampling, using nominal $N=10000$ and $c=1$ for the resolution criterion. Error bars report population standard deviations across the five observable families.

\subsection{Numerical construction and reliability conventions}
\label{sec:methods-construction}

Memory operators are propagated through the reduced channel, while temporal-feature dimensions and angles are evaluated in the Hilbert--Schmidt metric of the joint pre-measurement operator space. In the normalized memory-Pauli basis $F_\mu=P_\mu/\sqrt{d_M}$, with $F_0=I_M/\sqrt{d_M}$, the transfer matrix is $T_{\mu\nu}(u)=\Tr[F_\mu\Cmap_u(F_\nu)]$. Let $A(u)$ denote its traceless-to-traceless block and $\mathbf c(u)$ its identity-to-traceless column. The traceless memory coordinates obey
\begin{equation}
 \mathbf r_t=A(u_t)\mathbf r_{t-1}+\frac{\mathbf c(u_t)}{\sqrt{d_M}}.
 \label{eq:numerical-memory-transfer}
\end{equation}
At reference input $u_*=0$, the stationary coordinates and centered affine source satisfy
\begin{equation}
 \begin{aligned}
 [I-A(0)]\bar{\mathbf r}&=\frac{\mathbf c(0)}{\sqrt{d_M}},\\
 \mathbf b(u)&=A(u)\bar{\mathbf r}+\frac{\mathbf c(u)}{\sqrt{d_M}}-\bar{\mathbf r}.
 \end{aligned}
 \label{eq:numerical-centered-source}
\end{equation}
The joint reference is $\bar\sigma=U(0)(\bar\rho^M\otimes\rho_0^R)U(0)^\dagger$. Existence of a stationary state alone does not certify the uniform contraction assumptions in Supplementary Section~S1.2.

Ising Taylor coefficients are obtained from an augmented block exponential, while circuit coefficients are evaluated analytically from products of input rotations and gates. Mixed-lag kernels are composed from the recurrent channel coefficients and followed by the current joint-output map. Current-input derivatives of the output map, including its affine source, are retained. Equivalently, let $\mathcal F_L(v_0,\ldots,v_L)$ be the centered joint response with earlier inputs fixed at the reference. The kernel indexed by the multiplicity vector $\mathbf m=(m_0,\ldots,m_L)$ is
\begin{equation}
 \begin{gathered}
 \widehat h_{\mathbf m}=\frac{1}{\prod_{\tau=0}^{L}m_\tau!}\left.\frac{\partial^{|\mathbf m|}\mathcal F_L}{\partial v_0^{m_0}\cdots\partial v_L^{m_L}}\right|_{\mathbf v=0},\\
 1\leq|\mathbf m|\leq P.
 \end{gathered}
 \label{eq:numerical-joint-kernels}
\end{equation}
Here $v_0$ is the current centered input and $m_\tau$ counts occurrences of lag $\tau$ in the weakly ordered convention. The kernels are Hermitian and traceless in exact arithmetic. Small numerical residuals can be removed by Hermitian symmetrization and traceless projection before factorization.

Physical readout expectations can also be computed from memory through the effective observable
\begin{equation}
 \begin{aligned}
 E_j^P(u)=\Tr_R\bigl[&(I_M\otimes\rho_0^R)U(u)^\dagger\\
 &\times(I_M\otimes P_j)U(u)\bigr],
 \end{aligned}
 \label{eq:numerical-pulled-observable}
\end{equation}
which gives $f_{j,t}=\Tr[E_j^P(u_t)\rho^M_{t-1}]$. Its current-input dependence contributes to the measured kernel coefficients. This pullback is a device for evaluating expectations and does not replace the joint Hilbert--Schmidt metric by a memory-only metric.

For the direct construction, let $E_a$, $a=1,\ldots,d^2-1$, be a fixed Hilbert--Schmidt-orthonormal basis of traceless joint operators, and form
\begin{equation}
 H_{P,L}=\bigl[\operatorname{vec}_{\HS}(\widehat h_\lambda)\bigr]_{\lambda\in\Lambda_{P,L}}.
 \label{eq:numerical-kernel-matrix}
\end{equation}
Kernel columns are norm-equilibrated before block QR and SVD. Rank-revealing factorization yields an orthonormal basis $Q_{P,L}$ of the retained span; its column count is the numerical latent dimension. In the same joint coordinates, define
\begin{equation}
 \begin{aligned}
 R_{ja}&=\Tr[(I_M\otimes M_j)E_a],\\
 RR^{\mathsf T}&=I_J,\qquad O_{P,L}=RQ_{P,L}.
 \end{aligned}
 \label{eq:numerical-measurement-restriction}
\end{equation}
Latent rank and VVR retain singular values strictly above $10^{-9}$. These numerical cutoffs are separate from the operational threshold: the numerical rank of $O_{P,L}$ gives VVR, while its singular values above $2\epsilon$ give OVD. In the numerical benchmarks, the restriction matrix is computed from the known reservoir model.

An implicit construction uses the joint-kernel Gram matrix and measured coefficient matrix,
\begin{equation}
 \begin{aligned}
 G_{\lambda\mu}&=\Tr(\widehat h_\lambda\widehat h_\mu),\\
 C_{j\lambda}&=\Tr[(I_M\otimes M_j)\widehat h_\lambda].
 \end{aligned}
 \label{eq:numerical-gram-matrices}
\end{equation}
Let $V$ contain the retained eigenvectors of $G$, with positive eigenvalues collected in $\Lambda$. Then
\begin{equation}
 Q_{P,L}=H_{P,L}V\Lambda^{-1/2},\qquad O_{P,L}=CV\Lambda^{-1/2}.
 \label{eq:numerical-gram-restriction}
\end{equation}
The first identity specifies the orthonormal latent basis formally; the second evaluates its measurement restriction without assembling ambient coordinates. All overlaps use the same joint Hilbert--Schmidt metric. The equivalence of the direct and Gram representations on the same retained subspace is derived in Supplementary Section~S6.

Let $V_+$ contain the retained nonzero right singular vectors of $O_{P,L}$. Supported visibility statistics are computed from
\begin{equation}
 B^+=(O_{P,L}V_+)^{\mathsf T}(O_{P,L}V_+).
 \label{eq:numerical-supported-visibility}
\end{equation}
Its eigenvalues are the squared supported singular strengths, and the visibility angles satisfy $\theta_j^+=\arcsin\sigma_j(O_{P,L})$ for the supported indices. Angle summaries and isotropy defects exclude algebraically null directions. When the supported rank is zero, these summaries are undefined and the visible and operational dimension counts are zero.

\subsection{Resolution model and finite-shot calculations}
\label{sec:methods-resolution}

The operational threshold is distinct from the numerical tolerances used to retain kernel and measurement directions. If an estimated restricted matrix satisfies $\|\widehat O-O\|_{\mathrm{op}}\leq\epsilon$, singular-value perturbation theory~\cite{Bhatia1997} gives $|\sigma_j(\widehat O)-\sigma_j(O)|\leq\epsilon$. Thus $\sigma_j(O)>2\epsilon$ implies $\sigma_j(\widehat O)>\epsilon$, whereas an exactly null singular value can rise by at most $\epsilon$. This motivates the separation in Eq.~\eqref{eq:ovd}.

The reference resolution used in the numerical comparisons is
\begin{equation}
 \epsilon_{\mathrm{ref}}(N,\delta)=c\sqrt{\frac{\log(1/\delta)}{dN}},\qquad\delta=0.05.
 \label{eq:reference}
\end{equation}
Here $c$ is the reference coefficient in physical Pauli coordinates, $d=d_Md_R$, and the factor $1/\sqrt d$ converts the resolution to the normalized coordinates of Eq.~\eqref{eq:measurement}. Singular strengths and thresholds are rescaled together, preserving OVD under a change of coordinate normalization. At fixed register size, this factor can be absorbed into the coefficient; its size dependence must be retained when comparing registers at fixed physical sampling parameters. Fixing normalized $\epsilon$ instead defines a different resource comparison.

The reference model isolates inverse-square-root sampling dependence without assigning a protocol-specific confidence guarantee. The model-computed restriction matrices described in Section~\ref{sec:methods-construction} are distinct from the finite-shot task features described in Section~\ref{sec:methods-learning}. Entrywise and matrix concentration bounds for specified estimators are derived in Supplementary Section~S5.3~\cite{Hoeffding1963,Tropp2012}. A calibrated application must account for estimator variances, correlations, bias and any uncertainty in the latent basis.

\subsection{Learning experiments and statistical analysis}
\label{sec:methods-learning}

The classical readout is trained on physical Pauli feature estimates. For scalar STM and Mackey--Glass targets, we fit an affine ridge readout to the measured feature vector $\mathbf f_t$. With an unpenalized bias included in $\widetilde{\mathbf f}_t=(1,\mathbf f_t^{\mathsf T})^{\mathsf T}$, the prediction and fitted weights are
\begin{equation}
 \begin{aligned}
 \widehat s_t&=w^{\mathsf T}\widetilde{\mathbf f}_t,\\
 w&=\underset{w'}{\operatorname{argmin}}\biggl\{\sum_{t\in T_{\mathrm{train}}}\bigl(s_t-(w')^{\mathsf T}\widetilde{\mathbf f}_t\bigr)^2\\ 
 &\hspace{6em}+\lambda_{\mathrm{ridge}}\|w'_{\mathrm{feat}}\|_2^2\biggr\}.
 \end{aligned}
 \label{eq:ridge}
\end{equation}
Channel equalization uses an $L_2$-regularized four-class softmax readout with an unpenalized intercept. All scores are evaluated on held-out test data.

The observable-budget benchmarks use five seeded observable orderings, each defining nested observable subsets. At each observable budget, ten independent shot-sampling repeats are performed for each family, yielding 50 task-score evaluations. Points report the mean across these evaluations, and vertical error bars report their standard deviation. Horizontal error bars report the population standard deviation of OVD across the five observable families. The shot-budget and circuit-depth sweeps each use one fixed observable family and ten shot-sampling repeats per setting; their task-score means and standard deviations are calculated across these ten repeats.

The classical readout receives 128 feature slots plus bias. Features are repeated cyclically, with each copy divided by the square root of its observable's multiplicity; the first $128\bmod J$ observables receive one extra slot. Copies share the same measurement estimate and do not add independent information. Holding shots per distinct observable fixed means that the total number of sampled outcomes per timestep increases with the observable budget.

Additional STM scans vary the shot budget and circuit depth separately. The shot-budget sweep fixes depth $D=4$ and threshold coefficient $c=30$, with $N$ ranging from $10^3$ to $10^8$ shots per observable per timestep. Ten measurement realizations are used at each shot budget. The depth sweep varies the circuit depth up to $D=64$ at fixed $N=2^{18}=262144$ and $c=30$.

STM reconstructs delays $1$--$20$ from an independent uniform input stream on $[-1,1]$, with a separate linear output for each delay~\cite{Jaeger2002,Dambre2012}. Its reported capacity is $C_{\mathrm{STM}}=\sum_{d=1}^{20}\max\{0,R_d^2\}$. STM and Mackey--Glass use 200 reservoir washout steps, 600 training samples and 300 test samples.

The Mackey--Glass generator uses $(\beta,\gamma,\tau_{\mathrm{MG}},n)=(0.2,0.1,30,15)$, Euler step $1$, constant initial history $1.2$ and 200 signal warmup steps~\cite{MackeyGlass1977}. Training input indices are 200--799 and test input indices are 805--1104, with targets five steps ahead. Indices 800--804 are excluded from scoring. Performance is measured by test $R^2$.

Channel equalization uses symbols $s_t\in\{-3,-1,1,3\}$ and the causal filter $x_t=\sum_{k=0}^{7}a_ks_{t-k}$, with $(a_0,\ldots,a_7)=(1,0.18,-0.10,0.091,-0.05,0.04,0.03,0.01)$. The observed channel output is $x_t+0.06x_t^2-0.01x_t^3$ plus independent Gaussian noise of variance $0.1$. The target is the current symbol $s_t$. There are 20 training and 20 test messages of 100 symbols, with zero-padded channel history and the first 20 symbols discarded. Memory resets at message boundaries. A four-class softmax readout is scored by $1-\mathrm{SER}$.

For STM and Mackey--Glass, ridge regularization is selected separately per case and measurement realization from 37 logarithmically spaced values between $10^{-8}$ and $10^1$, using three purged chronological folds. Fitting minimizes summed squared error plus the ridge penalty in Eq.~\eqref{eq:ridge}, and validation uses mean squared error. Channel equalization uses L-BFGS-B softmax fitting with mean cross-entropy plus $\lambda\|W\|_{\mathrm F}^2/2$, selecting $\lambda\in\{10^{-6},10^{-4},10^{-2}\}$ using held-out training messages. Intercepts are unpenalized. Mackey--Glass and channel inputs are standardized using training data only; STM inputs are not standardized. Input streams, preprocessing, partitions and fitting procedures are held fixed within each sweep.

Memory-size angle bands show one population standard deviation across supported angles within each instance. Task-score vertical error bars combine the between-case variance of case means with within-case measurement-repeat variance. Horizontal error bars show the population standard deviation of OVD across cases.

\subsection{Truncation and finite-shot readout error}
\label{sec:methods-truncation}

A quantitative connection to the complete response requires control of the omitted Volterra terms. Fix a centered-input range $|v_t|\leq v_{\max}$, where $v_{\max}>0$ bounds the input amplitude. For each kernel with a Volterra index $\lambda$, let $p(\lambda)$ denote its nonlinear order and $\tau_{\max}(\lambda)$ its largest delay. We assume that the weighted coefficient sum
\begin{equation}
 K_{\varrho,\zeta}:=\sum_\lambda\|\widehat h_\lambda\|_{\HS}(\varrho v_{\max})^{p(\lambda)}\zeta^{\tau_{\max}(\lambda)}<\infty
 \label{eq:weighted-kernel-envelope}
\end{equation}
is finite for some dimensionless weights $\varrho>1$ and $\zeta>1$.

The factor $\varrho$ sets an amplitude margin: the coefficient sum remains finite at the larger input scale $\varrho v_{\max}$, beyond the actual input bound $v_{\max}$. The factor $\zeta^{\tau_{\max}(\lambda)}$ gives increasing weight to kernels involving longer delays. Finiteness of this sum therefore requires sufficient suppression of both high-order and long-delay contributions. These two margins yield the truncation factors $\varrho^{-(P+1)}$ and $\zeta^{-(L+1)}$ below. This weighted summability condition is stronger than absolute convergence on the actual input domain, which alone need not provide these geometric decay rates.

\begin{proposition}[Truncation and sampling guarantee]
\label{prop:truncation-sampling}
Under Eq.~\eqref{eq:weighted-kernel-envelope} and $|v_t|\leq v_{\max}$, the uniform truncation error satisfies
\begin{equation}
 \begin{aligned}
 \|Z_t-Z_t^{(P,L)}\|_{\HS}&\leq R_{P,L},\\
 R_{P,L}&:=K_{\varrho,\zeta}\max\{\varrho^{-(P+1)},\zeta^{-(L+1)}\}.
 \end{aligned}
 \label{eq:uniform-truncation-error}
\end{equation}
At one fixed input history, measure distinct nonidentity readout Pauli strings through $A_j=I_M\otimes P_j$ with $N_j$ independent shots each, independent also across observables. Let $d=d_Md_R$, $\bar f_j=\Tr(A_j\bar\sigma)$ be known, and define
\begin{equation}
 \begin{gathered}
 \widehat z_{j,t}=\frac{1}{\sqrt d}\left(\frac{1}{N_j}\sum_{a=1}^{N_j}X_{j,t}^{(a)}-\bar f_j\right),\\
 X_{j,t}^{(a)}\in\{-1,1\},
 \end{gathered}
 \label{eq:pauli-feature-estimator}
\end{equation}
where $\mathbb E[X_{j,t}^{(a)}]=\Tr(A_j\sigma_t)$. For a fixed readout $w$ and $0<\delta<1$, with probability at least $1-\delta$,
\begin{equation}
 \begin{aligned}
 &\left|w^{\mathsf T}\widehat z_t-w^{\mathsf T}\Mmap(Z_t^{(P,L)})\right|\\
 &\quad\leq\|w\|_2 R_{P,L}+\left[\frac{2\log(2/\delta)}{d}\sum_{j=1}^{J}\frac{w_j^2}{N_j}\right]^{1/2}.
 \end{aligned}
 \label{eq:truncation-sampling-bound}
\end{equation}
Here $I_M\otimes M_j=A_j/\sqrt d$, consistently with Eq.~\eqref{eq:measurement}.
\end{proposition}
The proof is given in Supplementary Section~S4.3.

For a learned readout, the sampling bound applies conditionally on training data independent of the evaluation shots. For guarantees at $T_{\mathrm{eval}}$ specified timesteps, a union bound replaces $\delta$ by $\delta/T_{\mathrm{eval}}$. Separately estimated reference values add their own uncertainty. The factor $d^{-1/2}$ is a coordinate normalization, not a reduction in physical shot cost: physical Pauli readout weights are $w/\sqrt d$. The total number of independently sampled outcomes per timestep is $B_{\mathrm{shot}}=\sum_jN_j$, which equals $JN$ for equal allocation. Equations~\eqref{eq:pauli-feature-estimator}--\eqref{eq:truncation-sampling-bound} certify feature and readout errors, not estimation of $O_{P,L}$; the latter additionally requires reconstruction of the latent basis and its measurement signatures, with estimator calibration discussed in Supplementary Section~S5.3.

These bounds also delimit the truncated capacity model. Centering the full and truncated responses about their respective input means makes their difference at most $2R_{P,L}$ in HS norm. For any fixed $w$ and target, the corresponding root-mean-square prediction errors therefore differ by at most $2\|w\|_2R_{P,L}$ when the same additive measurement noise is used in the comparison. This follows from the $L^2$ triangle inequality and does not assume that the omitted terms are unbiased noise.

\section*{Acknowledgments}

M.K. and T.K. have been supported by the National Research Foundation of Korea (NRF) grant (No. RS-2024-00442855, No. RS-2024-00413957), and the Institute of Information \& Communications Technology Planning \& Evaluation (IITP) grant (No. RS-2022-II221040), all of which are funded by the Korean government (MSIT). J.B. was supported by the Korean ARPA-H Project through the Korea Health Industry Development Institute (KHIDI), funded by the Ministry of Health \& Welfare, Republic of Korea (RS-2025-25456722), and the Ministry of Trade, Industry and Resources (MOTIR), Korea, under the project ``Industrial Technology Infrastructure Program'' (RS-2024-00466693). We acknowledge the Yonsei University Quantum Computing Project Group for providing support and access to the Quantum System One, which is operated at Yonsei University.

\section*{Author contributions}

M.K. conceived the project, identified the issues, proposed the algorithm, developed the mathematical framework, implemented the numerical simulations, and wrote the manuscript. J.B. contributed to developing and refining the theoretical results, particularly those concerning learning capacities, optimal readout observables, and truncation errors, and revised the manuscript. J.B. and T.K. supervised the project, including research planning and manuscript preparation.

\section*{Code availability}

The simulation and analysis code used in this study is available from the corresponding authors upon reasonable request.

\section*{Data availability}
The data supporting the findings of this study are available from the corresponding authors upon reasonable request.

\section*{Competing interests}
The authors declare no competing interests.

\section*{Use of generative AI}

The authors used generative AI for language refinement and for assistance with organizing and formatting research data and manuscript materials. All output was critically reviewed and verified by the authors, who take full responsibility for the manuscript and its reported results.

\bibliographystyle{apsrev4-2}
\bibliography{references}

@techreport{Jaeger2001,
  author        = {Jaeger, Herbert},
  title         = {The ``echo state'' approach to analysing and training recurrent neural networks},
  institution   = {German National Research Center for Information Technology},
  number        = {GMD Report 148},
  address       = {Sankt Augustin, Germany},
  year          = {2001}
}

@article{Maass2002,
  author        = {Maass, Wolfgang and Natschl{\"a}ger, Thomas and Markram, Henry},
  title         = {Real-time computing without stable states: A new framework for neural computation based on perturbations},
  journal       = {Neural Computation},
  volume        = {14},
  number        = {11},
  pages         = {2531--2560},
  year          = {2002},
  doi           = {10.1162/089976602760407955}
}

@article{Verstraeten2007,
  author        = {Verstraeten, David and Schrauwen, Benjamin and D'Haene, Michiel and Stroobandt, Dirk},
  title         = {An experimental unification of reservoir computing methods},
  journal       = {Neural Networks},
  volume        = {20},
  number        = {3},
  pages         = {391--403},
  year          = {2007},
  doi           = {10.1016/j.neunet.2007.04.003}
}

@article{BoydChua1985,
  author        = {Boyd, Stephen and Chua, Leon O.},
  title         = {Fading Memory and the Problem of Approximating Nonlinear Operators with {Volterra} Series},
  journal       = {IEEE Transactions on Circuits and Systems},
  volume        = {32},
  number        = {11},
  pages         = {1150--1161},
  year          = {1985},
  doi           = {10.1109/TCS.1985.1085649}
}

@article{Dambre2012,
  author        = {Dambre, Joni and Verstraeten, David and Schrauwen, Benjamin and Massar, Serge},
  title         = {Information Processing Capacity of Dynamical Systems},
  journal       = {Scientific Reports},
  volume        = {2},
  pages         = {514},
  year          = {2012},
  doi           = {10.1038/srep00514}
}

@article{Fujii2017,
  author        = {Fujii, Keisuke and Nakajima, Kohei},
  title         = {Harnessing Disordered-Ensemble Quantum Dynamics for Machine Learning},
  journal       = {Physical Review Applied},
  volume        = {8},
  pages         = {024030},
  year          = {2017},
  doi           = {10.1103/PhysRevApplied.8.024030}
}

@article{Ghosh2019,
  author        = {Ghosh, Sanjib and Opala, Andrzej and Matuszewski, Micha{\l} and Paterek, Tomasz and Liew, Timothy C. H.},
  title         = {Quantum reservoir processing},
  journal       = {npj Quantum Information},
  volume        = {5},
  pages         = {35},
  year          = {2019},
  doi           = {10.1038/s41534-019-0149-8}
}

@article{GrossRieser2026,
  author  = {Gross, Markus and Rieser, Hans-Martin},
  title   = {Theory and interpretability of quantum extreme learning machines: A {Pauli} transfer matrix approach},
  journal = {Physical Review A},
  volume  = {114},
  pages   = {042459},
  year    = {2026},
  doi     = {10.1103/6tm7-3tn4}
}

@article{Nakajima2019,
  author        = {Nakajima, Kohei and Fujii, Keisuke and Negoro, Makoto and Mitarai, Kosuke and Kitagawa, Masahiro},
  title         = {Boosting Computational Power through Spatial Multiplexing in Quantum Reservoir Computing},
  journal       = {Physical Review Applied},
  volume        = {11},
  pages         = {034021},
  year          = {2019},
  doi           = {10.1103/PhysRevApplied.11.034021}
}

@misc{Tran2020,
  author        = {Tran, Quoc Hoan and Nakajima, Kohei},
  title         = {Higher-Order Quantum Reservoir Computing},
  year          = {2020},
  eprint        = {2006.08999},
  archivePrefix = {arXiv},
  primaryClass  = {quant-ph}
}

@article{Chen2020,
  author        = {Chen, Jiayin and Nurdin, Hendra I. and Yamamoto, Naoki},
  title         = {Temporal Information Processing on Noisy Quantum Computers},
  journal       = {Physical Review Applied},
  volume        = {14},
  pages         = {024065},
  year          = {2020},
  doi           = {10.1103/PhysRevApplied.14.024065}
}

@article{Schuld2021,
  author        = {Schuld, Maria and Sweke, Ryan and Meyer, Johannes Jakob},
  title         = {Effect of data encoding on the expressive power of variational quantum-machine-learning models},
  journal       = {Physical Review A},
  volume        = {103},
  pages         = {032430},
  year          = {2021},
  doi           = {10.1103/PhysRevA.103.032430}
}

@article{MartinezPena2023a,
  author        = {Mart{\'i}nez-Pe{\~n}a, Rodrigo and Ortega, Juan-Pablo},
  title         = {Quantum Reservoir Computing in Finite Dimensions},
  journal       = {Physical Review E},
  volume        = {107},
  pages         = {035306},
  year          = {2023},
  doi           = {10.1103/PhysRevE.107.035306}
}

@article{MartinezPena2023b,
  author        = {Mart{\'i}nez-Pe{\~n}a, Rodrigo and Nokkala, Johannes and Giorgi, Gian Luca and Zambrini, Roberta and Soriano, Miguel C.},
  title         = {Information Processing Capacity of Spin-Based Quantum Reservoir Computing Systems},
  journal       = {Cognitive Computation},
  volume        = {15},
  number        = {5},
  pages         = {1440--1451},
  year          = {2023},
  doi           = {10.1007/s12559-020-09772-y}
}

@misc{Gonon2023,
  author        = {Gonon, Lukas and Grigoryeva, Lyudmila and Ortega, Juan-Pablo},
  title         = {Reservoir kernels and {Volterra} series},
  year          = {2022},
  eprint        = {2212.14641},
  archivePrefix = {arXiv},
  primaryClass  = {cs.LG}
}

@misc{Schutte2025,
  author        = {Sch{\"u}tte, Nils-Erik and G{\"o}tting, Niclas and M{\"u}ntinga, Hauke and List, Meike and Brunner, Daniel and Gies, Christopher},
  title         = {Expressivity of quantum reservoir computers},
  year          = {2025},
  eprint        = {2501.15528},
  archivePrefix = {arXiv},
  primaryClass  = {quant-ph}
}

@article{Mujal2023,
  author        = {Mujal, Pere and Mart{\'i}nez-Pe{\~n}a, Rodrigo and Giorgi, Gian Luca and Soriano, Miguel C. and Zambrini, Roberta},
  title         = {Time-series quantum reservoir computing with weak and projective measurements},
  journal       = {npj Quantum Information},
  volume        = {9},
  pages         = {16},
  year          = {2023},
  doi           = {10.1038/s41534-023-00682-z}
}

@article{Balasubramanian2022,
  author        = {Balasubramanian, Vijay and Caputa, Pawel and Magan, Javier M. and Wu, Qingyue},
  title         = {Quantum chaos and the complexity of spread of states},
  journal       = {Physical Review D},
  volume        = {106},
  pages         = {046007},
  year          = {2022},
  doi           = {10.1103/PhysRevD.106.046007}
}

@article{Ahmed2024,
  author        = {Ahmed, Osama and Tennie, Felix and Magri, Luca},
  title         = {Prediction of chaotic dynamics and extreme events: A recurrence-free quantum reservoir computing approach},
  journal       = {Physical Review Research},
  volume        = {6},
  pages         = {043082},
  year          = {2024},
  doi           = {10.1103/PhysRevResearch.6.043082}
}

@article{Domingo2024,
  author        = {Domingo, Laia and Borondo, F. and Scialchi, Gast{\'o}n and Roncaglia, Augusto J. and Carlo, Gabriel G. and Wisniacki, Diego A.},
  title         = {Quantum Reservoir Complexity by the {Krylov} Evolution Approach},
  journal       = {Physical Review A},
  volume        = {110},
  pages         = {022446},
  year          = {2024},
  doi           = {10.1103/PhysRevA.110.022446}
}

@book{Rudin1976,
  author        = {Rudin, Walter},
  title         = {Principles of Mathematical Analysis},
  edition       = {3},
  publisher     = {McGraw-Hill},
  address       = {New York},
  year          = {1976}
}

@article{Hu2024,
  author        = {Hu, Fangjun and Khan, Saeed A. and Bronn, Nicholas T. and Angelatos, Gerasimos and Rowlands, Graham E. and Ribeill, Guilhem J. and T{\"u}reci, Hakan E.},
  title         = {Overcoming the Coherence Time Barrier in Quantum Machine Learning on Temporal Data},
  journal       = {Nature Communications},
  volume        = {15},
  pages         = {7491},
  year          = {2024},
  doi           = {10.1038/s41467-024-51162-7}
}

@article{Cindrak2024,
  author        = {{\v C}indrak, Saud and Paschke, Adrian and Jaurigue, Lina C. and L{\"u}dge, Kathy},
  title         = {Measurable {Krylov} spaces and eigenenergy count in quantum state dynamics},
  journal       = {Journal of High Energy Physics},
  volume        = {2024},
  number        = {10},
  pages         = {083},
  year          = {2024},
  doi           = {10.1007/JHEP10(2024)083}
}

@article{MartinezPena2025,
  author        = {Mart{\'i}nez-Pe{\~n}a, Rodrigo and Ortega, Juan-Pablo},
  title         = {Input-Dependence in Quantum Reservoir Computing},
  journal       = {Physical Review E},
  volume        = {111},
  pages         = {065306},
  year          = {2025},
  doi           = {10.1103/3775-4hfd}
}

@article{Cindrak2025a,
  author        = {{\v C}indrak, Saud and Jaurigue, Lina C. and L{\"u}dge, Kathy},
  title         = {Engineering Quantum Reservoirs through {Krylov} Complexity, Expressivity, and Observability},
  journal       = {Physical Review Research},
  volume        = {7},
  pages         = {043190},
  year          = {2025},
  doi           = {10.1103/84f3-63mz}
}

@article{Cindrak2025b,
  author        = {{\v C}indrak, Saud and L{\"u}dge, Kathy and Jaurigue, Lina C.},
  title         = {From {Krylov} Complexity to Observability: Capturing Phase Space Dimension with Applications in Quantum Reservoir Computing},
  journal       = {Physical Review Research},
  volume        = {7},
  pages         = {L042039},
  year          = {2025},
  doi           = {10.1103/m9kq-wrln}
}

@misc{Sannia2025,
  author        = {Sannia, Antonio and Giorgi, Gian Luca and Zambrini, Roberta},
  title         = {Exponential Concentration and Symmetries in Quantum Reservoir Computing},
  year          = {2025},
  eprint        = {2505.10062},
  archivePrefix = {arXiv},
  primaryClass  = {quant-ph}
}

@article{Tropp2012,
  author        = {Tropp, Joel A.},
  title         = {User-Friendly Tail Bounds for Sums of Random Matrices},
  journal       = {Foundations of Computational Mathematics},
  volume        = {12},
  number        = {4},
  pages         = {389--434},
  year          = {2012},
  doi           = {10.1007/s10208-011-9099-z}
}

@techreport{Jaeger2002,
  author        = {Jaeger, Herbert},
  title         = {Short Term Memory in Echo State Networks},
  institution   = {German National Research Center for Information Technology},
  number        = {GMD Report 152},
  address       = {Sankt Augustin, Germany},
  year          = {2001}
}

@misc{Xiong2025,
  author        = {Xiong, Weijie and Holmes, Zo{\"e} and Angrisani, Armando and Suzuki, Yudai and Chotibut, Thiparat and Thanasilp, Supanut},
  title         = {Role of scrambling and noise in temporal information processing with quantum systems},
  year          = {2025},
  eprint        = {2505.10080},
  archivePrefix = {arXiv},
  primaryClass  = {quant-ph}
}

@article{MackeyGlass1977,
  author        = {Mackey, Michael C. and Glass, Leon},
  title         = {Oscillation and Chaos in Physiological Control Systems},
  journal       = {Science},
  volume        = {197},
  number        = {4300},
  pages         = {287--289},
  year          = {1977},
  doi           = {10.1126/science.267326}
}

@article{McClean2018,
  author        = {McClean, Jarrod R. and Boixo, Sergio and Smelyanskiy, Vadim N. and Babbush, Ryan and Neven, Hartmut},
  title         = {Barren plateaus in quantum neural network training landscapes},
  journal       = {Nature Communications},
  volume        = {9},
  pages         = {4812},
  year          = {2018},
  doi           = {10.1038/s41467-018-07090-4}
}

@article{GrigoryevaOrtega2019,
  author        = {Grigoryeva, Lyudmila and Ortega, Juan-Pablo},
  title         = {Differentiable Reservoir Computing},
  journal       = {Journal of Machine Learning Research},
  volume        = {20},
  number        = {179},
  pages         = {1--62},
  year          = {2019},
  url           = {https://jmlr.org/papers/v20/19-150.html}
}

@book{Bhatia1997,
  author        = {Bhatia, Rajendra},
  title         = {Matrix Analysis},
  series        = {Graduate Texts in Mathematics},
  volume        = {169},
  publisher     = {Springer},
  address       = {New York},
  year          = {1997},
  doi           = {10.1007/978-1-4612-0653-8},
  isbn          = {978-0-387-94846-1}
}

@article{Hoeffding1963,
  author        = {Hoeffding, Wassily},
  title         = {Probability Inequalities for Sums of Bounded Random Variables},
  journal       = {Journal of the American Statistical Association},
  volume        = {58},
  number        = {301},
  pages         = {13--30},
  year          = {1963},
  doi           = {10.1080/01621459.1963.10500830}
}

@article{Abbas2021,
  author        = {Abbas, Amira and Sutter, David and Zoufal, Christa and Lucchi, Aurelien and Figalli, Alessio and Woerner, Stefan},
  title         = {The power of quantum neural networks},
  journal       = {Nature Computational Science},
  volume        = {1},
  number        = {6},
  pages         = {403--409},
  year          = {2021},
  doi           = {10.1038/s43588-021-00084-1}
}

@article{Yarotsky2017,
  author        = {Yarotsky, Dmitry},
  title         = {Error bounds for approximations with deep {ReLU} networks},
  journal       = {Neural Networks},
  volume        = {94},
  pages         = {103--114},
  year          = {2017},
  doi           = {10.1016/j.neunet.2017.07.002}
}

@article{Thanasilp2024,
  author        = {Thanasilp, Supanut and Wang, Samson and Cerezo, M. and Holmes, Zo{\"e}},
  title         = {Exponential concentration in quantum kernel methods},
  journal       = {Nature Communications},
  volume        = {15},
  pages         = {5200},
  year          = {2024},
  doi           = {10.1038/s41467-024-49287-w}
}

@article{Barron1993,
  author        = {Barron, Andrew R.},
  title         = {Universal approximation bounds for superpositions of a sigmoidal function},
  journal       = {IEEE Transactions on Information Theory},
  volume        = {39},
  number        = {3},
  pages         = {930--945},
  year          = {1993},
  doi           = {10.1109/18.256500}
}

@article{Hu2023REC,
  author        = {Hu, Fangjun and Angelatos, Gerasimos and Khan, Saeed A. and Vives, Marti and T{\"u}reci, Esin and Bello, Leon and Rowlands, Graham E. and Ribeill, Guilhem J. and T{\"u}reci, Hakan E.},
  title         = {Tackling Sampling Noise in Physical Systems for Machine Learning Applications: Fundamental Limits and Eigentasks},
  journal       = {Physical Review X},
  volume        = {13},
  number        = {4},
  pages         = {041020},
  year          = {2023},
  doi           = {10.1103/PhysRevX.13.041020},
  month         = oct,
  publisher     = {American Physical Society}
}

% Supplemental Material starts on a new page in a one-column layout.
\clearpage
\onecolumngrid
\setcounter{section}{0}
\setcounter{subsection}{0}
\setcounter{subsubsection}{0}
\setcounter{equation}{0}
\setcounter{figure}{0}
\setcounter{table}{0}
\setcounter{suppthm}{0}
\setcounter{secnumdepth}{2}

\renewcommand{\thesection}{S\arabic{section}}
\renewcommand{\thesubsection}{\thesection.\arabic{subsection}}
\renewcommand{\thesubsubsection}{\thesubsection.\arabic{subsubsection}}
\numberwithin{equation}{section}
\renewcommand{\theequation}{S\arabic{section}.\arabic{equation}}
\renewcommand{\thefigure}{S\arabic{figure}}
\renewcommand{\thetable}{S\arabic{table}}

\makeatletter
\renewcommand{\p@subsection}{}
\renewcommand{\p@subsubsection}{}
\makeatother

% Distinct PDF destinations prevent collisions after counter resets.
\renewcommand{\theHsection}{supp.\arabic{section}}
\renewcommand{\theHsubsection}{\theHsection.\arabic{subsection}}
\renewcommand{\theHsubsubsection}{\theHsubsection.\arabic{subsubsection}}
\renewcommand{\theHequation}{supp.\arabic{section}.\arabic{equation}}
\renewcommand{\theHfigure}{supp.\arabic{figure}}
\renewcommand{\theHtable}{supp.\arabic{table}}
\providecommand{\theHsuppthm}{}
\renewcommand{\theHsuppthm}{supp.\arabic{section}.\arabic{suppthm}}

\pdfbookmark[0]{Supplemental Material}{supplement}
\begin{center}
{\large\bfseries Supplemental Material for}\\[0.4em]
{\large\bfseries The power of quantum reservoir computing}\\[0.7em]
Minwoo James Kim, Jeongho Bang, and Taehyun Kim
\end{center}

\begin{center}
\textbf{CONTENTS}
\end{center}
\begingroup
\small
\setlength{\parindent}{0pt}
\smallskip\noindent\hyperref[sec:supp-volterra-framework]{\ref*{sec:supp-volterra-framework}.\enspace Operator-valued Volterra representation of the reservoir}\nobreak\dotfill\pageref*{sec:supp-volterra-framework}\par
\noindent\hspace*{1.2em}\hyperref[subsec:supp-reservoir-operator-spaces]{\ref*{subsec:supp-reservoir-operator-spaces}.\enspace Measurement-reset dynamics and centered response}\nobreak\dotfill\pageref*{subsec:supp-reservoir-operator-spaces}\par
\noindent\hspace*{1.2em}\hyperref[subsec:supp-stable-memory-recursion]{\ref*{subsec:supp-stable-memory-recursion}.\enspace Stable causal memory response}\nobreak\dotfill\pageref*{subsec:supp-stable-memory-recursion}\par
\noindent\hspace*{1.2em}\hyperref[subsec:supp-joint-operator-volterra]{\ref*{subsec:supp-joint-operator-volterra}.\enspace Joint operator-valued Volterra expansion}\nobreak\dotfill\pageref*{subsec:supp-joint-operator-volterra}\par
\smallskip\noindent\hyperref[sec:supp-visibility-geometry]{\ref*{sec:supp-visibility-geometry}.\enspace Visibility geometry and alignment law}\nobreak\dotfill\pageref*{sec:supp-visibility-geometry}\par
\noindent\hspace*{1.2em}\hyperref[subsec:supp-observable-composition]{\ref*{subsec:supp-observable-composition}.\enspace Observable composition and visibility isotropy}\nobreak\dotfill\pageref*{subsec:supp-observable-composition}\par
\noindent\hspace*{1.2em}\hyperref[subsec:supp-dynamical-operator-sector]{\ref*{subsec:supp-dynamical-operator-sector}.\enspace Dynamically accessible operator sector}\nobreak\dotfill\pageref*{subsec:supp-dynamical-operator-sector}\par
\noindent\hspace*{1.2em}\hyperref[subsec:supp-operator-measurement-alignment]{\ref*{subsec:supp-operator-measurement-alignment}.\enspace Exact operator--measurement alignment law}\nobreak\dotfill\pageref*{subsec:supp-operator-measurement-alignment}\par
\noindent\hspace*{1.2em}\hyperref[subsec:supp-finite-resolution-consequence]{\ref*{subsec:supp-finite-resolution-consequence}.\enspace Finite-resolution consequence}\nobreak\dotfill\pageref*{subsec:supp-finite-resolution-consequence}\par
\noindent\hspace*{1.2em}\hyperref[subsec:supp-general-measurements]{\ref*{subsec:supp-general-measurements}.\enspace General measurement families}\nobreak\dotfill\pageref*{subsec:supp-general-measurements}\par
\smallskip\noindent\hyperref[sec:supp-haar-orientations]{\ref*{sec:supp-haar-orientations}.\enspace Haar-random latent spaces and supported visibility}\nobreak\dotfill\pageref*{sec:supp-haar-orientations}\par
\noindent\hspace*{1.2em}\hyperref[subsec:supp-haar-ensemble]{\ref*{subsec:supp-haar-ensemble}.\enspace Latent-space ensemble and support selection}\nobreak\dotfill\pageref*{subsec:supp-haar-ensemble}\par
\noindent\hspace*{1.2em}\hyperref[subsec:supp-haar-projector-moments]{\ref*{subsec:supp-haar-projector-moments}.\enspace First and second projector moments}\nobreak\dotfill\pageref*{subsec:supp-haar-projector-moments}\par
\noindent\hspace*{1.2em}\hyperref[subsec:supp-haar-supported-moments]{\ref*{subsec:supp-haar-supported-moments}.\enspace Supported mean, alignment and resolvable count}\nobreak\dotfill\pageref*{subsec:supp-haar-supported-moments}\par
\noindent\hspace*{1.2em}\hyperref[subsec:supp-haar-spectral-spread]{\ref*{subsec:supp-haar-spectral-spread}.\enspace Spread within the supported spectrum}\nobreak\dotfill\pageref*{subsec:supp-haar-spectral-spread}\par
\smallskip\noindent\hyperref[sec:supp-capacity-readout-truncation]{\ref*{sec:supp-capacity-readout-truncation}.\enspace Proofs of capacity, readout and truncation results}\nobreak\dotfill\pageref*{sec:supp-capacity-readout-truncation}\par
\noindent\hspace*{1.2em}\hyperref[subsec:supp-capacity-proof]{\ref*{subsec:supp-capacity-proof}.\enspace Population linear capacity and geometric bounds}\nobreak\dotfill\pageref*{subsec:supp-capacity-proof}\par
\noindent\hspace*{1.2em}\hyperref[subsec:supp-optimal-readout-proof]{\ref*{subsec:supp-optimal-readout-proof}.\enspace Optimal readout subspace}\nobreak\dotfill\pageref*{subsec:supp-optimal-readout-proof}\par
\noindent\hspace*{1.2em}\hyperref[subsec:supp-truncation-sampling-proof]{\ref*{subsec:supp-truncation-sampling-proof}.\enspace Truncation and finite-shot readout error}\nobreak\dotfill\pageref*{subsec:supp-truncation-sampling-proof}\par
\smallskip\noindent\hyperref[sec:supp-resolution-sampling]{\ref*{sec:supp-resolution-sampling}.\enspace Resolution models and sampling uncertainty}\nobreak\dotfill\pageref*{sec:supp-resolution-sampling}\par
\noindent\hspace*{1.2em}\hyperref[subsec:supp-singular-perturbations]{\ref*{subsec:supp-singular-perturbations}.\enspace Singular-value perturbations and the separation threshold}\nobreak\dotfill\pageref*{subsec:supp-singular-perturbations}\par
\noindent\hspace*{1.2em}\hyperref[subsec:supp-shot-normalization]{\ref*{subsec:supp-shot-normalization}.\enspace Physical Pauli sampling and coordinate normalization}\nobreak\dotfill\pageref*{subsec:supp-shot-normalization}\par
\noindent\hspace*{1.2em}\hyperref[subsec:supp-theoretical-operational-threshold]{\ref*{subsec:supp-theoretical-operational-threshold}.\enspace Statistical calibration of the resolution threshold}\nobreak\dotfill\pageref*{subsec:supp-theoretical-operational-threshold}\par
\smallskip\noindent\hyperref[sec:supp-gram-equivalence]{\ref*{sec:supp-gram-equivalence}.\enspace Equivalence of the direct and joint-Gram constructions}\nobreak\dotfill\pageref*{sec:supp-gram-equivalence}\par
\endgroup
\clearpage

\section{Operator-valued Volterra representation of the reservoir}
\label{sec:supp-volterra-framework}

This section establishes the dynamical representation used throughout the Supplementary Information. In the measurement-reset architecture, only the memory subsystem persists between timesteps, whereas the temporal features relevant to measurement-resolved expressivity are defined from the joint memory--readout state immediately before measurement. We therefore first establish a causal fading-memory response for the persistent memory and then lift this response to an operator-valued Volterra expansion of the joint pre-measurement state. The construction follows standard results for analytic fading-memory filters and Volterra representations~\cite{BoydChua1985,GrigoryevaOrtega2019,Hu2024}, specialized here to the measurement-reset reservoir used in this work.

\subsection{Measurement-reset dynamics and centered response}
\label{subsec:supp-reservoir-operator-spaces}

The reservoir Hilbert space is decomposed as
\begin{equation}
    \mathcal H
    =
    \mathcal H_M\otimes\mathcal H_R,
\end{equation}
where $\mathcal H_M$ is the persistent memory subsystem and $\mathcal H_R$ is the readout subsystem, with dimensions $d_M$ and $d_R$, respectively.

For a finite-dimensional Hilbert space \(\mathcal K\), let \(\mathcal B(\mathcal K)\) denote the complex vector space of linear operators on \(\mathcal K\), and let
\begin{equation}
    \mathcal B_{\rm h}(\mathcal K)
    :=
    \left\{
        X\in\mathcal B(\mathcal K):X=X^\dagger
    \right\}
\end{equation}
denote its real vector space of Hermitian operators. We define the traceless Hermitian spaces
\begin{align}
    \mathcal E_M
    &:=
    \left\{
        X\in\mathcal B_{\rm h}(\mathcal H_M):
        \Tr X=0
    \right\},
    \label{eq:supp-memory-deviation-space}
    \\
    \mathcal E_{MR}
    &:=
    \left\{
        Z\in\mathcal B_{\rm h}(\mathcal H_M\otimes\mathcal H_R):
        \Tr Z=0
    \right\}.
    \label{eq:supp-joint-deviation-space}
\end{align}
Their real dimensions are \(d_M^2-1\) and \((d_Md_R)^2-1\), respectively. Both are equipped with the Hilbert--Schmidt inner product
\begin{equation}
    \langle A,B\rangle_{\rm HS}
    :=
    \Tr(AB).
    \label{eq:supp-hs-inner-product}
\end{equation}
All orthogonality and projections below refer to this geometry.

At timestep $t$, a fixed readout state $\rho_0^R\in\mathcal D(\mathcal H_R)$ is coupled to the persistent memory state. For scalar input $u_t$, the joint state immediately before measurement is
\begin{equation}
    \sigma_t
    =
    \mathcal U_{u_t}
    \left(
        \rho_{t-1}^M\otimes\rho_0^R
    \right).
    \label{eq:supp-joint-premeasurement-state}
\end{equation}
For a readout measurement with outcome projectors
\(\{\Pi_a^{(j)}\}_a\), the corresponding unnormalized conditional memory state is
\begin{equation}
    \widetilde\rho_{t|a,j}^M
    =
    \Tr_R
    \left[
        (I_M\otimes\Pi_a^{(j)})
        \sigma_t
        (I_M\otimes\Pi_a^{(j)})
    \right].
\end{equation}
Since the measurement outcome is not fed back, the recurrent state is the
unconditional average
\begin{equation}
    \rho_t^M
    =
    \sum_a\widetilde\rho_{t|a,j}^M
    =
    \Tr_R\sigma_t.
    \label{eq:supp-memory-state-update}
\end{equation}
Thus, the averaged memory evolution is governed by the reduced channel
\begin{equation}
    \mathcal C_u(\rho^M)
    :=
    \Tr_R
    \left[
        \mathcal U_u
        \left(
            \rho^M\otimes\rho_0^R
        \right)
    \right],
\end{equation}
making only $\rho_t^M$ recurrent; the readout subsystem is measured and reset before the next input is applied.

The corresponding effective memory channel is
\begin{equation}
    \mathcal C(u)
    :=
    \Tr_R\circ\mathcal U_u \circ\mathcal I_{\rho_0^R},
    \qquad
    \mathcal I_{\rho_0^R}(\rho^M)
    :=
    \rho^M\otimes\rho_0^R,
    \label{eq:supp-effective-memory-channel}
\end{equation}
so that
\begin{equation}
    \rho_t^M
    =
    \mathcal C(u_t)(\rho_{t-1}^M).
    \label{eq:supp-effective-memory-update}
\end{equation}
For the unitary reservoirs considered numerically,
\begin{equation}
    \mathcal C(u)(\rho^M)
    =
    \Tr_R
    \left[
        U(u)
        \left(
            \rho^M\otimes\rho_0^R
        \right)
        U(u)^\dagger
    \right].
    \label{eq:supp-unitary-effective-memory-channel}
\end{equation}

Let \(u_\ast\) be a reference input, and let
\(\bar\rho^M\in\mathcal D(\mathcal H_M)\) denote a stationary memory state
satisfying
\begin{equation}
    \mathcal C(u_\ast)(\bar\rho^M)
    =
    \bar\rho^M.
    \label{eq:supp-reference-fixed-point}
\end{equation}
The corresponding reference joint state is
\begin{equation}
    \bar\sigma
    :=
    \mathcal U_{u_\ast}
    \left(
        \bar\rho^M\otimes\rho_0^R
    \right).
    \label{eq:supp-reference-joint-state}
\end{equation}

We center the input and reservoir states as
\begin{align}
    v_t
    &:=
    u_t-u_\ast,
    \label{eq:supp-centered-input}
    \\
    X_t
    &:=
    \rho_t^M-\bar\rho^M
    \in\mathcal E_M,
    \label{eq:supp-memory-deviation}
    \\
    Z_t
    &:=
    \sigma_t-\bar\sigma
    \in\mathcal E_{MR}.
    \label{eq:supp-joint-deviation}
\end{align}
The shifted recurrent update is therefore
\begin{equation}
    X_t
    =
    \widetilde{\mathcal C}_{u_t}(X_{t-1}),
    \qquad
    \widetilde{\mathcal C}_u(X)
    :=
    \mathcal C(u)(\bar\rho^M+X)-\bar\rho^M,
    \label{eq:supp-shifted-memory-update}
\end{equation}
with
\begin{equation}
    \widetilde{\mathcal C}_{u_\ast}(0)=0.
\end{equation}

The centered joint response is
\begin{equation}
    Z_t
    =
    G(v_t,X_{t-1}),
    \label{eq:supp-joint-response-from-memory}
\end{equation}
where
\begin{equation}
    G(v,X)
    :=
    \mathcal U_{u_\ast+v}
    \bigl(
    (\bar\rho^M+X)\otimes\rho_0^R
    \bigr)
    -
    \bar\sigma.
    \label{eq:supp-centered-joint-response-map}
\end{equation}

Throughout the text, $X_t\in\mathcal E_M$ is the traceless Hermitian deviation of the persistent memory state, while $Z_t\in\mathcal E_{MR}$ is the corresponding deviation of the joint memory--readout state before measurement. These spaces play different roles. Fading memory concerns the recurrent evolution of $X_t$ on the memory subsystem. The map $G$ then converts this memory response and the current input into the joint response $Z_t$, whose temporal expansion supplies the operator features analyzed below.

\subsection{Stable causal memory response}
\label{subsec:supp-stable-memory-recursion}

Choose a Hilbert--Schmidt orthonormal Hermitian basis
\begin{equation}
    P_0
    =
    \frac{I_M}{\sqrt{d_M}},
    \qquad
    P_1,\ldots,P_{d_M^2-1}\in\mathcal E_M,
    \qquad
    \Tr(P_\mu P_\nu)
    =
    \delta_{\mu\nu}.
    \label{eq:supp-memory-hs-basis}
\end{equation}
The traceless coordinates of $X\in\mathcal E_M$ are
\begin{equation}
    x_{\rm tr}(X)
    :=
    \bigl(
    \Tr(P_1X),\ldots,\Tr(P_{d_M^2-1}X)
    \bigr)^\top.
\end{equation}
Orthonormality gives
$\langle X,Y\rangle_{\rm HS}=x_{\rm tr}(X)^\top x_{\rm tr}(Y)$, so these coordinates identify $\mathcal E_M$ isometrically with $\mathbb R^{d_M^2-1}$.

Let $T(u)$ denote the Pauli-transfer-matrix representation of the effective channel on the memory operator space $\mathcal E_M$,
\begin{equation}
    T_{\mu\nu}(u)
    :=
    \Tr[
    P_\mu\mathcal C(u)(P_\nu)
    ].
    \label{eq:supp-ptm-definition}
\end{equation}
Trace preservation gives the block form
\begin{equation}
    T(u)
    =
    \begin{pmatrix}
        1 & 0\\
        c(u) & A(u)
    \end{pmatrix}.
    \label{eq:supp-ptm-block-form}
\end{equation}
Writing
\begin{equation}
    \delta x_t
    :=
    x_{\rm tr}(X_t),
\end{equation}
the recurrent dynamics reduce to
\begin{equation}
    \delta x_t
    =
    A(v_t)\delta x_{t-1}
    +
    b(v_t),
    \label{eq:supp-centered-affine-recursion}
\end{equation}
where the centered notation $A(v):=A(u_\ast+v)$ and $b(v):=b(u_\ast+v)$ has been used, with $b(0)=0$. Explicitly,
\begin{equation}
    b(u)
    =
    \bigl[
    T(u)x(\bar\rho)-x(\bar\rho)
    \bigr]_{\rm tr}.
    \label{eq:supp-affine-source-term}
\end{equation}

Let $V\subset\mathbb R$ be a compact centered input domain with $0\in V$.

\begin{suppassumption}[Uniform contractivity]
\label{ass:supp-uniform-contractivity}
There exists a norm $\|\cdot\|_*$ on $\mathbb R^{d_M^2-1}$ and a constant $q<1$ such that
\begin{equation}
    \|A(v)\|_{*\rightarrow *}
    \le
    q,
    \qquad
    v\in V.
    \label{eq:supp-uniform-contractivity}
\end{equation}
\end{suppassumption}

\begin{suppthm}[Unique bounded causal memory response]
\label{thm:supp-bounded-causal-memory-response}
Under Assumption~\ref{ass:supp-uniform-contractivity}, Eq.~\eqref{eq:supp-centered-affine-recursion} has a unique bounded causal solution,
\begin{equation}
    \delta x_t
    =
    \sum_{n=0}^{\infty}
    \Phi_t^{(n)}
    b(v_{t-n}),
    \label{eq:supp-causal-memory-solution}
\end{equation}
where
\begin{equation}
    \Phi_t^{(0)}
    :=
    I,
    \qquad
    \Phi_t^{(n)}
    :=
    A(v_t)A(v_{t-1})\cdots A(v_{t-n+1})
    \quad (n\ge1).
    \label{eq:supp-transition-product}
\end{equation}
The series converges absolutely and uniformly over admissible input histories.
\end{suppthm}

\begin{proof}
Let
\begin{equation}
    B
    :=
    \sup_{v\in V}\|b(v)\|_*<\infty.
\end{equation}
By uniform contractivity,
\begin{equation}
    \|\Phi_t^{(n)}\|_*
    \leq
    \prod_{k=0}^{n-1}\|A(v_{t-k})\|_*
    \leq
    q^n.
\end{equation}
Consequently,
\begin{equation}
    \left\|
        \Phi_t^{(n)}b(v_{t-n})
    \right\|_*
    \leq
    q^nB.
\end{equation}
Since \(\sum_{n=0}^{\infty}q^nB=B/(1-q)\), the Weierstrass \(M\)-test~\cite{Rudin1976} implies that Eq.~\eqref{eq:supp-causal-memory-solution} converges absolutely and uniformly over all admissible input histories. In particular, the resulting solution is bounded.
\begin{equation}
    \sup_t\|\delta x_t\|_*
    \leq
    \frac{B}{1-q}.
\end{equation}
Each term depends only on \(v_t,v_{t-1},\ldots\), so the solution is causal.

It remains to verify the recurrence. Using
\begin{equation}
    A(v_t)\Phi_{t-1}^{(n)}
    =
    \Phi_t^{(n+1)},
\end{equation}
uniform convergence permits termwise application of \(A(v_t)\), giving
\begin{align}
    A(v_t)\delta x_{t-1}+b(v_t)
    &=
    b(v_t)
    +
    \sum_{n=0}^{\infty}
    A(v_t)\Phi_{t-1}^{(n)}
    b(v_{t-1-n})
    \\
    &=
    b(v_t)
    +
    \sum_{n=0}^{\infty}
    \Phi_t^{(n+1)}
    b(v_{t-(n+1)})
    \\
    &=
    \sum_{n=0}^{\infty}
    \Phi_t^{(n)}
    b(v_{t-n})
    =
    \delta x_t.
\end{align}
Thus the series defines a bounded causal solution of Eq.~\eqref{eq:supp-centered-affine-recursion}.

For uniqueness, let \(\delta x_t\) and \(\delta x_t'\) be two bounded solutions and set \(z_t:=\delta x_t-\delta x_t'\). Their difference satisfies
\begin{equation}
    z_t=A(v_t)z_{t-1}.
\end{equation}
Iterating this relation \(m\) steps yields
\begin{equation}
    z_t
    =
    \Phi_t^{(m)}z_{t-m},
    \qquad
    \|z_t\|_*
    \leq
    q^m\|z_{t-m}\|_*.
\end{equation}
Because both solutions are bounded, there exists \(K<\infty\) such that \(\|z_s\|_*\leq K\) for every \(s\). Hence
\begin{equation}
    \|z_t\|_*
    \leq
    q^mK
\end{equation}
for every \(m\). Taking \(m\to\infty\) and using \(0\leq q<1\) gives \(z_t=0\). Therefore the bounded causal solution is unique.
\end{proof}

For a finite memory depth \(L\), define the truncated response
\begin{equation}
    \delta x_t^{(L)}
    :=
    \sum_{n=0}^{L}
    \Phi_t^{(n)}b(v_{t-n}),
\end{equation}
which depends only on the most recent inputs \(v_t,\ldots,v_{t-L}\). The omitted contribution satisfies
\begin{align}
    \left\|
        \delta x_t-\delta x_t^{(L)}
    \right\|_*
    &\leq
    \sum_{n=L+1}^{\infty}
    \left\|
        \Phi_t^{(n)}b(v_{t-n})
    \right\|_*
    \\
    &\leq
    B\sum_{n=L+1}^{\infty}q^n
    =
    \frac{Bq^{L+1}}{1-q}.
    \label{eq:supp-fading-memory-tail-bound}
\end{align}
The approximation error therefore decreases exponentially with \(L\), uniformly over admissible input histories. In particular, if two histories coincide over the latest \(L+1\) timesteps, their memory states at time \(t\) differ by at most \(2Bq^{L+1}/(1-q)\). This makes precise the fading influence of inputs earlier than \(t-L\).

\subsection{Joint operator-valued Volterra expansion}
\label{subsec:supp-joint-operator-volterra}

We now establish a Volterra expansion of the joint pre-measurement response $Z_t=G(v_t,X_{t-1})$, whose operator coefficients define the temporal feature space used in the main text. The causal memory solution supplies the dependence on past inputs, while $G$ incorporates the current input and the joint evolution before measurement. We impose analytic and summability conditions that allow this composition to be expanded directly in delayed-input monomials.

Consider centered inputs $v=u-u_\ast$ in a bounded domain $V$ containing the origin, and define
\begin{equation}
 v_{\max}:=\sup_{v\in V}|v|<\infty.
\end{equation}
A bounded input range can be shifted and rescaled before encoding, with corresponding changes in the expansion coefficients. Such normalization alone does not guarantee convergence of the temporal expansion.

\begin{suppassumption}[Analytic input dependence]
\label{ass:supp-analytic-input-dependence}
The memory coefficient maps admit absolutely convergent expansions
\begin{equation}
 A(v)=\sum_{k=0}^{\infty}v^kA_k,
 \qquad
 b(v)=\sum_{k=1}^{\infty}v^kb_k,
 \label{eq:supp-analytic-A-b-expansion}
\end{equation}
on a neighborhood of $V$, with
\begin{align}
 a&:=\sum_{k=0}^{\infty}
 v_{\max}^k\|A_k\|_{*\rightarrow *}<1,
 \label{eq:supp-A-majorant}\\
 \beta&:=\sum_{k=1}^{\infty}
 v_{\max}^k\|b_k\|_*<\infty.
 \label{eq:supp-b-majorant}
\end{align}
\end{suppassumption}

The bound $a<1$ implies $\|A(v)\|_{*\rightarrow *}\leq a<1$ uniformly over $V$, and hence ensures the contraction required for the causal memory solution. It also controls the sum of the Taylor-coefficient norms, allowing convergence to be established after expanding that solution in the inputs. This is a sufficient condition; more general analytic fading-memory conditions can be used in its place~\cite{BoydChua1985,GrigoryevaOrtega2019}.

To connect the memory coordinates to the joint response, equip $\mathcal E_M$ with the induced norm
\begin{equation}
 \left\|\sum_{\mu=1}^{d_M^2-1}x_\mu P_\mu\right\|_{M,*}
 :=\|x\|_*,
\end{equation}
where $\{P_\mu\}_{\mu=1}^{d_M^2-1}$ is the Hilbert--Schmidt-orthonormal memory basis used above. The joint-response map is affine in the memory deviation,
\begin{equation}
 G(v,X)=g(v)+\mathcal L(v)X,
 \label{eq:supp-joint-response-linear-decomposition}
\end{equation}
with
\begin{align}
 g(v)
 &:=
 \mathcal U_{u_\ast+v}
 \bigl(\bar\rho^M\otimes\rho_0^R\bigr)
 -\bar\sigma,\\
 \mathcal L(v)X
 &:=
 \mathcal U_{u_\ast+v}
 \bigl(X\otimes\rho_0^R\bigr).
\end{align}
By the definition of $\bar\sigma$, $g(0)=0$. The following assumption controls the analytic dependence of this joint map.

\begin{suppassumption}[Analytic joint response]
\label{ass:supp-analytic-joint-response}
The maps $g(v)$ and $\mathcal L(v)$ admit absolutely convergent expansions
\begin{equation}
 g(v)=\sum_{k=1}^{\infty}v^kg_k,
 \qquad
 \mathcal L(v)=\sum_{k=0}^{\infty}v^k\mathcal L_k
\end{equation}
on a neighborhood of $V$, with
\begin{align}
 \gamma&:=\sum_{k=1}^{\infty}
 v_{\max}^k\|g_k\|_{\mathrm{HS}}<\infty,\\
 \ell&:=\sum_{k=0}^{\infty}
 v_{\max}^k
 \|\mathcal L_k\|_{M,*\rightarrow\mathrm{HS}}<\infty.
\end{align}
\end{suppassumption}

For the input-linear Ising Hamiltonian and the finite products of input rotations used here, the one-step maps have Taylor series that converge on every bounded real input interval. This analyticity does not by itself establish convergence of the recurrent temporal expansion, which is ensured here by the separate majorant condition $a<1$.

\begin{suppprop}[Joint operator-valued Volterra expansion]
\label{prop:supp-joint-operator-valued-volterra-expansion}
Under Assumptions~\ref{ass:supp-analytic-input-dependence} and~\ref{ass:supp-analytic-joint-response}, the causal joint pre-measurement deviation admits the absolutely and uniformly convergent expansion
\begin{equation}
 Z_t
 =
 \sum_{p=1}^{\infty}
 \sum_{0\leq\tau_p\leq\cdots\leq\tau_1}
 \widehat h_p(\tau_1,\ldots,\tau_p)
 \prod_{r=1}^{p}v_{t-\tau_r},
 \label{eq:supp-joint-volterra-expansion}
\end{equation}
over admissible input histories, where each coefficient $\widehat h_p(\tau_1,\ldots,\tau_p)$ is a traceless Hermitian operator in $\mathcal E_{MR}$.
\end{suppprop}

\begin{proof}
Substitute the causal memory solution in Eq.~\eqref{eq:supp-causal-memory-solution} into
\begin{equation}
 Z_t=g(v_t)+\mathcal L(v_t)X_{t-1}.
\end{equation}
A memory contribution containing $n$ factors of $A$ followed by one factor of $b$ has, after Taylor expansion, a total weighted coefficient norm bounded by $a^n\beta$. Summing over $n\geq0$ gives
\begin{equation}
 \sum_{n=0}^{\infty}a^n\beta
 =\frac{\beta}{1-a}.
\end{equation}
Expanding $g(v_t)$ and $\mathcal L(v_t)$ therefore produces a joint-response series whose total weighted coefficient norm is at most
\begin{equation}
 \gamma+\ell\frac{\beta}{1-a}<\infty.
\end{equation}
This bound establishes absolute and uniform convergence and permits regrouping by delayed-input monomials.

Terms from $g(v_t)$ contain only zero-lag inputs. Multiplication by $v_t^k\mathcal L_k$ adds $k$ zero-lag factors to terms arising from the past memory response. Collecting equal monomials and ordering their lags gives Eq.~\eqref{eq:supp-joint-volterra-expansion}, with
\begin{equation}
 \sum_{p=1}^{\infty}
 \sum_{0\leq\tau_p\leq\cdots\leq\tau_1}
 v_{\max}^p
 \left\|\widehat h_p(\tau_1,\ldots,\tau_p)\right\|_{\mathrm{HS}}
 \leq
 \gamma+\ell\frac{\beta}{1-a}.
\end{equation}
The maps $g(v)$ and $\mathcal L(v)$ take values in $\mathcal E_{MR}$, so their Taylor coefficients and the resulting joint kernels are traceless and Hermitian.
\end{proof}

This construction gives the joint operator-valued temporal representation directly, using the recurrent memory solution only as an intermediate step. It is consistent with the quantum Volterra perspective developed in Ref.~\cite{Hu2024}. The kernels $\widehat h_p$ depend on the reservoir dynamics, input encoding, memory--readout coupling, reset state and expansion point, but not on the measured-observable family. They therefore define the generated temporal directions before measurement, whose classical accessibility is analyzed in the main text.

\section{Visibility geometry and alignment law}
\label{sec:supp-visibility-geometry}

We define the dynamically accessible operator sector and derive the alignment law used in the main text. This law separates the sector-average visibility from the contribution of the supported temporal space's alignment with measurement. The derivation assumes the orthonormal measurement operators specified in the main text. We treat general measurement families in Section~\ref{subsec:supp-general-measurements}.

\subsection{Observable composition and visibility isotropy}
\label{subsec:supp-observable-composition}

We examine how observable composition affects visibility isotropy at fixed reservoir dynamics, temporal truncation and observable count. The delay-line model, measurement-family construction and numerical settings are specified in the main-text Methods subsection ``Reservoir models and measurement protocol''.

\begin{figure}[tbp]
 \centering
 \includegraphics[width=.5\linewidth]{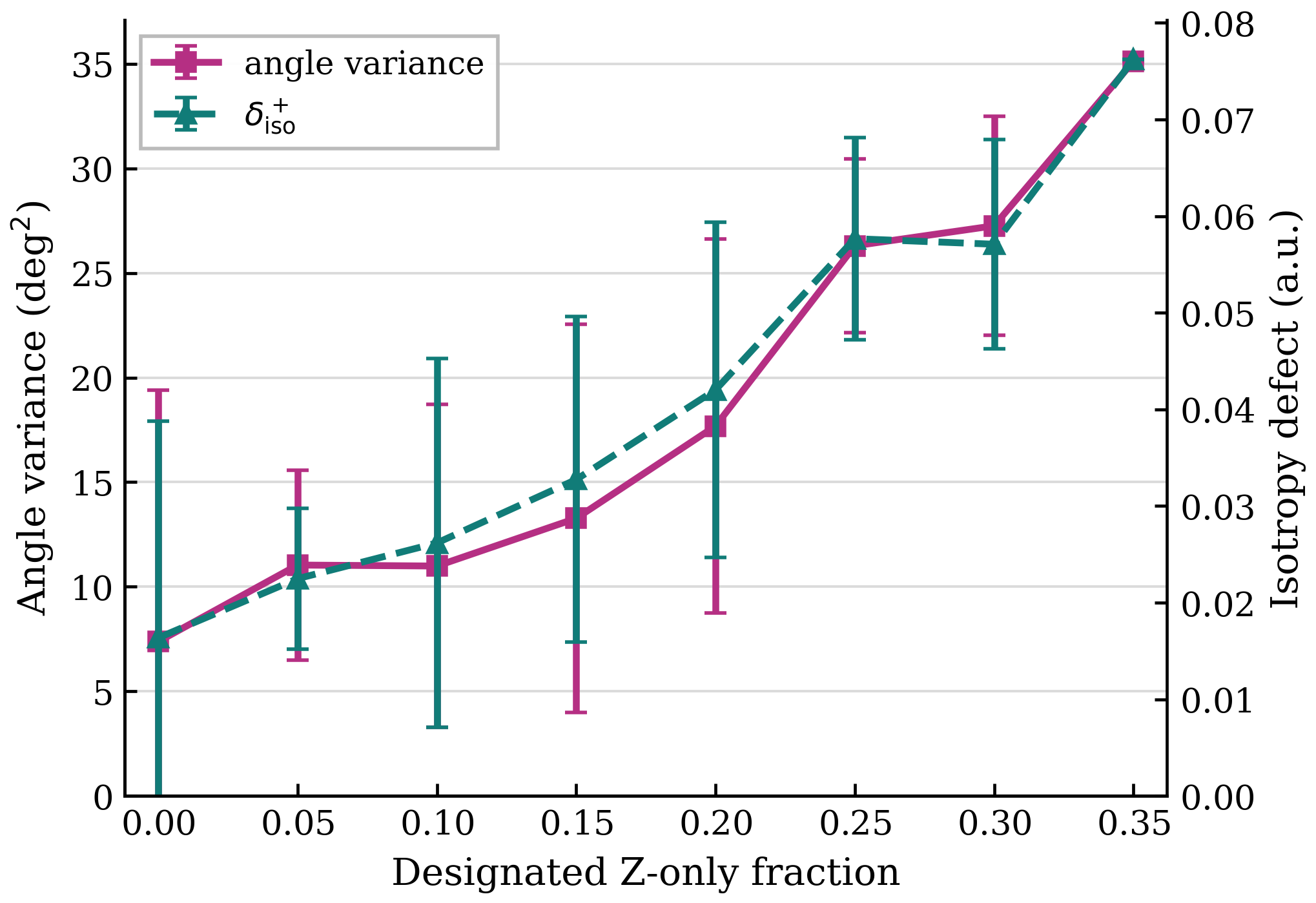}
 \caption{\textbf{Observable composition changes visibility isotropy at fixed budget.} Variance of the supported visibility angles and isotropy defect versus the designated fraction of $Z$-only observables. Reservoir dynamics, temporal truncation and observable count are held fixed. Points and error bars summarize the mean and population standard deviation across five observable families. The angular variance measures the spread within each supported spectrum, whereas the error bars describe variation between observable families.}
 \label{fig:supp-composition}
\end{figure}

Figure~\ref{fig:supp-composition} compares the variance of the supported angles, evaluated about each realization's own mean angle, with the isotropy defect $\delta_{\mathrm{iso}}^+$. The latter bounds the deviations of the squared visible weights from their mean, $|\sin^2\theta_j^+-\alpha^+|\leq\delta_{\mathrm{iso}}^+$. Their variation across observable compositions demonstrates that a fixed number of equally normalized measurements can expose the generated directions with different degrees of uniformity. Thus the observable budget alone does not determine the width of the visibility spectrum. This provides a numerical illustration of the role of isotropy in controlling deviations from the mean, complementing the main-text analysis of changes in the mean visibility itself.

\subsection{Dynamically accessible operator sector}
\label{subsec:supp-dynamical-operator-sector}

To distinguish measurement coverage of the accessible operator space from the alignment of a particular supported temporal space, we introduce the full dynamical response sector. Let $\mathscr U$ denote the admissible input histories in the operating domain, and let $\sigma[\mathbf u]$ be the corresponding joint pre-measurement response of the unique causal reservoir state. Define
\begin{equation}
 W_{\mathrm{dyn}}
 :=\Span\{\sigma[\mathbf u]-\bar\sigma
 \mid\mathbf u\in\mathscr U\}
 \subseteq\mathcal E_{MR}.
 \label{eq:supp-dynamical-sector-definition}
\end{equation}
The model, reset state, input encoding and parameters are held fixed. This sector is independent of the temporal truncation and measured-observable family. When admissible histories include a neighborhood of the reference history, each Volterra kernel is a limit of difference quotients of responses in $W_{\mathrm{dyn}}$. Since this finite-dimensional space is closed, $S^+\subseteq S\subseteq W_{\mathrm{dyn}}$.

Averaging the squared visible projections over an HS-orthonormal basis of $W_{\mathrm{dyn}}$ gives a geometric reference that weights operator directions equally, independently of the input-induced distribution of reservoir states. The alignment law compares the supported mean visibility with this sector average to quantify preferential exposure to measurement.

For any HS-orthonormal basis $F_1,\ldots,F_D$ of this sector, its orthogonal projector is
\begin{equation}
 P_{\rm dyn}(X)=\sum_{a=1}^{D}F_a\Tr(F_aX).
 \label{eq:supp-dynamical-projector}
\end{equation}
This definition is independent of the chosen basis. Then, define
\begin{equation}
 \begin{aligned}
 D_{\rm dyn}&=\Tr P_{\rm dyn}=\dim W_{\rm dyn},\\
 C_{\rm dyn}&=P_{\rm dyn}\Pi_{\rm vis}P_{\rm dyn},\\
 r_{\rm dyn}&=\Tr C_{\rm dyn}=\Tr(\Pi_{\rm vis}P_{\rm dyn}).
 \end{aligned}
 \label{eq:supp-dynamical-objects}
\end{equation}
The traces defining $D_{\mathrm{dyn}}$ and $r_{\mathrm{dyn}}$ are traces of linear maps acting on the real Hilbert space $\mathcal E_{MR}$, equipped with the Hilbert--Schmidt inner product. To establish the properties of $C_{\mathrm{dyn}}$, observe that for every $X\in\mathcal E_{MR}$,
\begin{equation}
 \begin{aligned}
 \langle X,C_{\mathrm{dyn}}X\rangle_{\HS}
 &=\|\Pi_{\mathrm{vis}}P_{\mathrm{dyn}}X\|_{\HS}^2\\
 &\leq\|P_{\mathrm{dyn}}X\|_{\HS}^2
 =\langle X,P_{\mathrm{dyn}}X\rangle_{\HS}.
 \end{aligned}
\end{equation}
Thus $0\preceq C_{\mathrm{dyn}}\preceq P_{\mathrm{dyn}}$. Since $P_{\mathrm{dyn}}$ acts as the identity on $W_{\mathrm{dyn}}$, the eigenvalues of $C_{\mathrm{dyn}}$ restricted to this sector lie in $[0,1]$. Moreover,
\begin{equation}
 \rank C_{\mathrm{dyn}}
 \leq\min\{\rank P_{\mathrm{dyn}},\rank\Pi_{\mathrm{vis}}\}
 =\min\{D_{\mathrm{dyn}},J\}.
\end{equation}
Its trace is the sum of its nonnegative eigenvalues, each at most one, so
\begin{equation}
 0\leq r_{\mathrm{dyn}}
 \leq\rank C_{\mathrm{dyn}}
 \leq\min\{D_{\mathrm{dyn}},J\}.
\end{equation}

To interpret this trace geometrically, let $\{F_a\}_{a=1}^{D_{\mathrm{dyn}}}$ be any Hilbert--Schmidt-orthonormal basis of $W_{\mathrm{dyn}}$. The map $C_{\mathrm{dyn}}$ vanishes on $W_{\mathrm{dyn}}^\perp$, and therefore
\begin{equation}
 r_{\mathrm{dyn}}
 =\sum_{a=1}^{D_{\mathrm{dyn}}}
 \langle F_a,C_{\mathrm{dyn}}F_a\rangle_{\HS}
 =\sum_{a=1}^{D_{\mathrm{dyn}}}
 \langle F_a,\Pi_{\mathrm{vis}}F_a\rangle_{\HS}
 =\sum_{a=1}^{D_{\mathrm{dyn}}}
 \|\Pi_{\mathrm{vis}}F_a\|_{\HS}^2.
\end{equation}
For $D_{\mathrm{dyn}}>0$, dividing by the sector dimension gives
\begin{equation}
 \frac{1}{D_{\mathrm{dyn}}}
 \sum_{a=1}^{D_{\mathrm{dyn}}}
 \|\Pi_{\mathrm{vis}}F_a\|_{\HS}^2
 =\frac{r_{\mathrm{dyn}}}{D_{\mathrm{dyn}}}
 \leq\min\left\{1,\frac{J}{D_{\mathrm{dyn}}}\right\}.
 \label{eq:supp-sector-average-visibility}
\end{equation}
The ratio is therefore the mean squared visible projection of an orthonormal sector basis. Trace invariance makes this mean independent of the chosen basis. Separately,
\begin{equation}
 P^+=Q^+(Q^+)^{\mathsf T}
 \label{eq:supp-supported-projector}
\end{equation}
projects onto the supported temporal space.

A finite retained kernel span is generally smaller than the complete response sector. Likewise, a containing ambient space need not be fully explored. However, when strong mixing makes the dynamically accessible sector approximately the full traceless joint operator space, its dimension grows exponentially with the total number of qubits, $D_{\mathrm{dyn}}\approx d^2-1=4^{n_M+n_R}-1$. The total visible weight remains bounded by the number of measured observables, $r_{\mathrm{dyn}}\leq J$. At fixed observable budget, the sector-average visibility therefore has an exponentially decreasing upper bound,
\begin{equation}
 \frac{r_{\mathrm{dyn}}}{D_{\mathrm{dyn}}}
 \leq\frac{J}{D_{\mathrm{dyn}}}
 \approx\frac{J}{4^{n_M+n_R}-1}.
\end{equation}
The full-space reference takes $D_a=d^2-1$ and $r_a=J$, giving the characteristic angle $\theta_a=\arcsin\sqrt{J/(d^2-1)}$. When preferential alignment is weak and the supported spectrum is relatively narrow, the supported angles concentrate near this small reference scale. Increasing the operator-space dimension can then leave the transmitted temporal directions collectively weakly exposed to measurement. The decreasing supported angles in the main-text size sweeps are consistent with this mechanism; the ambient reference describes the strong-mixing approximation rather than establishing that the dynamics explores the full operator space.

Whether the full-space approximation is appropriate also depends on the input encoding. A single-injection circuit provides an explicit restriction. If
\begin{equation}
 U(u)=U_0(R_{\mathrm{in}}^M(u)\otimes I_R),
\end{equation}
with input-independent $U_0$, the input rotation can be absorbed into the memory state. Every joint pre-measurement state therefore has the form $U_0(\rho^M\otimes\rho_0^R)U_0^\dagger$. Differences from the reference state belong to
\begin{equation}
 W_{\mathrm{dyn}}\subseteq
 U_0\bigl(\mathcal E_M\otimes\rho_0^R\bigr)U_0^\dagger,
 \qquad
 D_{\mathrm{dyn}}\leq d_M^2-1.
 \label{eq:supp-single-injection-sector-bound}
\end{equation}
Unitary conjugation preserves dimension, so increasing the depth or scrambling strength of $U_0$ can rotate this sector without enlarging it beyond the bound. For a nontrivial readout register, the full traceless joint space is therefore not an appropriate sector approximation for this protocol. Interleaving input rotations with circuit layers, or driving the Hamiltonian during evolution, generally removes this particular factorization and can permit a larger response sector. The encoding protocols used in the numerical experiments are specified in the main-text Methods subsection ``Reservoir models and measurement protocol''.

\subsection{Exact operator--measurement alignment law}
\label{subsec:supp-operator-measurement-alignment}

For $q>0$ and $D_{\mathrm{dyn}}>0$, subtracting the uniform components of the measurement compression and the supported-space projector defines the following quantities:
\begin{equation}
 C_{\mathrm{dyn}}^\circ
 =C_{\mathrm{dyn}}
 -\frac{r_{\mathrm{dyn}}}{D_{\mathrm{dyn}}}P_{\mathrm{dyn}},
 \qquad
 P_{\mathrm{aniso}}^+
 =P^+-\frac{q}{D_{\mathrm{dyn}}}P_{\mathrm{dyn}}.
 \label{eq:supp-anisotropic-components}
\end{equation}
Both have zero operator-space trace. The first subtracts the mean measurement exposure across the dynamical sector, while the second subtracts a uniform operator with the same trace as $P^+$.

The operator $C_{\mathrm{dyn}}^\circ$ describes how measurement exposure varies around the sector average. For every Hilbert--Schmidt-normalized $X\in W_{\mathrm{dyn}}$,
\begin{equation}
 \langle X,C_{\mathrm{dyn}}^\circ X\rangle_{\HS}
 =\|\Pi_{\mathrm{vis}}X\|_{\HS}^2
 -\frac{r_{\mathrm{dyn}}}{D_{\mathrm{dyn}}}.
\end{equation}
Positive values identify directions that are more visible than the sector average, and negative values identify less visible directions. It vanishes precisely when every normalized direction in the sector has the same visible weight.

The operator $P_{\mathrm{aniso}}^+$ describes which directions the supported space selects relative to a uniform allocation across the sector. For normalized $X\in W_{\mathrm{dyn}}$,
\begin{equation}
 \langle X,P_{\mathrm{aniso}}^+X\rangle_{\HS}
 =\|P^+X\|_{\HS}^2-\frac{q}{D_{\mathrm{dyn}}}.
\end{equation}
Its eigenvalues are $1-q/D_{\mathrm{dyn}}$ on $S^+$ and $-q/D_{\mathrm{dyn}}$ on the orthogonal complement of $S^+$ within $W_{\mathrm{dyn}}$. It records the placement of the supported space, whereas $\delta_{\mathrm{iso}}^+$ describes variation in measurement strength within that space.

Their overlap gives the alignment correction,
\begin{equation}
 \chi_{\mathrm{om}}
 =\frac1q\Tr(C_{\mathrm{dyn}}^\circ P_{\mathrm{aniso}}^+)
 =\frac1q\Tr(C_{\mathrm{dyn}}^\circ P^+).
 \label{eq:supp-chi-om-definition}
\end{equation}
The equality follows because $\Tr(C_{\mathrm{dyn}}^\circ P_{\mathrm{dyn}})=0$. A positive correction means that the supported space preferentially selects directions with above-average visibility; a negative correction means that it favors below-average directions. A zero correction indicates no net shift of the supported mean from the sector average, without requiring an isotropic supported spectrum.

\begin{suppthm}[Alignment law and supported visibility bound]
\label{thm:supp-operator-measurement-alignment-law}
For $S^+\subseteq W_{\rm dyn}$ with $q>0$,
\begin{equation}
 \alpha^+=\mu+\chi_{\rm om},\qquad
 |\sin^2\theta_j^+-\mu|\le|\chi_{\rm om}|+\delta_{\rm iso}^+.
 \label{eq:supp-alpha-alignment-law}
\end{equation}
\end{suppthm}

\begin{proof}
Using $P_{\rm dyn}P^+=P^+$ and cyclicity of trace,
\begin{equation}
 \begin{aligned}
 q\alpha^+&=\Tr(\Pi_{\rm vis}P^+)=\Tr(C_{\rm dyn}P^+)\\
 &=\mu\Tr P^++\Tr(C_{\rm dyn}^{\circ}P^+)
 =q\mu+q\chi_{\rm om}.
 \end{aligned}
\end{equation}
This proves the mean identity. By the definition of $\delta_{\mathrm{iso}}^+$, each supported visible weight satisfies $|\sin^2\theta_j^+-\alpha^+|\leq\delta_{\mathrm{iso}}^+$. Combining this with the alignment law and the triangle inequality gives
\begin{equation}
 \left|\sin^2\theta_j^+
 -\frac{r_{\mathrm{dyn}}}{D_{\mathrm{dyn}}}\right|
 \leq
 \delta_{\mathrm{iso}}^++|\chi_{\mathrm{om}}|,
 \qquad j=1,\ldots,q.
\end{equation}
\end{proof}

Small $\chi_{\rm om}$ concerns the relative orientation of two potentially structured operators. It does not require either operator to be individually isotropic. For a fixed sector, $\|P^+-(q/D_{\rm dyn})P_{\rm dyn}\|_{\HS}^2=q(1-q/D_{\rm dyn})$, which generally remains nonzero. Random orientation moments are given in Section~\ref{sec:supp-haar-orientations}.

\subsection{Finite-resolution consequence}
\label{subsec:supp-finite-resolution-consequence}

For $D_{\mathrm{dyn}}>0$, define
\begin{equation}
 \mu:=\frac{r_{\mathrm{dyn}}}{D_{\mathrm{dyn}}},
\end{equation}
the mean visible weight across the dynamically accessible operator sector. The alignment law then reads $\alpha^+=\mu+\chi_{\mathrm{om}}$.

\begin{suppthm}[Operational visibility bound under orthonormal measurement]
\label{thm:supp-finite-resolution-dilution}
Under the assumptions above,
\begin{equation}
 \sigma_{\max}(O_{P,L})
 \le\sqrt{\alpha^++\delta_{\rm iso}^+}
 \le\sqrt{\mu+|\chi_{\rm om}|+\delta_{\rm iso}^+}.
 \label{eq:supp-singular-dilution-bound}
\end{equation}
For any $\epsilon\ge0$, OVD vanishes whenever
\begin{equation}
 \sqrt{\mu+|\chi_{\rm om}|+\delta_{\rm iso}^+}\le2\epsilon.
 \label{eq:supp-zero-ovd-condition}
\end{equation}
\end{suppthm}

\begin{proof}
Since $(O^+)^{\mathsf T}O^+=B^+$ and $O^+$ has the same nonzero singular values as $O_{P,L}$, we have $\sigma_{\max}(O_{P,L})^2=\lambda_{\max}(B^+)$. Every eigenvalue is bounded by $\alpha^++\delta_{\rm iso}^+$, and $\alpha^+=\mu+\chi_{\rm om}\le\mu+|\chi_{\rm om}|$. If the resulting upper bound is at most $2\epsilon$, no singular value exceeds the strict OVD threshold.
\end{proof}

If $\alpha^+-\delta_{\rm iso}^+>4\epsilon^2$, all $q$ supported directions are resolved. If $\alpha^++\delta_{\rm iso}^+\le4\epsilon^2$, none is resolved. Individual spectral values determine intermediate counts. If $q=0$, VVR and OVD vanish and supported means and defects are left undefined. The bounds concern normalized directions at the specified resolution. 

The sector-average visibility $\mu=r_{\mathrm{dyn}}/D_{\mathrm{dyn}}$ describes measurement coverage of the accessible operator sector and can decrease as that sector grows at fixed observable budget. The alignment correction $\chi_{\mathrm{om}}$ describes whether the supported temporal space favors more or less visible directions within the sector. Positive alignment raises the supported mean above $\mu$, giving $\alpha^+=\mu+\chi_{\mathrm{om}}$. The defect $\delta_{\mathrm{iso}}^+$ controls the spread around this mean. When it is relatively small, the supported strengths cross the resolution threshold together; a broader spectrum can leave some directions resolved even when the mean is low.

\subsection{General measurement families}
\label{subsec:supp-general-measurements}

The main text assumes HS-orthonormal joint measurement operators. Here we extend the construction to arbitrary finite families of Hermitian measurement operators $F_1,\ldots,F_J$ on the joint memory--readout system. Let $d=d_Md_R$ and define their traceless parts by
\begin{equation}
 F_j^0=F_j-\frac{\Tr F_j}{d}I_{MR}.
 \label{eq:supp-general-centered-operators}
\end{equation}
Because the centered joint response is traceless, the measurement map on $E_{MR}$ is
\begin{equation}
 \mathcal M_{\mathrm g}(X)
 =\bigl(\Tr(F_jX)\bigr)_{j=1}^{J}
 =\bigl(\Tr(F_j^0X)\bigr)_{j=1}^{J}.
 \label{eq:supp-general-measurement-map}
\end{equation}
The identity components contribute only fixed offsets to uncentered expectations or probabilities.

In an HS-orthonormal basis $\{E_a\}$ of $\mathcal E_{MR}$, the general measurement map $\mathcal M_{\mathrm g}$ and its restriction to the latent space are represented by
\begin{equation}
 (R_{\mathrm g})_{ja}=\Tr(F_j^0E_a),\qquad
 O_{\mathrm g}=R_{\mathrm g}Q_{P,L},
 \label{eq:supp-general-restriction}
\end{equation}
where $Q_{P,L}$ contains the coordinates of an HS-orthonormal basis of the latent space. Each row of $R_{\mathrm g}$ represents one measurement functional, mapping joint operator coordinates to a measured signature.

The latent space and its dimension are independent of the measurement choice. Visible rank and OVD are obtained from $O_{\mathrm g}$ using the definitions in the main text, with resolution expressed in the coordinates of $\mathcal M_{\mathrm g}$.

For the general measurement map, write $K_{\mathrm g}=\ker\mathcal M_{\mathrm g}$ and denote the corresponding supported space by $S_{\mathrm g}^+$, using the same orthogonal decomposition as in the main text. Define the visible  $\Pi_\mathrm{vis,g}$ and measurement-functional Gram matrix $\Gamma_g$ as
\begin{equation}
 \begin{aligned}
 \Pi_{\mathrm{vis},\mathrm g}
 &=R_{\mathrm g}^{\mathrm{MP}}R_{\mathrm g},\\
 \Gamma_{\mathrm g}
 &=R_{\mathrm g}R_{\mathrm g}^{\mathsf T}, \quad
 (\Gamma_{\mathrm g})_{jk} =\Tr(F_j^0F_k^0).
 \end{aligned}
 \label{eq:supp-general-projector-gram}
\end{equation}
Here $\Pi_{\mathrm{vis},\mathrm g}$ projects orthogonally from the joint operator space onto $K_{\mathrm g}^{\perp}=\Span\{F_j^0\}$. It identifies the operator components accessible to measurement. By contrast, $\Gamma_{\mathrm g}$ acts on measurement-output coordinates.  It is an identity matrix for Pauli measurements, which we extend to a more general measurements. Its diagonal entries are the squared Hilbert--Schmidt norms of the measurement functionals, and its off-diagonal entries describe their overlaps. It is positive semidefinite and becomes singular when the functionals are linearly dependent.

For a reduced singular-value decomposition $R_{\mathrm g}=U_r\Sigma_rV_r^{\mathsf T}$, the columns of $U_r$ and $V_r$ are orthonormal, and $\Sigma_r$ contains the $r$ positive singular values. The Moore--Penrose inverse inverts these values while assigning zero to null directions, $R_{\mathrm g}^{\mathrm{MP}}=V_r\Sigma_r^{-1}U_r^{\mathsf T}$. Consequently,
\begin{equation}
 \Pi_{\mathrm{vis},\mathrm g}=V_rV_r^{\mathsf T},
 \qquad
 \Gamma_{\mathrm g}^{\mathrm{MP}}
 =U_r\Sigma_r^{-2}U_r^{\mathsf T}.
\end{equation}
The visible projector therefore depends on the row span of $R_{\mathrm g}$, independently of its nonzero singular values.

Using the reduced singular-value decomposition of $R_g$,
\begin{equation}
 \begin{aligned}
 R_{\mathrm g}^{\mathsf T}
 \Gamma_{\mathrm g}^{\mathrm{MP}}R_{\mathrm g}
 &=V_r\Sigma_r\Sigma_r^{-2}\Sigma_rV_r^{\mathsf T}\\
 &=V_rV_r^{\mathsf T}
 =\Pi_{\mathrm{vis},\mathrm g}.
 \end{aligned}
\end{equation}
Let $Q_{P,L}$ contain the coordinates of a Hilbert--Schmidt-orthonormal basis of $S$, and write $O_{\mathrm g}=R_{\mathrm g}Q_{P,L}$. The visibility matrix on the latent space is therefore
\begin{equation}
 \begin{aligned}
 B_{\mathrm g}
 &=Q_{P,L}^{\mathsf T}
 \Pi_{\mathrm{vis},\mathrm g}Q_{P,L}\\
 &=Q_{P,L}^{\mathsf T}R_{\mathrm g}^{\mathsf T}
 \Gamma_{\mathrm g}^{\mathrm{MP}}
 R_{\mathrm g}Q_{P,L}\\
 &=O_{\mathrm g}^{\mathsf T}
 \Gamma_{\mathrm g}^{\mathrm{MP}}O_{\mathrm g}.
 \end{aligned}
 \label{eq:supp-general-visibility-compression}
\end{equation}
Thus geometric visibility is obtained by weighting the measured signatures with the Gram pseudoinverse. Their ordinary squared strengths are instead described by $O_{\mathrm g}^{\mathsf T}O_{\mathrm g}$. The extremal positive eigenvalues of $\Gamma_{\mathrm g}^{\mathrm{MP}}$ bound the difference between these two matrices.

For nonzero measurement rank, define the minimum and maximum gains as the smallest positive and largest singular values of $R_{\mathrm g}$,
\begin{equation}
 m_-:=\sigma_{\min}^{+}(R_{\mathrm g}),
 \qquad
 m_+:=\sigma_{\max}(R_{\mathrm g}),
 \qquad 0<m_-\leq m_+.
 \label{eq:supp-general-gains}
\end{equation}
These quantify the smallest and largest amplification of Hilbert--Schmidt-normalized directions in $K_{\mathrm g}^{\perp}$. Since $\Gamma_{\mathrm g}^{\mathrm{MP}}=U_r\Sigma_r^{-2}U_r^{\mathsf T}$, its extremal positive eigenvalues are
\begin{equation}
 \lambda_{\min}^{+}(\Gamma_{\mathrm g}^{\mathrm{MP}})
 =\frac{1}{m_+^2},
 \qquad
 \lambda_{\max}(\Gamma_{\mathrm g}^{\mathrm{MP}})
 =\frac{1}{m_-^2}.
\end{equation}
Consequently, every $z\in\operatorname{range}R_{\mathrm g}$ satisfies
\begin{equation}
 \frac{\|z\|_2^2}{m_+^2}
 \leq z^{\mathsf T}\Gamma_{\mathrm g}^{\mathrm{MP}}z
 \leq\frac{\|z\|_2^2}{m_-^2}.
\end{equation}
The range restriction excludes null output directions when the measurement functionals are redundant.

For any latent coordinate vector $x$, the measured signature $z=O_{\mathrm g}x$ belongs to this range. Substituting it into the preceding inequality and using Eq.~\eqref{eq:supp-general-visibility-compression} gives
\begin{equation}
 \frac{1}{m_+^2}O_{\mathrm g}^{\mathsf T}O_{\mathrm g}
 \preceq B_{\mathrm g}
 \preceq
 \frac{1}{m_-^2}O_{\mathrm g}^{\mathsf T}O_{\mathrm g}.
\end{equation}
Equivalently,
\begin{equation}
 m_-^2B_{\mathrm g}
 \preceq O_{\mathrm g}^{\mathsf T}O_{\mathrm g}
 \preceq m_+^2B_{\mathrm g}.
 \label{eq:supp-general-loewner-bound}
\end{equation}

The positive eigenvalues of $B_{\mathrm g}$ are the supported squared visible weights $\sin^2\theta_{j,\mathrm g}^+$. The nonzero singular values of $O_{\mathrm g}$ equal those of its restriction $O_{\mathrm g}^+$ to the supported subspace. Applying the eigenvalue min--max principle to Eq.~\eqref{eq:supp-general-loewner-bound} therefore yields
\begin{equation}
 m_-^2\sin^2\theta_{j,\mathrm g}^+
 \leq\sigma_j(O_{\mathrm g}^+)^2
 \leq m_+^2\sin^2\theta_{j,\mathrm g}^+,
\end{equation}
and hence
\begin{equation}
 m_-\sin\theta_{j,\mathrm g}^+
 \leq\sigma_j(O_{\mathrm g}^+)
 \leq m_+\sin\theta_{j,\mathrm g}^+,
 \label{eq:supp-general-angle-strength}
\end{equation}
with consistently ordered spectra. For equal gain $m_-=m_+=g$, this becomes $\sigma_j(O_{\mathrm g}^+)=g\sin\theta_{j,\mathrm g}^+$. The orthonormal convention in the main text has $g=1$. For unequal gains, the canonical visibility directions need not coincide with the right singular directions of the measurement restriction.

For a nonzero supported space, the alignment law and isotropy bounds apply using the projector $\Pi_{\mathrm{vis},\mathrm g}$ and the corresponding supported space. Retaining the main-text notation for these geometric quantities gives
\begin{equation}
 \begin{aligned}
 \sigma_{\max}(O_{\mathrm g})
 &\leq m_+\sqrt{\alpha^++\delta_{\mathrm{iso}}^+}\\
 &\leq m_+\sqrt{
 \frac{r_{\mathrm{dyn}}}{D_{\mathrm{dyn}}}
 +|\chi_{\mathrm{om}}|+\delta_{\mathrm{iso}}^+}.
 \end{aligned}
 \label{eq:supp-general-operational-bound}
\end{equation}
Consequently, OVD vanishes whenever
\begin{equation}
 m_+\sqrt{
 \frac{r_{\mathrm{dyn}}}{D_{\mathrm{dyn}}}
 +|\chi_{\mathrm{om}}|+\delta_{\mathrm{iso}}^+}
 \leq2\epsilon.
 \label{eq:supp-general-zero-ovd}
\end{equation}
If the supported space is zero-dimensional, VVR and OVD already vanish. If the measurement rank itself is zero, the gain definitions are unnecessary.

\section{Haar-random latent spaces and supported visibility}
\label{sec:supp-haar-orientations}

The random-orientation model used in the main text samples the latent operator space before removing directions erased by measurement. This section derives its supported dimension, the first two moments of its supported mean visibility, and the expected finite-resolution count. We also distinguish fluctuations of the mean between random spaces from the spread of visible weights within each space.

\subsection{Latent-space ensemble and support selection}
\label{subsec:supp-haar-ensemble}

Work in the real Hilbert--Schmidt operator sector $W_{\mathrm{dyn}}$ and write
\begin{equation}
 \begin{gathered}
 D=D_{\mathrm{dyn}}\geq2,\qquad C=C_{\mathrm{dyn}},\qquad
 r=\Tr C,\qquad m=\rank C>0,\\
 \mu=\frac rD,\qquad C^\circ=C-\mu P_{\mathrm{dyn}}.
 \end{gathered}
 \label{eq:supp-haar-sector-data}
\end{equation}
Here $0\preceq C\preceq P_{\mathrm{dyn}}$, and $\ker C=\ker(\Mmap|_{W_{\mathrm{dyn}}})$ when both are regarded as maps on this sector. Traces and HS norms of endomorphisms in this section are taken over the real operator sector; $P_{\mathrm{dyn}}$ is its identity.

Let $P_0$ be a fixed rank-$s$ orthogonal projector in $W_{\mathrm{dyn}}$, with $1\leq s\leq D$. For a Haar-distributed orthogonal transformation $U$ of this sector, define
\begin{equation}
 \begin{gathered}
 P_U=UP_0U^{\mathsf T},\qquad S_U=\operatorname{ran}P_U,\\
 N_U=S_U\cap\ker C,\qquad S_U^+=S_U\ominus N_U.
 \end{gathered}
 \label{eq:supp-haar-supported-projector}
\end{equation}
The symbol $\ominus$ denotes the orthogonal complement within $S_U$. Let $P_U^+$ project onto $S_U^+$ and let $Q_U^+$ be an HS-orthonormal basis map for that space. Its supported compression and mean are
\begin{equation}
 \begin{gathered}
 B_U^+=(Q_U^+)^{\mathsf T}C Q_U^+,\qquad
 q=\dim S_U^+,\\
 \alpha_U^+=\frac1q\Tr B_U^+,\qquad
 \chi_{\mathrm{om},U}=\alpha_U^+-\mu.
 \end{gathered}
 \label{eq:supp-haar-supported-mean}
\end{equation}
The theorem below shows that $q=\min\{s,m\}>0$ almost surely, so these ensemble quantities are well defined outside a probability-zero exceptional set. Under the orthonormal measurement convention, the eigenvalues of $B_U^+$ are $b_{j,U}=\sin^2\theta_{j,U}^+=\sigma_j(O_U)^2$.

\subsection{First and second projector moments}
\label{subsec:supp-haar-projector-moments}

Orthogonal invariance and $\Tr P_U=s$ give
\begin{equation}
 \mathbb E[P_U]=\frac{s}{D}P_{\mathrm{dyn}}.
 \label{eq:supp-haar-first-moment}
\end{equation}
Indeed, the expectation commutes with every orthogonal transformation of $W_{\mathrm{dyn}}$, so it is a scalar multiple of the identity; its trace fixes that scalar.

In an orthonormal coordinate basis of the sector, invariance and symmetry of $P_U$ also imply
\begin{equation}
 \mathbb E[(P_U)_{ij}(P_U)_{kl}]
 =a_s\delta_{ij}\delta_{kl}
  +b_s(\delta_{ik}\delta_{jl}+\delta_{il}\delta_{jk}).
 \label{eq:supp-haar-projector-entry-moment}
\end{equation}
The identities $(\Tr P_U)^2=s^2$ and $\Tr(P_U^2)=s$ give the two equations
\begin{equation}
 a_sD^2+2b_sD=s^2,\qquad
 a_sD+b_s(D^2+D)=s.
\end{equation}
Solving them yields
\begin{equation}
 a_s=\frac{s[s(D+1)-2]}{D(D-1)(D+2)},\qquad
 b_s=\frac{s(D-s)}{D(D-1)(D+2)}.
 \label{eq:supp-haar-projector-coefficients}
\end{equation}
Consequently, for arbitrary symmetric endomorphisms $A,B$ of $W_{\mathrm{dyn}}$,
\begin{equation}
 \mathbb E[\Tr(AP_U)\Tr(BP_U)]
 =a_s\Tr A\Tr B+2b_s\Tr(AB).
 \label{eq:supp-haar-projector-trace-moment}
\end{equation}
Subtracting the product of the means gives
\begin{equation}
 \operatorname{Cov}\!\left(\Tr(AP_U),\Tr(BP_U)\right)
 =2b_s\left[\Tr(AB)-\frac{\Tr A\Tr B}{D}\right].
 \label{eq:supp-haar-orbit-second-moment}
\end{equation}
These identities hold for every $1\leq s\leq D$, including $s>m$. They do not assume that the sampled latent space is already supported.

\subsection{Supported mean, alignment and resolvable count}
\label{subsec:supp-haar-supported-moments}

\begin{suppthm}[Supported visibility moments]
\label{thm:supp-haar-alignment-concentration}
For the ensemble in Eq.~\eqref{eq:supp-haar-supported-projector},
\begin{equation}
 q=\min\{s,m\}\quad\text{almost surely},\qquad
 \mathbb E[\alpha_U^+]=\frac{sr}{Dq}.
 \label{eq:supp-haar-supported-first-moment}
\end{equation}
The supported mean and its alignment correction have variance
\begin{equation}
 \operatorname{Var}(\alpha_U^+)
 =\operatorname{Var}(\chi_{\mathrm{om},U})
 =\frac{2s(D-s)}{q^2D(D-1)(D+2)}\|C^\circ\|_{\HS}^2,
 \label{eq:supp-haar-alignment-variance}
\end{equation}
while
\begin{equation}
 \mathbb E[\chi_{\mathrm{om},U}]
 =\left(\frac{s}{q}-1\right)\frac rD.
 \label{eq:supp-haar-selected-alignment-mean}
\end{equation}
For every $\epsilon>0$, orthonormal measurement gives
\begin{equation}
 \mathbb E[\omega_U(\epsilon)]
 \leq\min\left\{q,\frac{sr}{4D\epsilon^2}\right\}.
 \label{eq:supp-haar-supported-count}
\end{equation}
\end{suppthm}

\begin{proof}
A uniformly random $s$-plane can be generated as the column span of a real $D\times s$ matrix of independent standard Gaussian entries. In coordinates diagonalizing $C$, its projection onto $\operatorname{ran}C$ has the rank of an $m\times s$ Gaussian matrix, namely $\min\{s,m\}$ almost surely. Thus $\dim N_U=s-q$ and $q=\min\{s,m\}$.

Let $P_{N_U}$ project onto the invisible intersection. The orthogonal decomposition $P_U=P_U^++P_{N_U}$ and $CP_{N_U}=0$ imply
\begin{equation}
 q\alpha_U^+
 =\Tr(CP_U^+)=\Tr(CP_U).
 \label{eq:supp-haar-support-trace-identity}
\end{equation}
Taking expectations and using Eq.~\eqref{eq:supp-haar-first-moment} gives $\mathbb E[\alpha_U^+]=sr/(Dq)$. Taking $A=B=C$ in Eq.~\eqref{eq:supp-haar-orbit-second-moment} gives
\begin{equation}
 \operatorname{Var}(\Tr(CP_U))
 =2b_s\left[\Tr(C^2)-\frac{r^2}{D}\right]
 =2b_s\|C^\circ\|_{\HS}^2.
\end{equation}
Division by $q^2$ proves Eq.~\eqref{eq:supp-haar-alignment-variance}. Since $\chi_{\mathrm{om},U}=\alpha_U^+-r/D$, it has the same variance and the mean in Eq.~\eqref{eq:supp-haar-selected-alignment-mean}.

Finally, $\Tr(CP_U)=\sum_{j=1}^q\sigma_j(O_U)^2$. Every strength counted by OVD has square greater than $4\epsilon^2$, so
\begin{equation}
 4\epsilon^2\omega_U(\epsilon)\leq\Tr(CP_U).
\end{equation}
Taking expectations and combining with $\omega_U\leq q$ proves Eq.~\eqref{eq:supp-haar-supported-count}.
\end{proof}

When $s\leq m$, no invisible direction is removed almost surely, so $q=s$ and the supported mean is centered at $\mu=r/D$. When $s>m$, the support is selected using the measurement kernel; its distribution is generally not uniform on the rank-$m$ Grassmannian. The positive expected alignment correction in Eq.~\eqref{eq:supp-haar-selected-alignment-mean} records this selection. It does not require additional dynamical alignment.

For $\eta>0$, Chebyshev's inequality gives
\begin{equation}
 \Pr\!\left(\left|\alpha_U^+-\frac{sr}{Dq}\right|\geq\eta\right)
 \leq
 \frac{2s(D-s)\|C^\circ\|_{\HS}^2}
 {q^2D(D-1)(D+2)\eta^2}.
 \label{eq:supp-haar-alignment-tail}
\end{equation}
The contraction property also gives
\begin{equation}
 \|C^\circ\|_{\HS}^2
 =\Tr(C^2)-\frac{r^2}{D}
 \leq r\left(1-\frac rD\right).
 \label{eq:supp-haar-anisotropy-bound}
\end{equation}
These bounds describe fluctuations between random latent spaces. Relative concentration requires the variance to be small compared with the square of the expected mean; increasing $D$ alone does not guarantee this. Likewise, finite-resolution collapse follows in probability along a sequence of ensembles if $sr/(D\epsilon^2)\to0$, because $\Pr(\omega_U(\epsilon)\geq1)\leq\mathbb E[\omega_U(\epsilon)]$. The dependence of $s$ and $\epsilon$ on system size must be included in this condition.

\subsection{Spread within the supported spectrum}
\label{subsec:supp-haar-spectral-spread}

The preceding variance concerns the scalar $\alpha_U^+$ across realizations. The population variance of the supported visible weights within one realization is instead
\begin{equation}
 V_U^+:=\frac1q\sum_{j=1}^q(b_{j,U}-\alpha_U^+)^2
 =\frac1q\Tr[(B_U^+)^2]-(\alpha_U^+)^2.
 \label{eq:supp-haar-within-spectrum-variance}
\end{equation}
Put $c_2=\Tr(C^2)$ and use $a_s,b_s$ from Eq.~\eqref{eq:supp-haar-projector-coefficients}. The exact expected spread is
\begin{equation}
 \mathbb E[V_U^+]
 =\frac{(a_s+b_s)c_2+b_s r^2}{q}
  -\frac{a_s r^2+2b_s c_2}{q^2}.
 \label{eq:supp-haar-expected-spectral-spread}
\end{equation}
To verify this expression, removal of the null directions preserves both nonzero spectral moments, giving
\begin{equation}
 \Tr[(B_U^+)^2]=\Tr(CP_UCP_U).
\end{equation}
Diagonalizing $C$ and expanding $\Tr(CP_UCP_U)=\sum_{i,j}c_ic_j(P_U)_{ij}^2$ in Eq.~\eqref{eq:supp-haar-projector-entry-moment} yields
\begin{equation}
 \mathbb E[\Tr(CP_UCP_U)]
 =(a_s+b_s)c_2+b_s r^2.
\end{equation}
Equation~\eqref{eq:supp-haar-projector-trace-moment} similarly gives $\mathbb E[(\Tr(CP_U))^2]=a_s r^2+2b_s c_2$. Substitution proves Eq.~\eqref{eq:supp-haar-expected-spectral-spread}.

If $C$ is an orthogonal projector of rank $m$, then $r=c_2=m$. Writing $h=\max\{s,m\}$ and $q=\min\{s,m\}$ simplifies the results to
\begin{equation}
 \begin{aligned}
 \mathbb E[\alpha_U^+]&=\frac hD,\\
 \operatorname{Var}(\alpha_U^+)
 &=\frac{2sm(D-s)(D-m)}{q^2D^2(D-1)(D+2)},\\
 \mathbb E[V_U^+]
 &=\frac{h(D-h)(q-1)(q+2)}{qD(D-1)(D+2)}.
 \end{aligned}
 \label{eq:supp-haar-projector-specialization}
\end{equation}
In particular, full-sector rank-$J$ measurement has $m=J$, so the supported mean is $\max\{s,J\}/D$, not generally $J/D$. The relative expected spread obeys
\begin{equation}
 \frac{\mathbb E[V_U^+]}{\mathbb E[\alpha_U^+]^2}
 =\frac{D(D-h)(q-1)(q+2)}{qh(D-1)(D+2)}.
 \label{eq:supp-haar-relative-spectral-spread}
\end{equation}
For fixed $s,m$ and large $D$, the absolute spread decreases, but this ratio approaches $(q-1)(q+2)/(qh)$. A small absolute spread around a vanishing mean is therefore not sufficient for relative isotropic compression.

For each realization, the spectral variance and the isotropy defect are related by
\begin{equation}
 V_U^+\leq(\delta_{\mathrm{iso},U}^+)^2\leq qV_U^+.
 \label{eq:supp-haar-spread-isotropy-comparison}
\end{equation}
Thus the expected spread supplies a bound on the expected squared defect, but does not certify a small defect in every realization. The angular scale $\arcsin\sqrt{\mathbb E[\alpha_U^+]}$ is a reference derived from the mean squared visibility; it is not the arithmetic mean of the angles. Statements about individual angles or their relative concentration require control of the spectral spread and the nonlinear transformation.

All results above concern a Haar ensemble of real operator subspaces. Haar-distributed quantum gates, physical scrambling, or a random basis of readout observables do not by themselves establish this distribution for the generated latent space.

\section{Proofs of capacity, readout and truncation results}
\label{sec:supp-capacity-readout-truncation}

This section proves the main-text population linear-capacity theorem, optimal-readout proposition, and truncation and sampling proposition. The arguments use the same HS-normalized measurement coordinates as the main text. Each result is restated before its proof; the supplementary numbering is independent of the main-text numbering.

\subsection{Population linear capacity and geometric bounds}
\label{subsec:supp-capacity-proof}

The following restates main-text Theorem~3 together with its OVD consequence.

\begin{suppthm}[Population linear capacity]
\label{thm:supp-capacity-restatement}
Let $z\in\mathbb R^s$ be the centered coordinates of the truncated response in an HS-orthonormal latent basis, and let $O\in\mathbb R^{J\times s}$. Write
\begin{equation}
 y=Oz+\eta,\qquad \mathbb E[\eta\mid z]=0,\qquad
 \Sigma=\mathbb E[zz^{\mathsf T}]\succ0,\qquad
 V=\mathbb E[\eta\eta^{\mathsf T}]\succ0.
 \label{eq:supp-capacity-model}
\end{equation}

For $x=\Sigma^{-1/2}z$ and $A=V^{-1/2}O\Sigma^{1/2}$, the total optimal linear reconstruction capacity is
\begin{equation}
 \begin{aligned}
 C_{\mathrm{lin}}
 &:=\sum_{\ell=1}^{s}
 \left[1-\min_{w\in\mathbb R^J}
 \mathbb E\!\left[(x_\ell-w^{\mathsf T}y)^2\right]\right]\\
 &=\sum_{\ell=1}^{s}\frac{\lambda_\ell}{1+\lambda_\ell},
 \end{aligned}
 \label{eq:supp-capacity-identity}
\end{equation}
where $\lambda_\ell$ are the eigenvalues of $A^{\mathsf T}A$, including zeros. If $aI_s\preceq\Sigma\preceq bI_s$ and $v_-I_J\preceq V\preceq v_+I_J$, with $a,v_->0$, the decreasingly ordered spectra satisfy
\begin{equation}
 \kappa_-\sigma_\ell(O)^2\leq\lambda_\ell
 \leq\kappa_+\sigma_\ell(O)^2,\qquad
 \kappa_-=\frac{a}{v_+},\quad
 \kappa_+=\frac{b}{v_-},
 \label{eq:supp-capacity-spectral-comparison}
\end{equation}
with zeros padded to length $s$.

Under these covariance bounds and orthonormal measurement, put $q=\rank O$, $\omega=\#\{\ell:\sigma_\ell(O)>2\epsilon\}$ for $\epsilon\geq0$, $t_\epsilon=\min\{4\epsilon^2,1\}$, and $F_\kappa(t)=\kappa t/(1+\kappa t)$. Then
\begin{equation}
 \omega F_{\kappa_-}(t_\epsilon)
 \leq C_{\mathrm{lin}}
 \leq\omega F_{\kappa_+}(1)
 +(q-\omega)F_{\kappa_+}(t_\epsilon).
 \label{eq:supp-capacity-ovd-bounds}
\end{equation}
\end{suppthm}

\begin{proof}
The conditional mean assumption gives $\mathbb E[z\eta^{\mathsf T}]=0$; independence or Gaussian statistics are not required. Put $B=O\Sigma^{1/2}$, so $A=V^{-1/2}B$. Then $\mathbb E[xx^{\mathsf T}]=I_s$, $y=Bx+\eta$, $\mathbb E[xy^{\mathsf T}]=B^{\mathsf T}$ and $R_y:=\mathbb E[yy^{\mathsf T}]=BB^{\mathsf T}+V\succ0$.

For a linear estimator $\widehat x=Wy$, completing the square in its residual covariance gives
\begin{equation}
 \begin{aligned}
 \mathbb E[(x-Wy)(x-Wy)^{\mathsf T}]
 &=I_s-B^{\mathsf T}R_y^{-1}B\\
 &\quad +(W-W_*)R_y(W-W_*)^{\mathsf T},\\
 W_*&=B^{\mathsf T}R_y^{-1}.
 \end{aligned}
 \label{eq:supp-capacity-least-squares}
\end{equation}
The last term is positive semidefinite, so $W_*$ minimizes each target's mean squared error. The optimal residual covariance is
\begin{equation}
 \begin{aligned}
 E_{\mathrm{lin}}
 &=I_s-B^{\mathsf T}(BB^{\mathsf T}+V)^{-1}B\\
 &=I_s-A^{\mathsf T}(I_J+AA^{\mathsf T})^{-1}A\\
 &=(I_s+A^{\mathsf T}A)^{-1}.
 \end{aligned}
 \label{eq:supp-capacity-residual-covariance}
\end{equation}
The second equality follows by factoring $V^{1/2}$ on both sides of $BB^{\mathsf T}+V$; the last follows from a singular-value decomposition of $A$. Since each $x_\ell$ has unit variance, the total optimal reconstruction capacity is
\begin{equation}
 C_{\mathrm{lin}}=\Tr(I_s-E_{\mathrm{lin}})
 =\sum_{\ell=1}^{s}\frac{\lambda_\ell}{1+\lambda_\ell},
 \label{eq:supp-capacity-trace-identity}
\end{equation}
where $\lambda_\ell$ are the eigenvalues of $A^{\mathsf T}A$, including zeros. This proves the capacity identity in the main text.

For the geometric comparison, assume $aI_s\preceq\Sigma\preceq bI_s$ and $v_-I_J\preceq V\preceq v_+I_J$, with $a,v_->0$. Put $K=O^{\mathsf T}V^{-1}O$. Then
\begin{equation}
 \begin{gathered}
 v_+^{-1}O^{\mathsf T}O\preceq K
 \preceq v_-^{-1}O^{\mathsf T}O,\\
 aK\preceq K^{1/2}\Sigma K^{1/2}\preceq bK.
 \end{gathered}
 \label{eq:supp-capacity-gram-comparison}
\end{equation}
The matrices $A^{\mathsf T}A=\Sigma^{1/2}K\Sigma^{1/2}$ and $K^{1/2}\Sigma K^{1/2}$ have identical eigenvalues, including zeros: they are $G^{\mathsf T}G$ and $GG^{\mathsf T}$ for $G=K^{1/2}\Sigma^{1/2}$. The eigenvalue min--max principle therefore proves Eq.~\eqref{eq:supp-capacity-spectral-comparison}.

For the OVD bounds, orthonormal measurement gives $\sigma_\ell(O)\leq1$. When $4\epsilon^2<1$, the $\omega$ resolved squared strengths lie above $t_\epsilon$, the remaining $q-\omega$ nonzero strengths lie at or below $t_\epsilon$, and the other strengths vanish. Applying the increasing functions $F_{\kappa_-}$ and $F_{\kappa_+}$ to these groups proves Eq.~\eqref{eq:supp-capacity-ovd-bounds}. When $4\epsilon^2\geq1$, $\omega=0$ and $t_\epsilon=1$; the same bounds follow from $0\leq\sigma_\ell(O)^2\leq1$.

At fixed response statistics, if $V=V_1/N$ with $V_1\succ0$, then $A^{\mathsf T}A=N\Sigma^{1/2}O^{\mathsf T} V_1^{-1}O\Sigma^{1/2}$. Thus $C_{\mathrm{lin}}(N)=\sum_\ell N\lambda_\ell^{(1)}/ (1+N\lambda_\ell^{(1)})$, where $\lambda_\ell^{(1)}$ are the eigenvalues at $N=1$. If $\Sigma$ is singular, the capacity identity applies after restricting the latent coordinates to its covariance support; the geometric rank and OVD in the comparison must then be computed on that same support. A singular noise covariance requires separate treatment of noiseless directions.
\end{proof}

\subsection{Optimal readout subspace}
\label{subsec:supp-optimal-readout-proof}

The following restates main-text Proposition~2.

\begin{suppprop}[Optimal readout subspace]
\label{prop:supp-optimal-readout-restatement}
Let $S$ be the generated traceless joint operator space, with $s=\dim S$, and define
\begin{equation}
 T_R:S\longrightarrow\operatorname{Herm}_0(\mathcal H_R),
 \qquad T_R(X)=\frac{\Tr_M X}{\sqrt{d_M}}.
\end{equation}
Both spaces carry their HS inner products. Let $\tau_1\geq\tau_2\geq\cdots\geq0$ be the singular values of $T_R$, padded with zeros when needed. For $1\leq J\leq d_R^2-1$, maximize over traceless Hermitian readout observables $M_1,\ldots,M_J$ satisfying
\begin{equation}
 \Tr[(I_M\otimes M_j)(I_M\otimes M_k)]=\delta_{jk}.
\end{equation}
In an HS-orthonormal basis $\{H_\ell\}_{\ell=1}^s$ of $S$, put $O_{j\ell}=\Tr[(I_M\otimes M_j)H_\ell]$ and $\omega_{P,L}(\epsilon)=\#\{k:\sigma_k(O)>2\epsilon\}$. At a common geometric resolution $\epsilon\geq0$,
\begin{equation}
 \max_{\{M_j\}}\omega_{P,L}(\epsilon)
 =\min\{J,\#\{k:\tau_k>2\epsilon\}\}.
 \label{eq:supp-optimal-readout-maximum}
\end{equation}
The maximum is attained by $M_j=B_j/\sqrt{d_M}$, where the $B_j$ are the first $J$ orthonormal left singular operators of $T_R$, completed in the readout space when its rank is smaller than $J$.
\end{suppprop}

\begin{proof}
Fix the generated joint operator space $S$, with $s=\dim S$, and put $a=d_R^2-1$. Choose HS-orthonormal bases $\{H_\ell\}_{\ell=1}^{s}$ of $S$ and $\{F_b\}_{b=1}^{a}$ of traceless Hermitian readout operators. For a qubit readout, the latter can be constructed from the nonidentity Pauli strings divided by $\sqrt{d_R}$. The restriction of the normalized partial trace to $S$ is represented by $T\in\mathbb R^{a\times s}$, with
\begin{equation}
 T_{b\ell}
 =\langle F_b,T_R(H_\ell)\rangle_{\HS}
 =\frac{\Tr[(I_M\otimes F_b)H_\ell]}{\sqrt{d_M}}.
 \label{eq:supp-readout-restriction-matrix}
\end{equation}
Its singular values $\tau_k$ depend on $S$, even though the full partial-trace operation is fixed.

On the full traceless joint operator space, the normalized partial trace has adjoint $B\mapsto I_M\otimes B/\sqrt{d_M}$. This adjoint preserves the HS norm, because
\begin{equation}
 \left\|\frac{I_M\otimes B}{\sqrt{d_M}}\right\|_{\HS}^2
 =\frac{\Tr I_M}{d_M}\Tr(B^2)=\|B\|_{\HS}^2.
\end{equation}
The full map therefore has norm one; restriction of its domain gives $\|T_R\|_{\mathrm{op}}\leq1$.

For any allowed observable family, define $B_j=\sqrt{d_M}M_j$. Orthonormality of the joint extensions $I_M\otimes M_j$ implies $\Tr(B_jB_k)=\delta_{jk}$. Expand
\begin{equation}
 B_j=\sum_{b=1}^{a}U_{bj}F_b,\qquad
 U^{\mathsf T}U=I_J.
\end{equation}
The measured restriction satisfies
\begin{equation}
 \begin{aligned}
 O_{j\ell}
 &=\Tr[(I_M\otimes M_j)H_\ell]\\
 &=\langle B_j,T_R(H_\ell)\rangle_{\HS}
 =\sum_b U_{bj}T_{b\ell}.
 \end{aligned}
\end{equation}
Consequently,
\begin{equation}
 O=U^{\mathsf T}T,\qquad
 O^{\mathsf T}O=T^{\mathsf T}UU^{\mathsf T}T
 \preceq T^{\mathsf T}T,
 \label{eq:readout-proof-compression}
\end{equation}
since $UU^{\mathsf T}$ is an orthogonal projector. The eigenvalue min--max principle gives $\sigma_k(O)\leq\tau_k$, and $\rank O\leq J$. Thus every allowed measurement family satisfies
\begin{equation}
 \omega_{P,L}(\epsilon)
 \leq\min\{J,\#\{k:\tau_k>2\epsilon\}\}.
\end{equation}

To attain this bound, choose $B_j$ along the first $J$ orthonormal left singular operators of $T_R$ and set $M_j=B_j/\sqrt{d_M}$. If $X_k$ are the right singular operators, then
\begin{equation}
 T_R(X_k)=\tau_kB_k,\qquad
 \Tr[(I_M\otimes M_j)X_k]=\tau_k\delta_{jk}
\end{equation}
for the nonzero singular directions. Hence $O$ retains the largest $J$ singular values of $T_R$, or all of them if its rank is smaller than $J$. In the latter case, complete the family with orthonormal readout operators orthogonal to $\operatorname{ran}T_R$; these add zero measurement strength. The upper bound is attained, proving the optimal-readout proposition.
\end{proof}

The optimization fixes the geometric normalization and resolution; it does not impose equal physical sampling costs for different observable choices.

\subsection{Truncation and finite-shot readout error}
\label{subsec:supp-truncation-sampling-proof}

The following restates main-text Proposition~4.

\begin{suppprop}[Truncation and sampling]
\label{prop:supp-truncation-sampling-restatement}
Let $Z_t$ be the complete centered Volterra response, with kernel orders $p(\lambda)$ and largest delays $\tau_{\max}(\lambda)$. Let $Z_t^{(P,L)}$ retain the kernels with order at most $P$ and largest delay at most $L$. Assume $v_{\max}>0$, $|v_t|\leq v_{\max}$ and the weighted summability condition
\begin{equation}
 K_{\varrho,\zeta}
 :=\sum_\lambda\|\widehat h_\lambda\|_{\HS}
 (\varrho v_{\max})^{p(\lambda)}
 \zeta^{\tau_{\max}(\lambda)}<\infty,
 \qquad \varrho>1,\quad\zeta>1.
 \label{eq:supp-truncation-weighted-envelope}
\end{equation}

Then, uniformly over admissible input histories,
\begin{equation}
 \begin{gathered}
 \|Z_t-Z_t^{(P,L)}\|_{\HS}\leq R_{P,L},\\
 R_{P,L}:=K_{\varrho,\zeta}
 \max\{\varrho^{-(P+1)},\zeta^{-(L+1)}\}.
 \end{gathered}
 \label{eq:supp-truncation-uniform-bound}
\end{equation}
For the sampling statement, fix an input history and let $P_j$ be distinct nonidentity readout Pauli strings. Write $A_j=I_M\otimes P_j$, $d=d_Md_R$, and $M_j=P_j/\sqrt d$, so the joint measurement operators $I_M\otimes M_j=A_j/\sqrt d$ are HS-orthonormal. From $N_j$ independent shots per observable, independent also across observables, form
\begin{equation}
 \widehat z_{j,t}
 =\frac1{\sqrt d}
 \left(\frac1{N_j}\sum_{a=1}^{N_j}X_{j,t}^{(a)}-\bar f_j\right),
 \qquad X_{j,t}^{(a)}\in\{-1,1\},
 \label{eq:supp-truncation-pauli-estimator}
\end{equation}
where $\mathbb E[X_{j,t}^{(a)}]=f_{j,t}:=\Tr(A_j\sigma_t)$ and the reference $\bar f_j=\Tr(A_j\bar\sigma)$ is known. Then $\mathbb E[\widehat z_t]=\Mmap(Z_t)$.

For every fixed readout vector $w\in\mathbb R^J$ and $0<\delta<1$, with probability at least $1-\delta$,
\begin{equation}
 \left|w^{\mathsf T}\widehat z_t
       -w^{\mathsf T}\Mmap(Z_t^{(P,L)})\right|
 \leq\|w\|_2R_{P,L}
 +\left[\frac{2\log(2/\delta)}d
 \sum_{j=1}^J\frac{w_j^2}{N_j}\right]^{1/2}.
 \label{eq:supp-truncation-readout-guarantee}
\end{equation}
\end{suppprop}

\begin{proof}
Every omitted kernel has $p(\lambda)\geq P+1$ or $\tau_{\max}(\lambda)\geq L+1$. Its inverse weight is therefore bounded by
\begin{equation}
 \varrho^{-p(\lambda)}\zeta^{-\tau_{\max}(\lambda)}
 \leq\max\{\varrho^{-(P+1)},\zeta^{-(L+1)}\}.
\end{equation}
Since $|\phi_\lambda(\mathbf v_t)|\leq v_{\max}^{p(\lambda)}$, the triangle inequality gives
\begin{equation}
 \begin{aligned}
 \|Z_t-Z_t^{(P,L)}\|_{\HS}
 &\leq\sum_{\lambda\notin\Lambda_{P,L}}
 \|\widehat h_\lambda\|_{\HS}v_{\max}^{p(\lambda)}\\
 &\leq R_{P,L},\\
 R_{P,L}&:=K_{\varrho,\zeta}
 \max\{\varrho^{-(P+1)},\zeta^{-(L+1)}\}.
 \end{aligned}
 \label{eq:supp-truncation-tail-estimate}
\end{equation}
The bound is uniform over admissible input histories.

By the estimator definition, $\mathbb E[\widehat z_t]=\Mmap(Z_t)$. For a fixed readout vector $w$, its sampling error is
\begin{equation}
 \xi_t:=w^{\mathsf T}[\widehat z_t-\Mmap(Z_t)]
 =\sum_{j=1}^J\sum_{a=1}^{N_j}
 \frac{w_j}{\sqrt d\,N_j}
 (X_{j,t}^{(a)}-f_{j,t}).
\end{equation}
The summands are independent and centered, with range lengths $2|w_j|/(\sqrt d\,N_j)$. The sum of their squared range lengths is $4d^{-1}\sum_jw_j^2/N_j$. Hoeffding's inequality therefore gives, for $t_0>0$,
\begin{equation}
 \Pr(|\xi_t|\geq t_0)
 \leq2\exp\!\left[-\frac{d\,t_0^2}
 {2\sum_j w_j^2/N_j}\right].
 \label{eq:supp-truncation-hoeffding-tail}
\end{equation}
If $w=0$, the error is zero and the following bound is immediate. Otherwise, choosing the right-hand side to equal $\delta$ gives, with probability at least $1-\delta$,
\begin{equation}
 |\xi_t|\leq
 \left[\frac{2\log(2/\delta)}d
 \sum_{j=1}^J\frac{w_j^2}{N_j}\right]^{1/2}.
\end{equation}

Orthonormal measurement satisfies $\|\Mmap\|_{\HS\to2}\leq1$, so the deterministic truncation contribution obeys
\begin{equation}
 \left|w^{\mathsf T}\Mmap(Z_t-Z_t^{(P,L)})\right|
 \leq\|w\|_2 R_{P,L}.
\end{equation}
Combining the two errors yields
\begin{equation}
 \left|w^{\mathsf T}\widehat z_t
       -w^{\mathsf T}\Mmap(Z_t^{(P,L)})\right|
 \leq\|w\|_2R_{P,L}
 +\left[\frac{2\log(2/\delta)}d
 \sum_{j=1}^J\frac{w_j^2}{N_j}\right]^{1/2},
 \label{eq:supp-truncation-proof-bound}
\end{equation}
which proves the main-text truncation and sampling proposition.

For a learned readout, the result holds conditionally on training data independent of the evaluation shots, with the input history also fixed. A union bound over $T_{\mathrm{eval}}$ specified evaluation times replaces $\delta$ by $\delta/T_{\mathrm{eval}}$; independence between these times is unnecessary. An estimated reference adds its own error. The factor $d^{-1/2}$ is a coordinate normalization, not a reduction in physical sampling cost: the physical Pauli readout weights are $w/\sqrt d$, and the shot budget per timestep is $\sum_jN_j$. This result certifies a fixed readout of the response, not estimation of the restricted matrix $O$, which additionally requires construction of the latent basis and its measured signatures.

Finally, put $r_t=Z_t-Z_t^{(P,L)}$. Centering the full and truncated responses about their respective input means gives the difference $r_t-\mathbb E[r_t]$, whose HS norm is at most $2R_{P,L}$. For the same fixed readout $w$, target and additive measurement noise, let $e_t$ and $e_t^{(P,L)}$ denote the two prediction errors. The $L^2$ triangle inequality yields
\begin{equation}
 \left|\|e_t\|_{L^2}-\|e_t^{(P,L)}\|_{L^2}\right|
 \leq\|e_t-e_t^{(P,L)}\|_{L^2}
 \leq2\|w\|_2R_{P,L}.
 \label{eq:supp-truncation-centered-rms-comparison}
\end{equation}
This comparison does not assume that omitted Volterra terms are unbiased noise or that the readout is refitted.
\end{proof}

\section{Resolution models and sampling uncertainty}
\label{sec:supp-resolution-sampling}

The geometric analysis defines OVD for any prescribed $\epsilon$. A physical sampling interpretation additionally requires an estimator and an error model. The reference threshold used in the numerical comparisons specifies a common sampling scale and is not an estimator-independent confidence guarantee.

\subsection{Singular-value perturbations and the separation threshold}
\label{subsec:supp-singular-perturbations}

If $\widehat O=O+E$ and $\|E\|_{\rm op}\le\epsilon$, singular-value perturbation theory gives~\cite{Bhatia1997}
\begin{equation}
 |\sigma_j(\widehat O)-\sigma_j(O)|\le\epsilon.
 \label{eq:supp-singular-value-perturbation}
\end{equation}
An exact strength above $2\epsilon$ remains above $\epsilon$ after the worst downward perturbation. A null strength can rise only to $\epsilon$. This explains the factor of two in OVD. A nonzero direction below the threshold can still be useful under a different estimator or task, so the count is specific to the stated resolution criterion.

\subsection{Physical Pauli sampling and coordinate normalization}
\label{subsec:supp-shot-normalization}

Physical Pauli outcomes are $x_{j,t}^{(a)}\in\{-1,1\}$, with mean $f_{j,t}=\Tr[(I_M\otimes P_j)\sigma_t]$. For $N$ independent outcomes, the sample mean satisfies
\begin{equation}
 \widehat f_{j,t}=\frac1N\sum_{a=1}^{N}x_{j,t}^{(a)},\qquad
 \operatorname{Var}(\widehat f_{j,t})=\frac{1-f_{j,t}^2}{N}.
 \label{eq:supp-pauli-feature-estimator}
\end{equation}
With $d=d_Md_R$, the HS-normalized measurement coordinate is $z_{j,t}=f_{j,t}/\sqrt d$. Its estimator $\widehat z_{j,t}=\widehat f_{j,t}/\sqrt d$ therefore has variance $(1-f_{j,t}^2)/(dN)$. Subtracting a known reference expectation leaves this variance unchanged. If the reference is separately estimated, its uncertainty and any covariance with the feature estimate must also be included.

For a restriction computed in physical Pauli coordinates, denote its matrix and resolution by $O_{\rm P}$ and $\epsilon_{\rm P}$. The main-text convention uses
\begin{equation}
 \begin{gathered}
 O=\frac{O_{\rm P}}{\sqrt d},\qquad
 \epsilon=\frac{\epsilon_{\rm P}}{\sqrt d},\\
 \#\{j\mid\sigma_j(O)>2\epsilon\}
 =\#\{j\mid\sigma_j(O_{\rm P})>2\epsilon_{\rm P}\}.
 \end{gathered}
 \label{eq:supp-pauli-ovd-normalization}
\end{equation}
Thus common rescaling preserves OVD when both the strengths and threshold are converted. The latent space and measurement kernel are unchanged, so latent dimension, algebraic rank and geometric visibility angles are also unchanged. For supported directions, the angles satisfy
\begin{equation}
 \sin\theta_j^+=\sigma_j(O)
 =\frac{\sigma_j(O_{\rm P})}{\sqrt d}.
 \label{eq:supp-pauli-angle-normalization}
\end{equation}
Absolute numerical cutoffs on the measured spectrum require the same conversion. This normalization changes coordinate magnitudes, not the physical sampling cost.

The reference-resolution model and its use in comparisons between different register sizes are specified in the main-text Methods subsection ``Resolution model and finite-shot calculations''.

\subsection{Statistical calibration of the resolution threshold}
\label{subsec:supp-theoretical-operational-threshold}

A statistically calibrated resolution depends on the estimator used for $O_{P,L}$. Relevant quantities include estimator ranges, variances, correlations and uncertainty in the latent basis. We first treat a fixed Hilbert--Schmidt-orthonormal latent basis of dimension $s$ and derive two confidence thresholds for the resulting $J\times s$ restriction matrix. Throughout, $0<\delta<1$.

Suppose each entry is estimated from $N_{\mathrm{shot}}$ independent unbiased samples $Z_{j\ell}^{(n)}\in[-1,1]$, satisfying
\begin{equation}
 \mathbb E[Z_{j\ell}^{(n)}]=(O_{P,L})_{j\ell}.
\end{equation}
Define
\begin{equation}
 (\widehat O_{P,L})_{j\ell}
 =\frac{1}{N_{\mathrm{shot}}}
 \sum_{n=1}^{N_{\mathrm{shot}}}Z_{j\ell}^{(n)},
 \qquad
 E_{P,L}:=\widehat O_{P,L}-O_{P,L}.
\end{equation}
Hoeffding's inequality~\cite{Hoeffding1963} gives, for every $t>0$,
\begin{equation}
 \mathbb P\left[
 |(E_{P,L})_{j\ell}|>t
 \right]
 \leq
 2\exp\left(-\frac{N_{\mathrm{shot}}t^2}{2}\right).
\end{equation}
Taking a union bound over all $Js$ entries,
\begin{equation}
 \begin{aligned}
 \mathbb P\left[\|E_{P,L}\|_{\max}>t\right]
 &\leq
 \sum_{j=1}^{J}\sum_{\ell=1}^{s}
 \mathbb P\left[|(E_{P,L})_{j\ell}|>t\right]\\
 &\leq
 2Js\exp\left(-\frac{N_{\mathrm{shot}}t^2}{2}\right).
 \end{aligned}
\end{equation}
Independence between different entries is not required for this union bound.

The norm inequalities
\begin{equation}
 \|E_{P,L}\|_{\mathrm{op}}
 \leq\|E_{P,L}\|_F
 \leq\sqrt{Js}\,\|E_{P,L}\|_{\max}
\end{equation}
then imply
\begin{equation}
 \mathbb P\left[\|E_{P,L}\|_{\mathrm{op}}>\epsilon\right]
 \leq
 2Js\exp\left(
 -\frac{N_{\mathrm{shot}}\epsilon^2}{2Js}
 \right).
\end{equation}
Setting the right-hand side equal to $\delta$ yields
\begin{equation}
 \mathbb P\left[
 \|\widehat O_{P,L}-O_{P,L}\|_{\mathrm{op}}
 >\tau_{\mathrm H}(N_{\mathrm{shot}},\delta)
 \right]\leq\delta,
\end{equation}
where
\begin{equation}
 \tau_{\mathrm H}(N_{\mathrm{shot}},\delta)
 :=
 \sqrt{\frac{2Js\log(2Js/\delta)}{N_{\mathrm{shot}}}}.
 \label{eq:supp-hoeffding-threshold}
\end{equation}
This threshold is conservative because it first controls every entry and then converts that bound into an operator-norm bound. For samples bounded in $[-b,b]$, the same derivation multiplies $\tau_{\mathrm H}$ by $b$.

The bounded-sample assumption concerns the estimator of each matrix entry. It does not follow solely from bounded physical Pauli outcomes. A latent basis element $H_\ell$ is a traceless operator rather than a density matrix, so estimating its measurement signature requires an appropriate preparation or reconstruction protocol.

A direct matrix bound is available when one sampling repetition produces an unbiased random matrix $X_n\in\mathbb R^{J\times s}$,
\begin{equation}
 \widehat O_{P,L}
 =\frac{1}{N_{\mathrm{shot}}}
 \sum_{n=1}^{N_{\mathrm{shot}}}X_n,
 \qquad
 \mathbb E[X_n]=O_{P,L}.
\end{equation}
Assume that the matrices are independent and identically distributed. Their centered fluctuations are
\begin{equation}
 Y_n:=X_n-O_{P,L},
 \qquad
 \mathbb E[Y_n]=0.
\end{equation}
Suppose a finite constant $L_B$ bounds every sampling outcome,
\begin{equation}
 \|Y_n\|_{\mathrm{op}}\leq L_B.
\end{equation}
Define the per-sample matrix variance parameter
\begin{equation}
 v_B:=
 \max\left\{
 \left\|\mathbb E[Y_nY_n^{\mathsf T}]\right\|_{\mathrm{op}},
 \left\|\mathbb E[Y_n^{\mathsf T}Y_n]\right\|_{\mathrm{op}}
 \right\}.
 \label{eq:supp-bernstein-variance-proxy}
\end{equation}
These two terms describe fluctuations in measurement-output and latent-coordinate space, respectively.

To apply a self-adjoint matrix concentration inequality, introduce the dilation
\begin{equation}
 \mathcal D(Y_n):=
 \begin{pmatrix}
 0&Y_n\\
 Y_n^{\mathsf T}&0
 \end{pmatrix}.
\end{equation}
Its nonzero eigenvalues occur in pairs equal to the positive and negative singular values of $Y_n$. Consequently,
\begin{equation}
 \lambda_{\max}(\mathcal D(Y_n))
 =\|\mathcal D(Y_n)\|_{\mathrm{op}}
 =\|Y_n\|_{\mathrm{op}}.
\end{equation}
Moreover,
\begin{equation}
 \mathcal D(Y_n)^2=
 \begin{pmatrix}
 Y_nY_n^{\mathsf T}&0\\
 0&Y_n^{\mathsf T}Y_n
 \end{pmatrix},
\end{equation}
so
\begin{equation}
 \left\|\mathbb E[\mathcal D(Y_n)^2]\right\|_{\mathrm{op}}
 =v_B.
\end{equation}

Let
\begin{equation}
 S_N:=\sum_{n=1}^{N_{\mathrm{shot}}}Y_n.
\end{equation}
Linearity of the dilation gives
\begin{equation}
 \mathcal D(S_N)
 =\sum_{n=1}^{N_{\mathrm{shot}}}\mathcal D(Y_n),
 \qquad
 \lambda_{\max}(\mathcal D(S_N))=\|S_N\|_{\mathrm{op}}.
\end{equation}
The variance parameter of this sum is
\begin{equation}
 \begin{aligned}
 \sigma_B^2
 &:=
 \left\|
 \sum_{n=1}^{N_{\mathrm{shot}}}
 \mathbb E[\mathcal D(Y_n)^2]
 \right\|_{\mathrm{op}}\\
 &=N_{\mathrm{shot}}v_B.
 \end{aligned}
\end{equation}
Applying the upper-tail self-adjoint matrix Bernstein inequality to the $(J+s)$-dimensional dilation yields~\cite{Tropp2012}
\begin{equation}
 \mathbb P\left[\|S_N\|_{\mathrm{op}}>t\right]
 \leq
 (J+s)\exp\left[
 -\frac{t^2}{
 2N_{\mathrm{shot}}v_B+\frac23L_Bt}
 \right].
 \label{eq:supp-bernstein-tail}
\end{equation}
The prefactor is $J+s$ because the largest eigenvalue of the dilation already equals the rectangular operator norm.

To obtain an explicit confidence threshold, set
\begin{equation}
 \eta:=\log\left(\frac{J+s}{\delta}\right),
 \qquad
 t:=\sqrt{2N_{\mathrm{shot}}v_B\eta}
 +\frac{2L_B\eta}{3}.
\end{equation}
This choice satisfies
\begin{equation}
 t^2-\eta\left(
 2N_{\mathrm{shot}}v_B+\frac23L_Bt
 \right)=
 \frac{2L_B\eta}{3}
 \sqrt{2N_{\mathrm{shot}}v_B\eta}
 \geq0.
\end{equation}
Therefore the exponent in Eq.~\eqref{eq:supp-bernstein-tail} is at most $-\eta$, giving
\begin{equation}
 \mathbb P\left[
 \|S_N\|_{\mathrm{op}}>
 \sqrt{2N_{\mathrm{shot}}v_B\eta}
 +\frac{2L_B\eta}{3}
 \right]\leq\delta.
\end{equation}
Since $\widehat O_{P,L}-O_{P,L}=S_N/N_{\mathrm{shot}}$, division by $N_{\mathrm{shot}}$ gives
\begin{equation}
 \mathbb P\left[
 \|\widehat O_{P,L}-O_{P,L}\|_{\mathrm{op}}
 >\tau_{\mathrm B}(N_{\mathrm{shot}},\delta)
 \right]\leq\delta,
\end{equation}
with
\begin{equation}
 \tau_{\mathrm B}(N_{\mathrm{shot}},\delta)
 :=
 \sqrt{
 \frac{2v_B\log[(J+s)/\delta]}{N_{\mathrm{shot}}}
 } +
 \frac{2L_B\log[(J+s)/\delta]}{3N_{\mathrm{shot}}}.
 \label{eq:supp-bernstein-threshold}
\end{equation}
The first term is the variance-controlled $N_{\mathrm{shot}}^{-1/2}$ contribution, while the second is a bounded-fluctuation correction of order $N_{\mathrm{shot}}^{-1}$. The matrix bound can improve on the entrywise estimate when the variance parameter is substantially smaller than the worst-case entrywise scale.

Correlations between entries within one matrix sample are allowed, but must be included in $v_B$. Independence is required between the matrix samples $X_n$. Shared measurements and repeated feature slots must therefore be represented consistently when constructing these samples. Also, $N_{\mathrm{shot}}$ counts samples per entry in the first protocol and complete matrix samples in the second; their physical measurement costs need not be equal.

These inequalities calibrate a resolution threshold for the specified estimator with a fixed latent basis. Additional uncertainty in the latent basis or estimator bias must be controlled separately. The model-computed spectra and reference resolution used in the numerical comparisons are described in the main-text Methods.

\section{Equivalence of the direct and joint-Gram constructions}
\label{sec:supp-gram-equivalence}

The direct and Gram representations describe the same measurement geometry when they use the same retained joint-kernel subspace. Let $H_{P,L}$ contain the joint-kernel coordinates in a Hilbert--Schmidt-orthonormal ambient basis, and let $R$ represent the normalized measurement map in that basis. The kernel Gram matrix and measured coefficient matrix satisfy
\begin{equation}
 G=H_{P,L}^{\mathsf T}H_{P,L},\qquad C=RH_{P,L}.
 \label{eq:supp-gram-factorization}
\end{equation}
Let $V$ contain $s$ orthonormal eigenvectors of $G$ associated with the retained positive eigenvalues in $\Lambda$, so $GV=V\Lambda$ and $V^{\mathsf T}V=I_s$. Define
\begin{equation}
 H_\ell=\frac{1}{\sqrt{\Lambda_{\ell\ell}}}\sum_\lambda V_{\lambda\ell}\widehat h_\lambda,\qquad \ell=1,\ldots,s.
 \label{eq:supp-gram-basis}
\end{equation}
Their inner products are
\begin{equation}
 \begin{aligned}
 \Tr(H_\ell H_k)&=\frac{\sum_{\lambda,\mu}V_{\lambda\ell}G_{\lambda\mu}V_{\mu k}}{\sqrt{\Lambda_{\ell\ell}\Lambda_{kk}}}\\
 &=\bigl[\Lambda^{-1/2}V^{\mathsf T}GV\Lambda^{-1/2}\bigr]_{\ell k}=\delta_{\ell k}.
 \end{aligned}
 \label{eq:supp-gram-orthonormality}
\end{equation}
Thus the operators $H_\ell$ form an orthonormal basis of the retained kernel subspace. The coordinate matrix of this basis and its measurement restriction are
\begin{equation}
 Q_{P,L}=H_{P,L}V\Lambda^{-1/2},\qquad RQ_{P,L}=CV\Lambda^{-1/2}.
 \label{eq:supp-gram-restriction}
\end{equation}
Any other orthonormal basis of the same subspace has the form $Q_{P,L}W$ for an orthogonal matrix $W$. Its restricted measurement matrix is $O_{P,L}W$, which has the same singular values as $O_{P,L}$. The direct and Gram representations therefore give the same latent dimension, visibility spectrum, VVR and OVD for the same retained subspace and resolution criterion.

\end{document}